\documentclass[traditabstract]{aa}  
\usepackage{graphicx}
\usepackage{textcomp}
\usepackage{lscape}
\usepackage[varg]{txfonts}
\usepackage{mathrsfs}
\usepackage{amssymb}
\usepackage{natbib}
\usepackage[colorlinks, citecolor={blue}]{hyperref}
\usepackage{color}
\authorrunning{S.~Duarte~Puertas et al.,}
\titlerunning{Star formation regions in Stephan's Quintet: I. Structures and kinematics}

\begin{document} 

   \title{Searching for intergalactic star forming regions in Stephan's Quintet with SITELLE. I. Ionised gas structures and kinematics}

\author{
S.~Duarte~Puertas\inst{1}
\and 
J.~Iglesias-P\'{a}ramo\inst{1,2}
\and
J.~M.~Vilchez\inst{1}
\and
L. Drissen\inst{3,4}
\and
C.~Kehrig\inst{1}
\and
T. Martin\inst{3,4}
}
\institute{
Instituto de Astrof\'{\i}sica de Andaluc\'{\i}a - CSIC, Glorieta de la Astronom\'{\i}a s.n., 18008 Granada, Spain\label{inst1} \\ \email{salvini@iaa.es}
\and
Estaci\'{o}n Experimental de Zonas \'{A}ridas - CSIC, Ctra. de Sacramento s.n., La Ca\~{n}ada, Almer\'{\i}a, Spain\label{inst2}
\and
D\'epartement de physique, de g\'enie physique et d'optique, Universit\'e Laval, Qu\'ebec (QC), G1V 0A6, Canada\label{inst3}
\and
Centre de recherche en astrophysique du Qu\'ebec\label{inst4}
}

   \date{Received \today; accepted \today}

 \abstract 
{Stephan's Quintet (SQ), the prototypical compact group of galaxies in the local Universe, has been observed with the imaging Fourier transform spectrometer SITELLE, attached to the Canada-France-Hawaii-Telescope, to perform a deep search for intergalactic star-forming emission. In this paper we present the extended ionised gaseous structures detected and analyse their kinematical properties. The large field of view (11'x11') and the spectral ranges of SITELLE have allowed a thorough study of the entire galaxy system, its interaction history and the main properties of the ionised gas. The observations have revealed complex three-dimensional strands in SQ seen for the first time, as well as the spatially resolved velocity field for a new SQ dwarf galaxy (M82-like) and the detailed spectral map of NGC7320c, confirming its AGN nature. A total of 175 SQ H$\alpha$ emission regions have been found, 22 of which present line profiles with at least two kinematical components. We studied 12 zones and 28 sub-zones in the SQ system in order to define plausible physical spatial connections between its different parts in the light of the kinematical information gathered. In this respect we have found five velocity systems in SQ: i) v=[5600-5900] $km\, s^{-1}$ associated with the new intruder and the southern debris region; ii) v=[5900-6100] $km\, s^{-1}$, associated with the north starburst A and south starburst A and the strands connected to these zones; iii) v=[6100-6600] $km\, s^{-1}$, associated with the strands from the large-scale shock region (LSSR); iv) v=[6600-6800] $km\, s^{-1}$, associated with the young tidal tail, the starburst A (SQA), NGC7319, and the NGC7319 north lobe; and v) v=[6800-7000] $km\, s^{-1}$, associated with the strands seen connecting LSSR with SQA. We fail to detect ionised gas emission in the old tail, neither in the vicinity of NGC7318A nor in NGC7317, and the connection between NGC7319 north lobe and SQA cannot be confirmed. Conversely, a clear gaseous bridge has been confirmed both spatially and kinematically between the LSSR zone and the NGC7319 AGN nucleus. Finally, a larger scale, outer rim winding the NGC7318B/A system clockwise north-west to south-east has been highlighted in continuum and in H$\alpha$. This structure may be reminiscent of a sequence of a previously proposed scenario for SQ a sequence of individual interactions.}


   \maketitle
\section{Introduction}
Stephan's Quintet \citep[SQ,][]{1877MNRAS..37..334S} is one of the most studied galaxy aggregates and it represents a unique laboratory for understanding how a compact group evolves. SQ is composed of elliptical (NGC7317 and NGC7318A) and spiral galaxies (NGC7319, NGC7320c, and NGC7318B). NGC7318B is a gas-rich galaxy that has experienced several interactions throughout its evolution \citep[e.g.][]{1997ApJ...485L..69M,2001AJ....122.2993S}. One foreground spiral galaxy (NGC7320) with a discordant redshift \citep{1961ApJ...134..244B} can also be seen in the field of view (FoV) of SQ.

\cite{1997ApJ...485L..69M} and \cite{2001AJ....122.2993S} suggested that compact groups of galaxies are formed by acquiring intruder galaxies associated with a large-scale structure. All the galaxies of SQ, except NGC7318B, have evolved dynamically, losing their interstellar medium (ISM). \cite{1997ApJ...485L..69M} proposed an evolutionary model for SQ based on two intruders, an old intruder (OI, NGC7320c) and a new intruder (NI, NGC7318B). SQ is at an advanced state of interaction, evidenced by the existence of, for example, at least young and old tidal tails with ages of 150-200 Myr and 400-500 Myr, respectively \citep{2011AJ....142...42F}; a large-scale shock region \citep[LSSR, e.g.][]{1972Natur.239..324A,1998ApJ...492L..25O,2012A&A...539A.127I} produced by an ongoing interaction between NGC7318B and both NGC7319 and debris material from previous interactions; a diffuse gas halo close to NGC7317 composed by old stars that indicates that the formation of SQ took place several Gyr ago \citep{2018MNRAS.475L..40D}.

From the kinematical point of view, SQ covers a large radial velocity range ($\sim$5500 -- $\sim$6600 $km\, s^{-1}$). The core of SQ (i.e. NGC7317, NGC7318A, and NGC7319) and NGC7320c show a mild radial velocity dispersion, whereas NGC7318B has an unusually high radial velocity dispersion compared to the others ($\Delta V\sim$1000 $km\, s^{-1}$), indicating that it is not bound to the SQ \citep{2005A&A...444..697T}. Classically, two radial velocity structures were detected by \cite{2001AJ....122.2993S}: the gas associated with the SQ and the gas associated with NI, where most of the emission detected in the ISM comes from the NI. \cite{2012A&A...539A.127I} observed three radial velocity components but could only focus on the lower part of the shock. The LSSR spans a wide range of redshift \citep[e.g.][]{2014ApJ...784....1K}. Its structure and gas distribution have not been entirely examined until now due to the inability to fully study the system (from NGC7320c to NGC7317).

SQ has been widely researched with different wavelengths by many authors: with radio \citep[e.g.][]{2002AJ....123.2417W,2004A&A...426..471L,2012ApJ...749..158G}; ultraviolet \citep[e.g.][]{2005ApJ...619L..95X,2012MNRAS.426.2441D}; X-ray \citep[e.g.][]{2003A&A...401..173T,2012MNRAS.424.1563H}; infrared \citep[e.g.][]{2006ApJ...639L..51A,2017ApJ...836...76A,2010ApJ...710..248C,2009A&A...502..515G}; and visible \citep[e.g.][]{1998A&A...334..473M,2001ApJ...550..204I,2001AJ....122.2993S,2004ApJ...605L..17M,2012A&A...539A.127I,2012ApJ...748..102T,2014ApJ...784....1K,2014MNRAS.442..495R}. There are also photometric studies \citep[e.g.][]{2001AJ....122..163G,2011AJ....142...42F,2018MNRAS.475L..40D} and simulations \citep[e.g.][]{2010ApJ...724...80R,2012MNRAS.419.1780H}. Finally, several previous interferometric Fabry-Perot studies of the SQ \citep[e.g.][]{2001AJ....122.2993S,2001AJ....121.2524M} have been done with diverse spatial coverages in wavelength-velocity resolution, though sometimes making difficult the discrimination between the emission of nearby lines at different velocities. 

This work is based on the spectroscopic data from SITELLE \citep{2012SPIE.8446E..0UG}, an imaging Fourier transform spectrometer (IFTS), attached to the Canada-France-Hawaii Telescope (CFHT). SITELLE allows us to study, for the first time, a large FoV considering the optical spectrum ranging from the UV atmosphere limit to the red, in order to study the physical, chemical, and kinematical properties of the entire SQ area. Our study presents an extensive spectral mapping that covers from the inner SQ 'core' to the large gas halo surrounding the SQ. SITELLE observations offer us a great opportunity to perform an unbiased search for the star forming and tidal emission regions (e.g. tidal dwarf galaxies, TDG) present in the SQ in order to study their origin and properties. These data also provide a detailed spectroscopic mapping of the extended shock zone and its associated emission, with a seeing limited spatial resolution ($\sim$0.8$^{\prime\prime}$). This is relevant to obtain a more complete view of the ionisation structure of the shock and its environment, discriminating in velocity space the shock ionised gas from other intervening gaseous material (e.g. HII regions from the arms of the central spiral galaxies, gas associated with the AGN main galaxy).

The structure of this paper is organised as follows: in Sect.~\ref{sec:data} we describe the data and the methodology used to select the H$\alpha$ emission regions from the SQ. We detail our main results in Sect.~\ref{sec:results}. Finally, the discussion and main conclusions of our work are presented in Sect.~\ref{sec:conclu}. Line fluxes within the SN1, SN2, and SN3 data cubes, and the study of the physical properties and excitation conditions of the gas, will be presented in a future paper (Duarte Puertas et al. paper II, in prep). Throughout the paper, we assume a Friedman-Robertson-Walker cosmology with $\Omega_{\Lambda 0}=0.7$, $\Omega_{\rm m 0}=0.3$, and $\rm H_0=70\,km\,s^{-1}\,Mpc^{-1}$.

\section{Observations and data analysis}
\label{sec:data}

\begin{figure}
    \centering
    \includegraphics[width=\columnwidth]{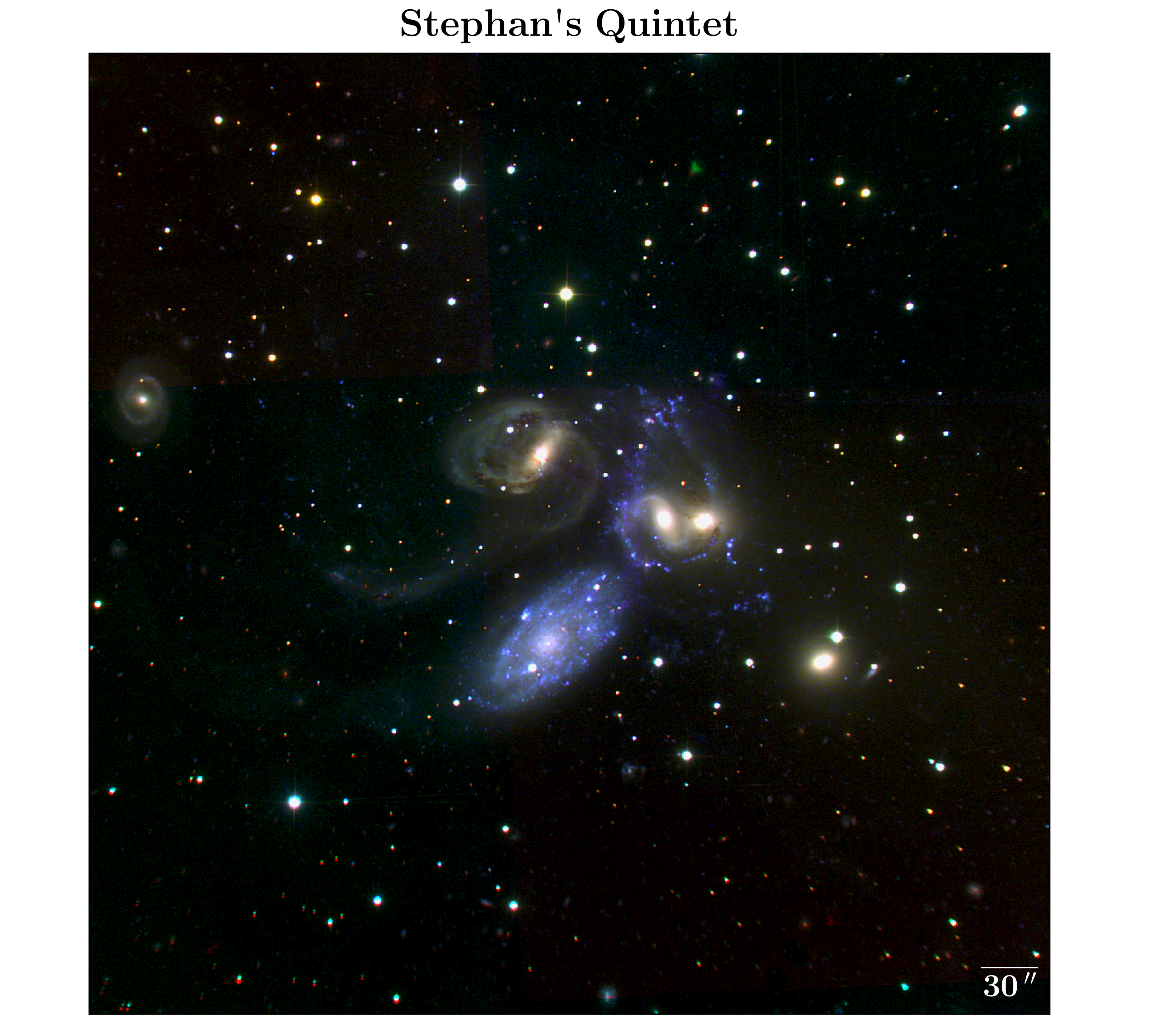}
    \caption{SITELLE deep-colour image of SQ composed using SN1, SN2, and SN3 data cubes. North is top and east is left. The distance considered for SQ in this paper is d = 88.6 Mpc (from the NASA/IPAC Extragalactic Database known as NED). At the distance of SQ, 30$^{\prime\prime}$ corresponds to $\sim$13.04 kpc.}
    \label{fig:SQ}
\end{figure}

\subsection{Observations}
\label{subsec:observ}
SITELLE is an IFTS that aims to study the spatially-resolved spectra of extended sources. It has a FoV of 11$^\prime$x11$^\prime$ \citep[0.32$^{\prime\prime}$ per pixel,][]{2019MNRAS.485.3930D}, suitable for investigating complex and extensive systems of galaxies such as SQ. The observations were carried out in August 2015 and July 2016 (P.I. Drissen) under an average seeing of $\sim$0.8$^{\prime\prime}$. The three filters SN1, SN2, and SN3 were used to capture the emission lines of our interest: [\ion{O}{ii}]$\lambda$3727, [\ion{O}{iii}]$\lambda\lambda$4959,5007, H$\beta$, [\ion{N}{ii}]$\lambda\lambda$6548,6583, [\ion{S}{ii}]$\lambda\lambda$6716,6731, and H$\alpha$. Table~\ref{table:table1} shows the observing parameters (right ascension or RA, declination or DEC, observing date, spectral range, number of steps, mean resolution, and the total exposure time) considered for each filter. Figure~\ref{fig:SQ} shows a SITELLE deep-colour image of the SQ field composed using the integrated emission on the SN1, SN2, and SN3 data cubes \citep[see][Sect. 5.4]{2019MNRAS.485.3930D}. We worked the data cubes in wavenumbers ($cm^{-1}$), which is SITELLE's natural units.

\begin{table}
\tiny
\caption{Observing parameters.}
\label{table:table1} 
\centering
\begin{tabular}{c | c c c }
\hline\hline  \\[-2ex]
(1) & (2) & (3) & (4) \\[0.5ex] 
Filter & SN1 & SN2 & SN3 \\[0.5ex] 
\hline\\[-2ex]
RA (h m s) & 22:36:05.30 & 22:36:05.30 & 22:36:05.84\\[0.5ex]
DEC ($^o\ ^\prime\ ^{\prime\prime}$) & 33:58:59.9 & 33:58:59.9 & 33:59:09.6\\[0.5ex]
Observing date & 2016-07-09 & 2016-07-10 & 2015-08-07\\[0.5ex]
 &  &  & 2015-08-08\\[0.5ex]
Spectral range T > 90\% (nm) & 363-386 & 482-513 & 648-685\\[0.5ex]
Number of steps & 85 & 169 & 342\\[0.5ex]
Exp. time/step (s) & 75 & 65 & 25\\[0.5ex]
Total exp. time (h) & 1.77 & 3.05 & 2.37\\[0.5ex]
Mean resolution, R & 500 & 760 & 1560\\
\hline
\end{tabular}
\end{table}

\begin{figure}
    \centering
    \includegraphics[width=\columnwidth]{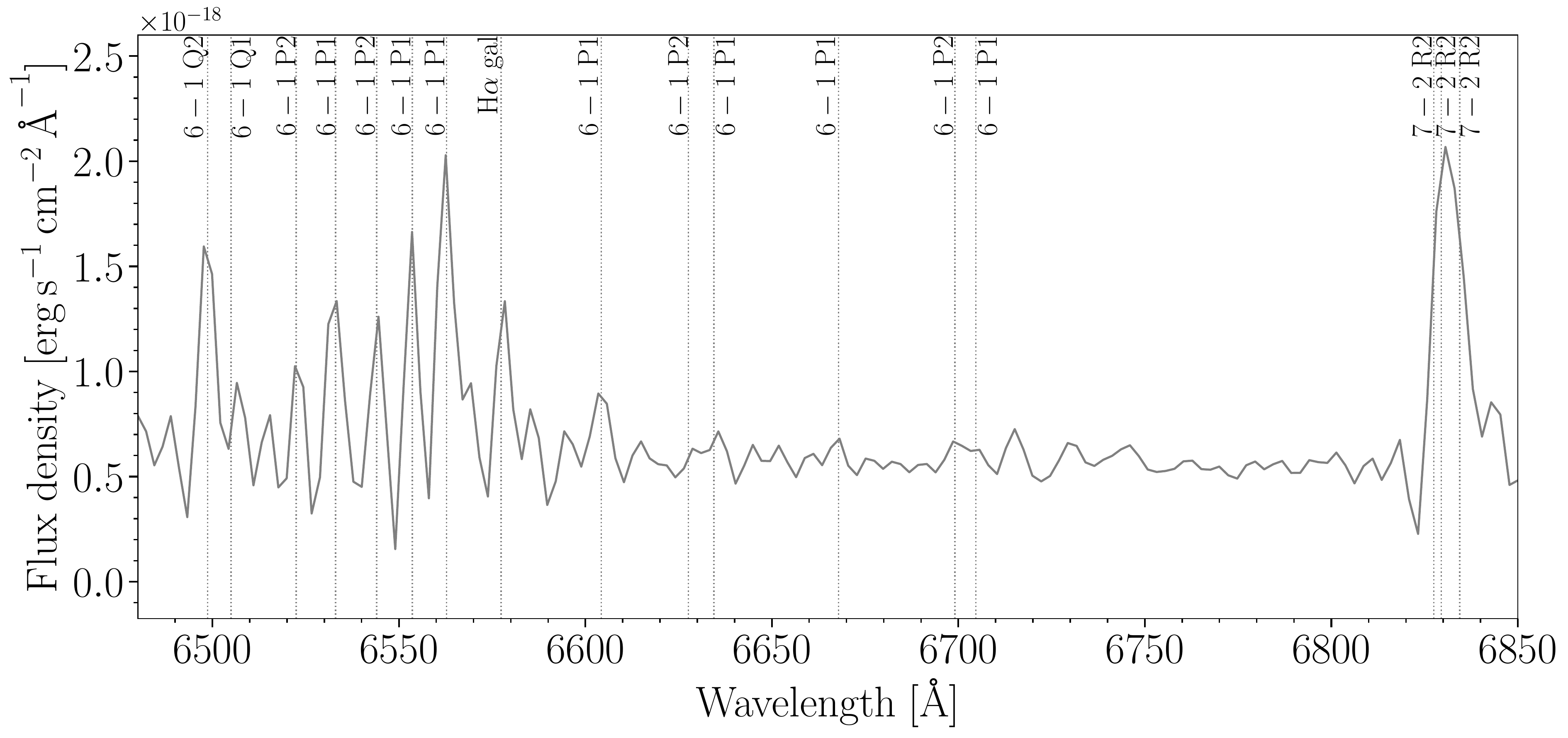}
    \caption{Median sky spectrum in SN3 data cube (2700 pixels have been combined). The dashed vertical lines show multiple known OH sky lines.}
    \label{fig:sky}
\end{figure}

\subsection{Reduction, calibration, and measurements}
\label{subsec:reduc}
The data were fitted using the Python-based software \textsc{ORCS} \citep{2015ASPC..495..327M}. Data were recalibrated in wavelength using the OH sky lines that are visible in most parts of the FoV. We fitted the OH sky lines using the function \textsc{SpectralCube.map\_sky\_velocity} and the resulting corrections were implemented to the cube using \textsc{SpectralCube.correct\_wavelength} \citep{2018MNRAS.473.4130M}. For each data cube, we subtracted the sky using the median sky spectrum from three dark regions (2700 pixels were used). For illustrative purposes, we show the median sky spectrum in the SN3 data cube in Fig.~\ref{fig:sky}. In order to obtain a more precise radial velocity correction we used the function \textsc{get\_radial\_velocity\_correction}. Through this function, we found a shift of 15 $km\, s^{-1}$ that was added to the measured velocity. The flux calibration was performed using the data cube of the standard star GD71. The Galactic extinction is very small, E(B - V) = 0.07, according to the NASA/IPAC Extragalactic Database known as NED\footnote{\texttt{http://ned.ipac.caltech.edu/}}, and we did not correct for it. The methodology used to analyse the data involved several steps that we describe below.

\subsubsection{Searching for H$\alpha$ emission}
We considered a binning 3x3 to amplify the signal-to-noise ratio ($\sim$450 000 pixels). In addition a study was done only in H$\alpha$ for a binning 6x6 (see Sect.~\ref{subsec:diff})\footnote{Binning 6x6 was chosen after an iterative process to maximise the the signal-to-noise ratio while minimising the contribution of the background sky.} in order to unveil the lowest surface brightness gaseous emission in SQ. Subsequently, we searched for the wavenumber position ($k_{max}$) where the maximum intensity peak of the H$\alpha$ emission line ($A_{max}$) is found in the typical wavenumber range of SQ ($\Delta K_{SQ} = \left[14820, 15021 \right]cm^{-1}$) and NGC7320 ($\Delta K_{NGC7320} = \left[15147, 15239 \right]cm^{-1}$), selecting the largest $A_{max}$ between them. Once we found the position of the maximum intensity, we derived the standard deviation when the transmission curve is higher than 90\% for the SN3 filter ($std_{90}$). To compute $std_{90}$, the typical wavenumber range for the emission lines H$\alpha$, [\ion{N}{ii}]$\lambda\lambda$6548,6583, and [\ion{S}{ii}]$\lambda\lambda$6716,6731 of each pixel have not been considered. We selected those pixels with $\frac{A_{max}}{std_{90}} \geq 2.5$ ($\sim$25 800 pixels). We assumed that the maximum value corresponds to the H$\alpha$ position. We know that the maximum in the AGN galaxy nucleus corresponds to [\ion{N}{ii}]$\lambda$6583 and this effect will be corrected later. We used the $k_{max}$ value to derive the possible initial velocity for H$\alpha$ ($v_{ini}(H\alpha)$) according to the following equation: $v_{ini}(H\alpha)=\left[\left(\left(\frac{1e7}{k_{max}\cdot \lambda(H\alpha)\cdot 0.1}\right)-1\right) \cdot c\right]$, where $\lambda(H\alpha)= 6562.8$ nm and $c=299792\,km\, s^{-1}$.

\subsubsection{Identifying H$\alpha$ emission regions}
\begin{figure*}
    \centering
    \vspace{-2cm}
    \includegraphics[width=\textwidth]{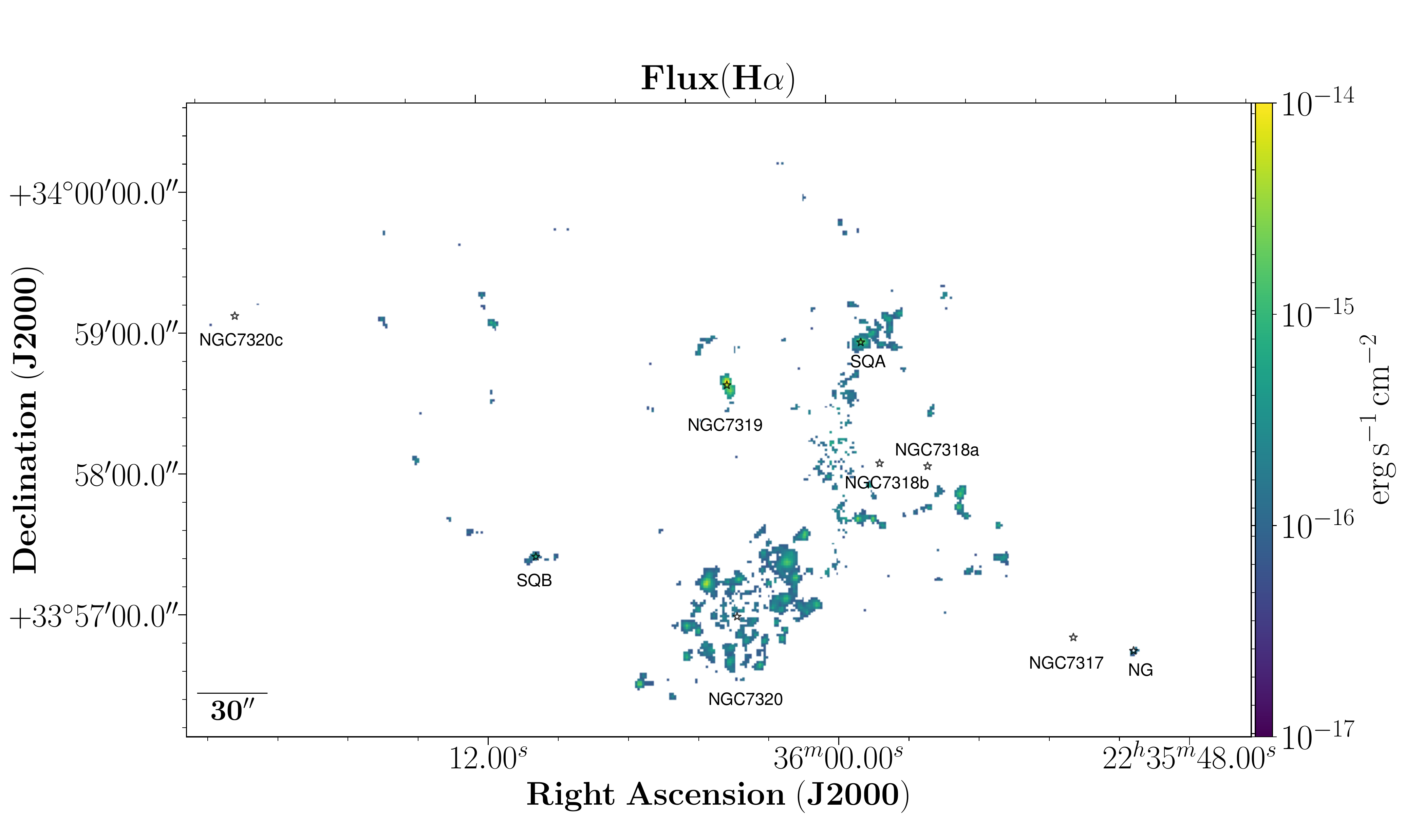}
    \includegraphics[width=\textwidth]{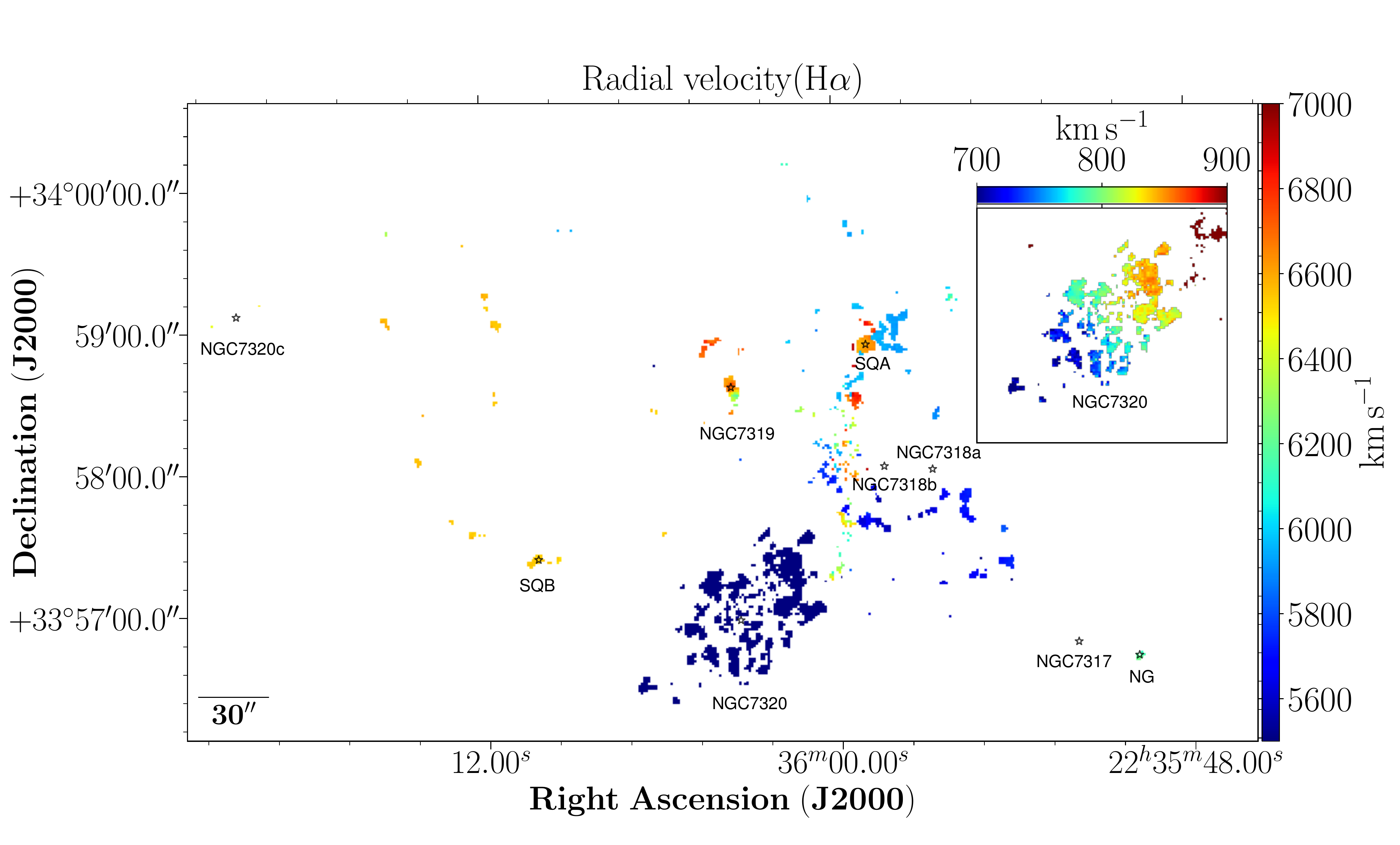}
    \caption{(Top panel) H$\alpha$ flux map of SQ considering pixels with H$\alpha$ contrast $\geq$ 5. (Bottom panel) H$\alpha$ radial velocity map of SQ considering pixels with contrast(H$\alpha$) $\geq$ 5. The inset plot in the right panel indicates the radial velocity map of NGC7320, the foreground galaxy, with contrast(H$\alpha$) $\geq$ 5 (see the text for details). NG (new dwarf galaxy) indicates the position of the new dwarf galaxy in SQ (see Sect.\ref{subsec:glx_SQ}).}
    \label{fig:velocity_map_Ha}
\end{figure*}

We fitted the spectrum for every selected pixel using the $v_{ini}(H\alpha)$ derived above. Initially, we fitted each spectrum to a sincgauss function (the convolution of a Gaussian with a sinc function) for H$\alpha$ and [\ion{N}{ii}]$\lambda\lambda$6548,6583 simultaneously. The theoretical relation [\ion{N}{ii}]$\lambda$6548/6583=0.333 is considered. When the broadening of the H$\alpha$ line was lower than 70 $km\, s^{-1}$, we adjusted the spectrum using a sinc function \citep[the instrumental line shape,][]{2016MNRAS.463.4223M}. When no emission was detected in [\ion{N}{ii}]$\lambda$6583, we refitted the spectrum only for the H$\alpha$ emission line. The output parameters are the radial velocity, broadening, intensity peak, flux, and the corresponding uncertainties, as well as the standard deviation of each spectrum. After fitting the H$\alpha$ line, we selected the pixels with $contrast(H\alpha)=\frac{A_{fit}}{std_{90}} \geq 5$, where $A_{fit}$ is the intensity peak of the H$\alpha$ emission line obtained from the fitting, and $v_{ini}(H\alpha) \geq 5000\,km\, s^{-1}$ ($\sim$950 pixels). Figure~\ref{fig:velocity_map_Ha} shows the H$\alpha$ flux (upper panel) and H$\alpha$ radial velocity (lower panel) maps for the $\sim$2100 pixels (binning 3x3) with contrast(H$\alpha$) $\geq$ 5 and without velocity restrictions. We detect H$\alpha$ emission in several galaxies: NGC7319, NGC7320c, NGC7320 (the foreground galaxy with an average radial velocity of 786 $km\, s^{-1}$, see inset plot in the lower panel for more details), and the new dwarf galaxy in SQ (NG, see Sect.\ref{subsec:glx_SQ} and ID. 176 in Table~\ref{table:table2}). Also, we detect H$\alpha$ emission from the starbursts A and B \citep[SQA and SQB, e.g.][]{1999ApJ...512..178X}. Additionally, this figure shows emission from the young tidal tail \citep[YTT, e.g.][]{2002A&A...394..823L}, LSSR, and the NI \citep[e.g.][]{1997ApJ...485L..69M}. We also detect emission to the left of NGC7317; this region is discussed in depth in Sect.~\ref{subsec:glx_SQ}. From the sample of pixels selected above, we searched for all the H$\alpha$ emission region candidates associated with the SQ. To do so, we first grouped the spatially associated emitting pixels. Then, the individual H$\alpha$ regions associated with each group of pixels are defined searching for the local emission peaks in the H$\alpha$ map using the Python-package \textsc{astrodendro}\footnote{This tool estimates the background in our images and we searched the regions that have FWHM $\sim$9 pixels and have peaks $\sim\,3\sigma$ above the background, from our sample of selected pixels.}; The membership of each group of pixels to a given individual H$\alpha$ region has been defined from the segmentation method using the Python packages \textsc{scipy.ndimage} and \textsc{skimage.morphology.watershed}. This method allows us to define a population of 209 SQ H$\alpha$ region candidates.

\subsubsection{One velocity component analysis for each SQ H$\alpha$ emission region}
\begin{figure}
    \centering
    \includegraphics[width=\columnwidth]{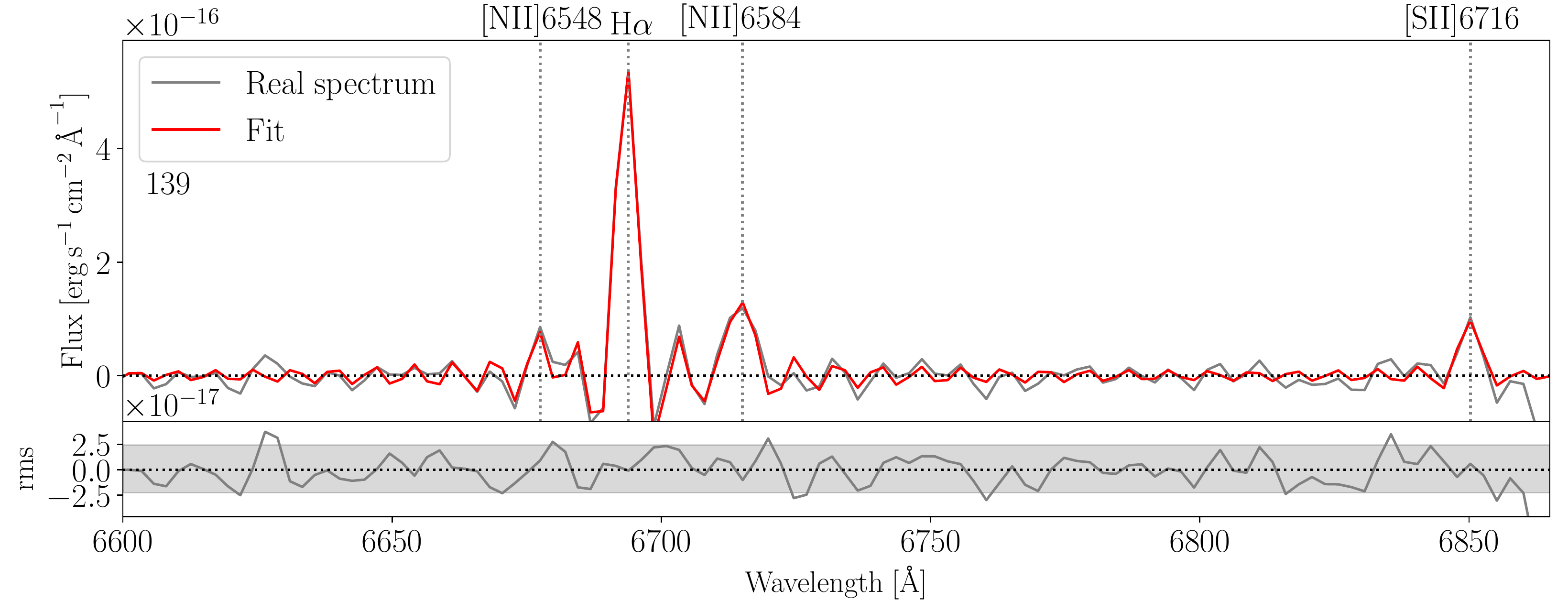}\\\vspace{-0.3cm}
    \includegraphics[width=\columnwidth]{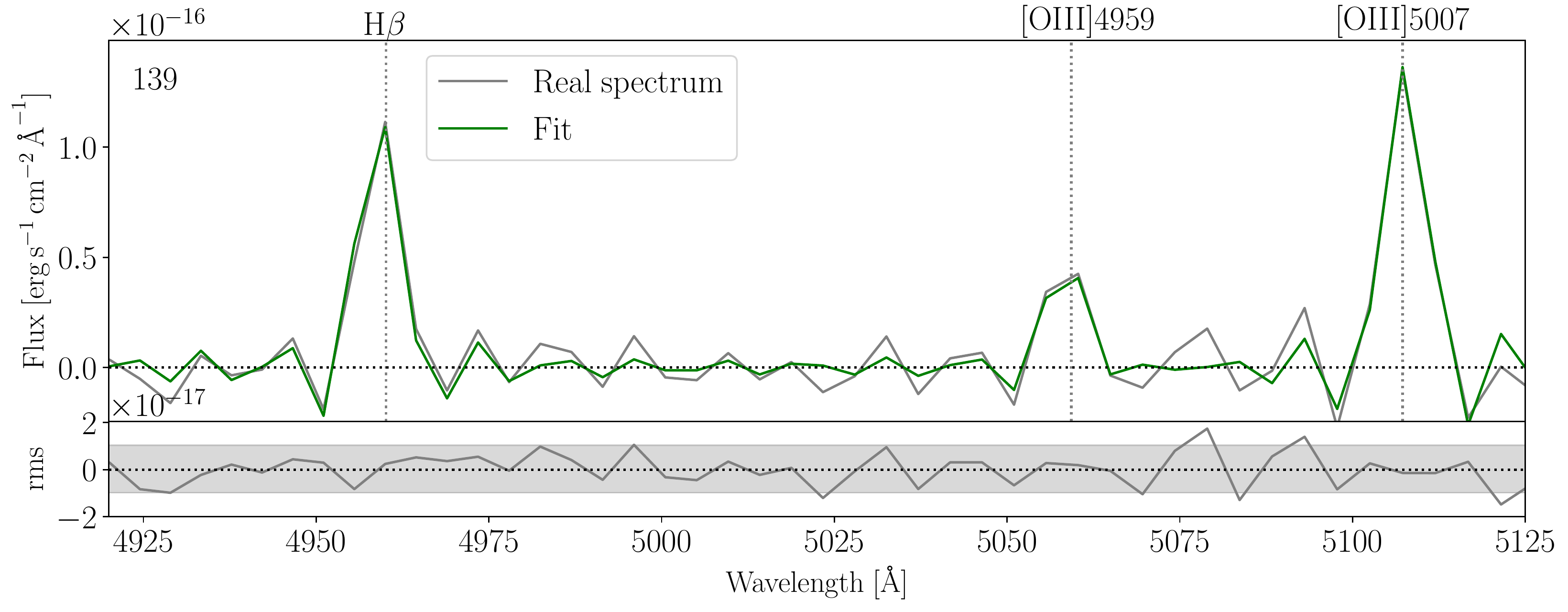}\\\vspace{-0.3cm}
    \includegraphics[width=\columnwidth]{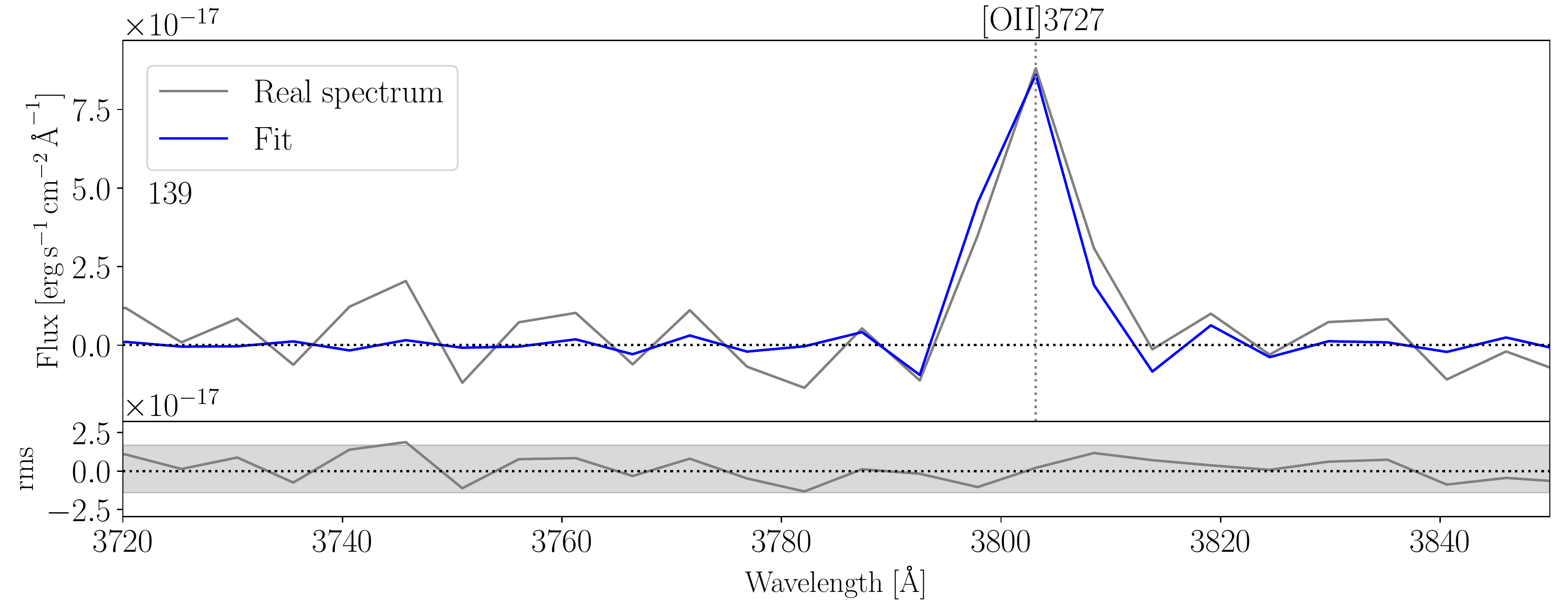}\\
    \caption{Fitting example: Spectra from region 139. In each upper panel, the grey line shows the real spectrum and the red, green, and blue coloured lines show the fit obtained with ORCS in the SN3, SN2, and SN1 filters, respectively. In the upper panels, the grey dotted lines show the location of the emission lines studied for this region. In the lower panels, the grey line shows the residual after the fit. The horizontal grey band indicates the 3$\sigma$ scatter.}
    \label{fig:spectro139}
\end{figure}

For each H$\alpha$ emission region we obtained the integrated spectrum adding the flux of its pixels; then all integrated spectra were fitted using the \textsc{ORCS} function \textsc{orcs.fit\_lines\_in\_integrated\_region}. The spectral lines H$\alpha$ and [\ion{N}{ii}]$\lambda\lambda$6548,6583 are fitted to a sincgauss function simultaneously, and when broadening of the H$\alpha$ line is smaller than 70 $km\, s^{-1}$, a sinc function was used, as detailed above. As the beginning of the SN3 transmission curve coincides with the emission lines [\ion{S}{ii}]$\lambda\lambda$6716,6731 at SQ's redshift, we only take into account in the fit the spectral line [\ion{S}{ii}]$\lambda$6716 for those H$\alpha$ emission regions with radial velocities lower than 6260 $km\, s^{-1}$, and [\ion{S}{ii}]$\lambda$6731 for those H$\alpha$ emission regions with velocities lower than 5760 $km\, s^{-1}$. The output parameters are the radial velocity, broadening, intensity peak, total flux, and the corresponding uncertainties, as well as the standard deviation of each spectrum. The emission lines from the SN1 and SN2 data cubes were fitted using the velocity found in the previous step for H$\alpha$ as a reference. It is important to note that we assumed the H$\alpha$ radial velocity as the representative velocity for every H$\alpha$ emission region. We followed the same methodology as previously described. For SN1 we fitted the spectral line [\ion{O}{ii}]$\lambda$3727. For SN2 we fitted the lines $H\beta$ and [\ion{O}{iii}]$\lambda\lambda$4959,5007 in the wavenumber range [19493, 20746]cm$^{-1}$ (assuming the theoretical relation [\ion{O}{iii}]$\lambda$4959/5007=0.333). The standard deviation for the SN1 and SN2 filters ($std_{90}$) was derived as above. Figure~\ref{fig:spectro139} illustrates examples of the line fits for the SN1, SN2, and SN3 filters. All the line fluxes fitted in the SN1, SN2, and SN3 data cubes will be presented in a future paper (Duarte Puertas et al. paper II, in prep).

\subsubsection{Complex kinematics spectral fitting of emission regions}
The one-velocity-component spectral fitting procedure produced unsatisfactory results for a selection of emission regions. A close visual inspection of these fits showed complex kinematics, with large velocity dispersion for the emission regions associated with the AGN galactic nucleus of NGC7319 and the shock region (see Fig.~\ref{fig:velocity_map_Ha} lower panel). Taking into account our spectral resolution and the signal-to-noise ratio, we studied only two velocity components for those emission regions with the velocity components resolved, as follows: 

The H$\alpha$, [\ion{N}{ii}]$\lambda\lambda$6548,6583, and [\ion{S}{ii}]$\lambda\lambda$6716,6731 emission lines of the spectrum of the SQ shock region were fitted simultaneously in the wavenumber range [14820, 15021]cm$^{-1}$ and with a sincgauss function. When one or both velocity components presented contrast in the H$\alpha$ line lower than three, we assumed that they fit the criterion to be considered as one velocity components. Finally, for the rest of the emission lines ([\ion{O}{ii}]$\lambda$3727, H$\beta$, and [\ion{O}{iii}]$\lambda\lambda$4959,5007), we performed a two-component fit using the velocities and line widths derived from the H$\alpha$ fit as input. We found 22 SQ H$\alpha$ emission regions with line profiles that can be fit with two velocity components.

\subsection{Sample of SQ H$\alpha$ emission regions}
\label{subsec:sample}
Following the methodology described above, we found 209 candidate emission regions in the SQ field. However, in this work we only focus on those regions that belong to SQ (hereafter, 'SQ H$\alpha$ emission regions') according to the following criteria: i) the radial velocity of the region is within the radial velocity range of SQ (between $\sim$5600 and $\sim$7000 $km\, s^{-1}$); and ii) at least one additional emission line besides H$\alpha$ has been detected in the data cubes. According to this criteria, 175 SQ H$\alpha$ emission regions were found. From these regions, 22 present broad H$\alpha$ profiles that have been fit using two velocity components. There may be more than two velocity components, but for our spectral resolution, the residuals of the fit produced are found to be within a band of width 3 sigma rms of the defined continuum window. We selected the minimum number of components obtained, considering the resolution and the signal-to-noise ratio of our data cubes.

\begin{figure*}
    \centering
    \includegraphics[width=\textwidth]{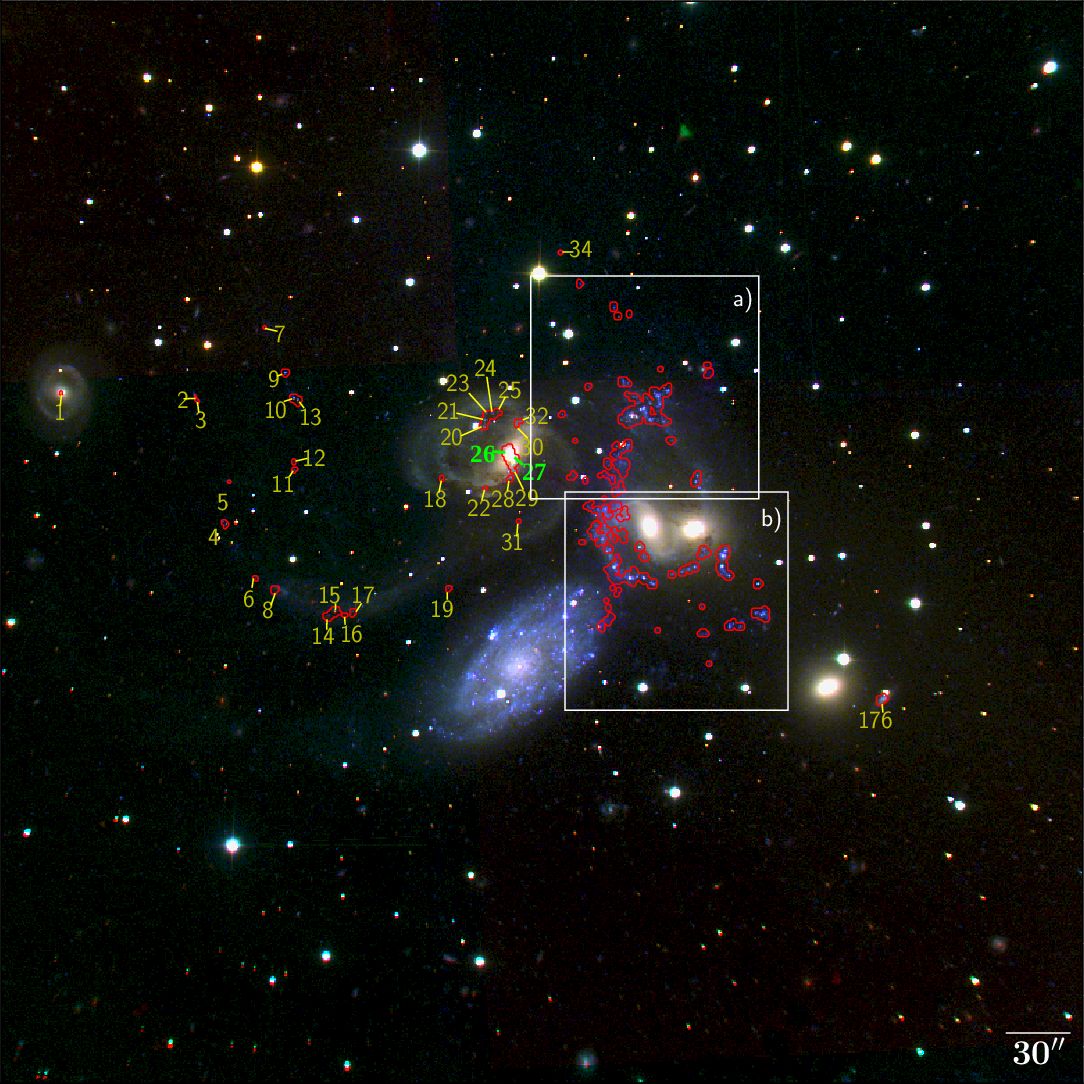}
    \caption{SITELLE composite image from deep SN1, SN2 and SN3 data cubes of SQ. The red lines represent the H$\alpha$ contours corresponding to all 175 SQ H$\alpha$ emission regions defined in Sect.~\ref{subsec:sample}. The H$\alpha$ regions from YTT, NGC7319, NGC7320c, and NG are labelled according to the nomenclature from Table~\ref{table:table2}. Green and yellow labels identify the SQ H$\alpha$ regions showing a broad and narrow line profile, respectively. A zoomed-in view of the areas marked with the white rectangles are shown in Fig~\ref{fig:HIIa}.}
    \label{fig:HII}
\end{figure*}

\begin{figure*}
    \centering
    \includegraphics[width=.625\textwidth]{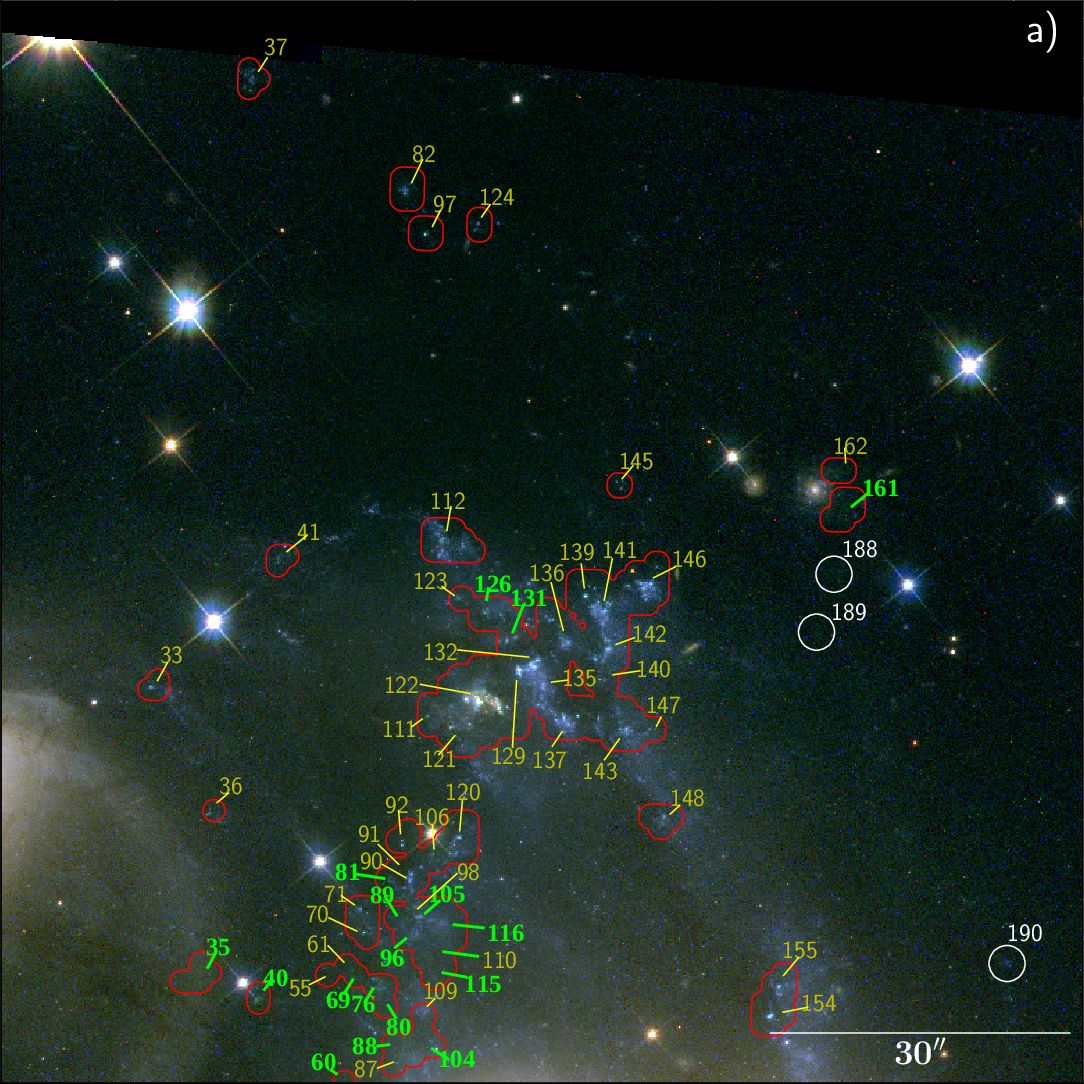}\\
    \includegraphics[width=.625\textwidth]{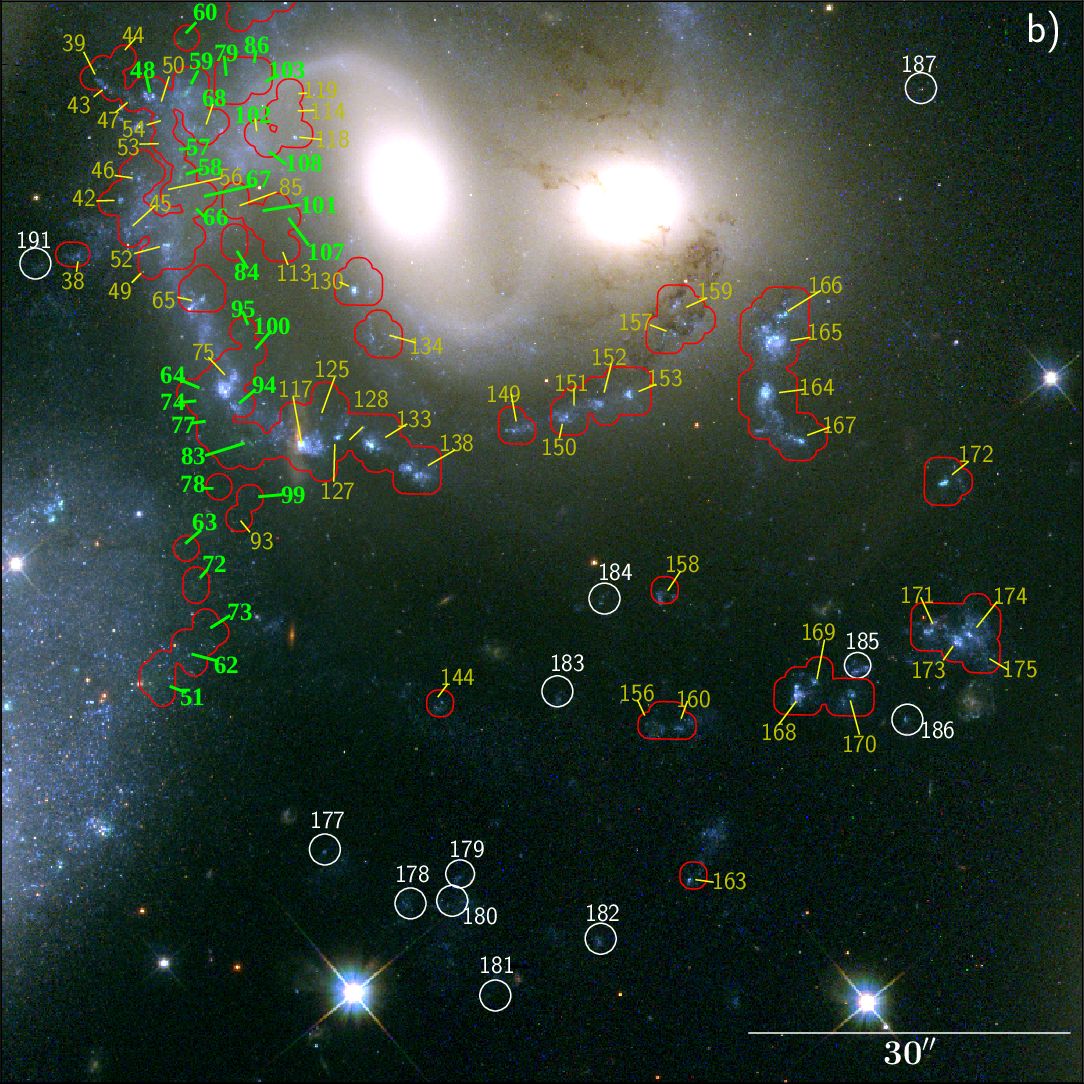}
    \caption{Zoomed-in view of image from Hubble Space Telescope Wide Field Camera 3, also known as HST/WFC3 (PID 11502, PI Keith S. Noll) showing regions a) and b) from Fig.~\ref{fig:HII}. The red circles and labels are defined as in Fig.~\ref{fig:HII}. White circles show the location of the 15 extra H$\alpha$ emitter regions presented in Table~\ref{table:table3}. Green and yellow labels identify the SQ H$\alpha$ regions showing a broad and narrow line profile, respectively.}
    \label{fig:HIIa}
\end{figure*}

In Figs.~\ref{fig:HII} and \ref{fig:HIIa} we show the sample of 175 SQ H$\alpha$ emission regions found in this work. The location of the 15 extra H$\alpha$ emitter regions presented in Table~\ref{table:table3}. are also shown. In Table~\ref{table:table2} we present the catalogue of these SQ H$\alpha$ emission regions. In Column 1 the region name is presented, Column 2 shows the RA and Dec coordinates, Columns 3 and 4 show the radial velocity and the area, Column 5 indicates the corresponding subzone. We add the information for NGC7320c which is discussed in Sect.~\ref{subsec:glx_SQ} (ID 1 in Table~\ref{table:table2}). We have compared our results with the recent works by \cite{2014ApJ...784....1K} and \cite{2012A&A...539A.127I}. From the 40 H$\alpha$ emitting regions presented in \cite{2014ApJ...784....1K}, 34 have the same $\alpha, \delta$ as our regions. Twenty-five of these regions can be directly compared since they present a single value of radial velocity. When compared with \cite{2012A&A...539A.127I}, which has lower spatial resolution, only six regions were identified (pointings M and S). The mean of the differences between our H$\alpha$ radial velocities and the ones in these previous works, $\Delta$v, is $\Delta$v=$1.1\pm 14.9\ km s^{-1}$, which is consistent to within the errors.

For the sake of the analysis of the sample, in the following we separate the SQ emission regions in two sub-samples: lower radial velocity sub-sample (LV) composed of those regions where the radial velocity $\leq$ 6160 $km\, s^{-1}$ (for the two velocity components fit, the lowest radial velocity was considered); conversely, the higher radial velocity sub-sample (HV) is defined including all the regions with radial velocity > 6160 $km\, s^{-1}$ (for the two velocity components fit, the greatest radial velocity was considered). The nominal radial velocity of 6160 $km\, s^{-1}$ was chosen as a limit value to highlight the discrete emission features associated with NI (NGC7318B). 

Figure~\ref{fig:regions} presents the different zones and subzones defined by the 175 SQ H$\alpha$ emission regions (see below). In Fig.~\ref{fig:velocity} we show the radial velocity versus RA diagram in the upper right panel; Dec versus radial velocity diagram in the upper right panel; and a three-dimensional view (i.e. RA -- Dec -- radial velocity diagram) of SQ H$\alpha$ emission regions in the lower panel. From the three-dimensional information obtained considering Figs.~\ref{fig:regions} and \ref{fig:velocity}, we have defined 12 zones and 28 subzones in SQ. We have divided the zones (and thus the subzones) in the following way: i) YTT \citep[e.g.][]{2002A&A...394..823L}, so that north and south strands are respectively YTTN and YTTS, and NGC7320c is the old intruder (OI); ii) NGC7319 (NGC7319 nucleus, NGC7319 'arm', north lobe); iii) H$\alpha$ 'bridge'; iv) high radial velocity strands, Hs (H1 and H2); v) SQA \citep[e.g.][]{1999ApJ...512..178X}; vi) low radial velocity strands, Ls (L1, L2, L3, and L4); vii) shock strands, Shs (Sh1, Sh2, Sh3, and Sh4); viii) north and south of SQA (NSQA and SSQA, respectively); ix) tidal tail at north of NSQA \citep[NW, e.g.][]{2010ApJ...724...80R}; x) NI \citep[e.g.][]{1997ApJ...485L..69M} strands, NIs (NI1, NI2, NI3, NI4, and NI5); xi) southern debris region \citep[SDR, e.g.][]{2011AJ....142...42F}; and xii) NG (see Sect.~\ref{subsec:glx_SQ}). Also, the 15 extra H$\alpha$ emitter regions from Appendix~\ref{append:app1} are shown. Section~\ref{subsec:diff} explains in more detail the connections between each zone and subzone. For each zone a detailed study was made to understand better the kinematic and chemical processes acting on SQ (see Sect.~\ref{sec:results} and Duarte Puertas et al. paper II, in prep).

\begin{landscape}
\begin{figure}
    \centering
    \includegraphics[width=\columnwidth]{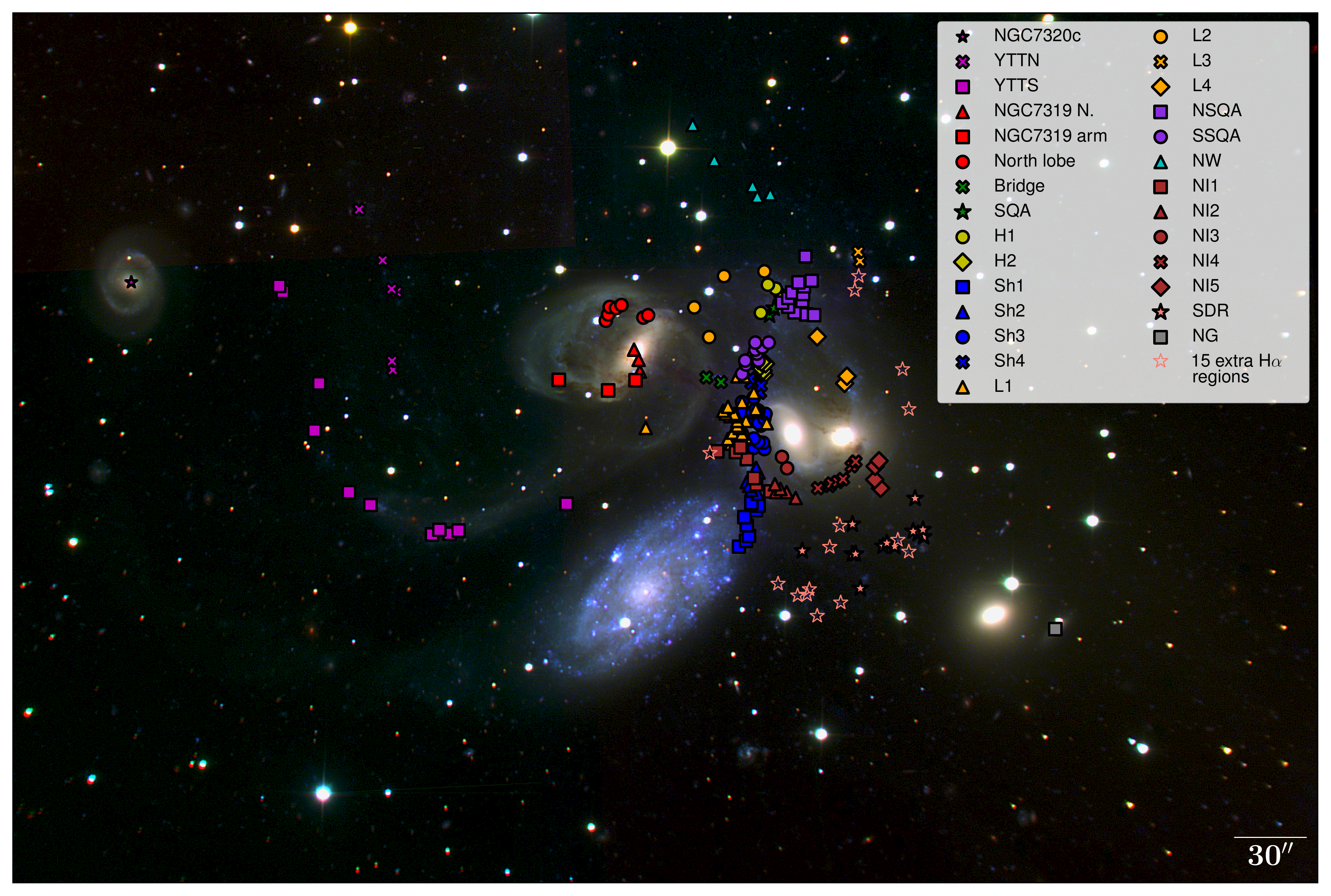}
    \caption{Different systems of emission line objects defined here and indicated on SITELLE deep-colour image of SQ. Shown are the positions of YTTN and YTTS (magenta crosses and squares, respectively); NGC7319 nucleus, 'arm', and north lobe (red triangles, squares, and circles, respectively); bridge (green crosses); SQA (green stars); Hs (H1: yellow circles; H2: yellow diamonds); Shs (Sh1: blue squares; Sh2: blue triangles; Sh3: blue circles; Sh4: blue crosses); Ls (L1: orange triangles; L2: orange circles; L3: orange crosses; L4: orange diamonds); NSQA and SSQA (violet squares and circles); NW (cyan triangles); NIs (NI1: brown squares; NI2: brown triangles; NI3: brown circles; NI4: brown crosses; NI5: brown diamonds); SDR (salmon stars); and NG (grey squares). In addition, the salmon unfilled stars show the position of the 15 extra H$\alpha$ emitter regions in Appendix~\ref{append:app1} from Table~\ref{table:table3}.}
    \label{fig:regions}
\end{figure}
\end{landscape}

\begin{figure*}
    \centering
    \includegraphics[width=\textwidth]{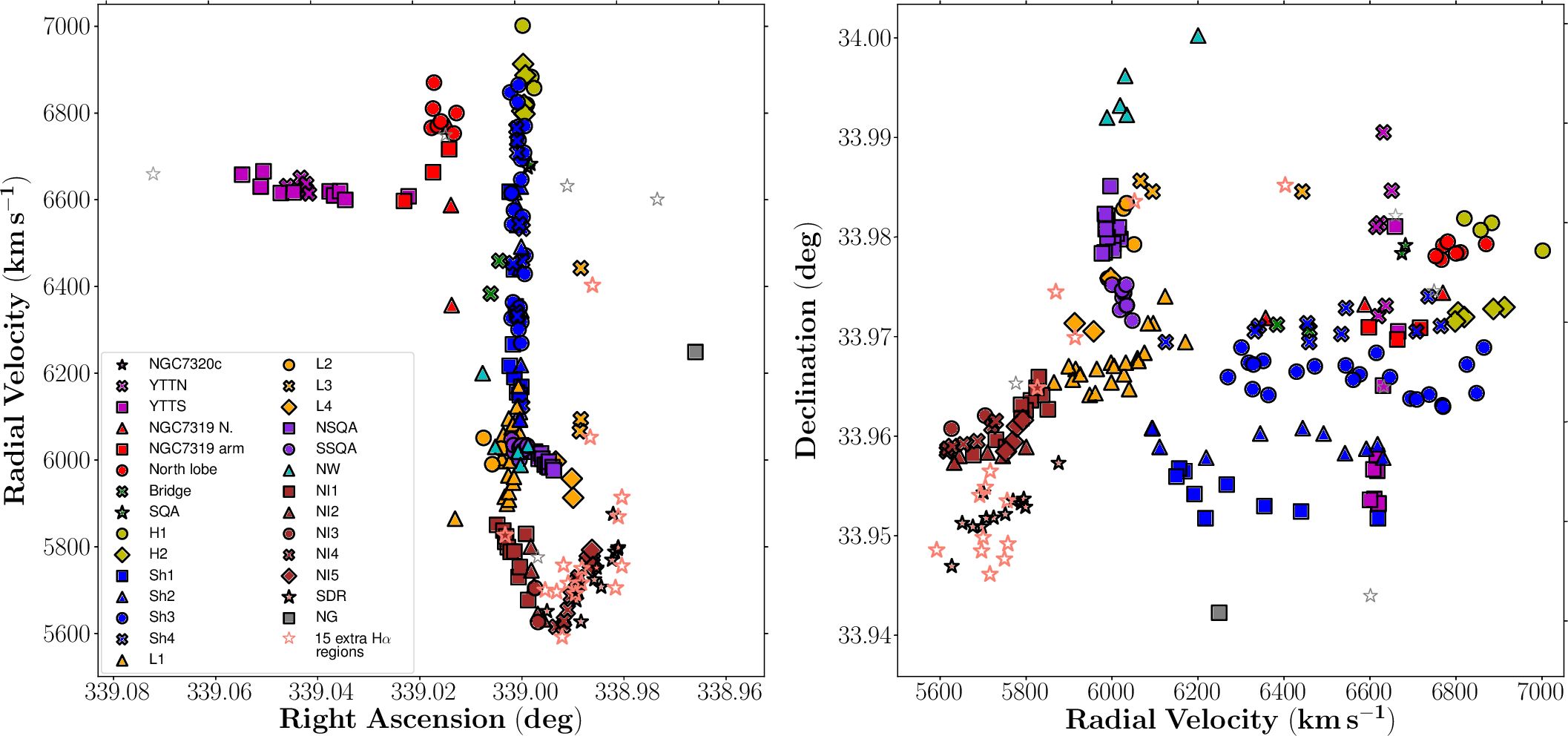}\\
    \includegraphics[width=0.75\textwidth]{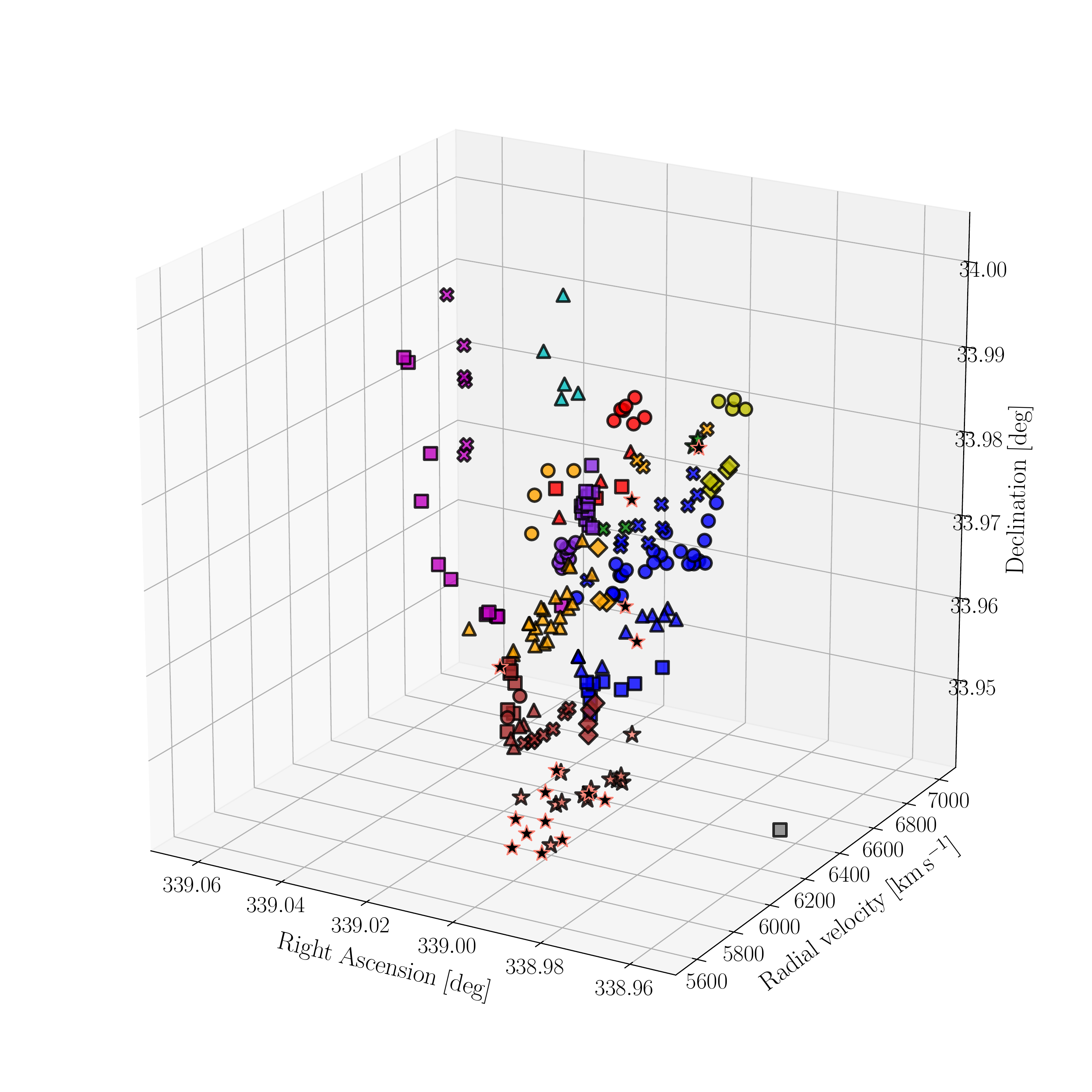}
    \caption{Shown in the upper left panel is the radial velocity versus RA diagram; in the upper right panel: Dec versus radial velocity diagram; in the lower panel: three-dimensional view of SQ H$\alpha$ emission regions (RA, Dec, and radial velocity diagram). All the points in the figures have the same colours and markers as Fig.~\ref{fig:regions}. In both upper panels, the nuclei of NGC7320c, NGC7319, NGC7318B, NGC7318A, and NGC7317 are marked with an unfilled grey star.}
    \label{fig:velocity}
\end{figure*}

\begin{table*}[t]
\begin{minipage}[t]{.48\linewidth}
\tiny
\caption{Relevant information of H$\alpha$ emission regions in Stephan's Quintet.}\vspace{-2.75ex}
\label{table:table2}
\centering 
\begin{tabular}[t]{c | l l c c }
\hline\hline  \\[-2ex]
(1) & $\ \ \ \ \ \ \ \ \ \ \ \ \ \ \ $ (2) & (3) & (4) & (5) \\[0.5ex] 
Region & $\ \ \ \ \ \ \ \ \alpha\ \ \ \ \ \ \ \ \ \ \ \ \ \ \ \ \delta$ & Velocity & Area & Subzone\\[0.5ex]
ID & $\ \ $ (h m s) $\ \ \ \ \ \ $ ($^o\ ^\prime\ ^{\prime\prime}$) & ($km\, s^{-1}$) & (kpc$^2$) &  \\[0.5ex] 
\hline\\[-2ex]
1$^\dagger$ & 22 36 20.3 +33 59 04.4    & 6657$\pm$24 & --   & NGC7320c      \\[0.5ex]
2 & 22 36 15.3  +33 59 1.9              & 6644$\pm$3  & 1.0  & YTT           \\[0.5ex] 
3 & 22 36 15.2  +33 58 59.9             & 6656$\pm$6  & 0.6  & YTT           \\[0.5ex]
4 & 22 36 14.1  +33 58 1.0              & 6628$\pm$2  & 1.3  & YTT           \\[0.5ex]
5 & 22 36 14.0  +33 58 21.4             & 6664$\pm$11 & 0.2  & YTT           \\[0.5ex]
6 & 22 36 13.0  +33 57 36.5             & 6613$\pm$5  & 0.5  & YTT           \\[0.5ex]
7 & 22 36 12.6  +33 59 33.3             & 6630$\pm$15 & 0.2  & YTT           \\[0.5ex]
8 & 22 36 12.4  +33 57 30.2             & 6614$\pm$4  & 1.3  & YTT           \\[0.5ex]
9 & 22 36 11.8  +33 59 11.5             & 6648$\pm$3  & 1.1  & YTT           \\[0.5ex]
10 & 22 36 11.5 +33 58 59.6             & 6622$\pm$1  & 1.3  & YTT           \\[0.5ex]
11 & 22 36 11.5 +33 58 26.0             & 6618$\pm$5  & 0.5  & YTT           \\[0.5ex]
12 & 22 36 11.5 +33 58 30.3             & 6635$\pm$6  & 0.3  & YTT           \\[0.5ex]
13 & 22 36 11.4 +33 58 58.6             & 6612$\pm$1  & 1.3  & YTT           \\[0.5ex]
14 & 22 36 10.2 +33 57 18.5             & 6618$\pm$2  & 2.3  & YTT           \\[0.5ex]
15 & 22 36 10.0 +33 57 20.0             & 6608$\pm$1  & 1.7  & YTT           \\[0.5ex]
16 & 22 36 9.6  +33 57 17.7             & 6618$\pm$5  & 0.3  & YTT           \\[0.5ex]
17 & 22 36 9.4  +33 57 19.2             & 6598$\pm$4  & 0.8  & YTT           \\[0.5ex]
18 & 22 36 6.0  +33 58 20.0             & 6595$\pm$7  & 0.3  & NGC7319 'Arm' \\[0.5ex]
19 & 22 36 5.7  +33 57 29.9             & 6606$\pm$5  & 0.5  & YTT           \\[0.5ex]
20 & 22 36 4.4  +33 58 44.9             & 6764$\pm$4  & 1.1  & North lobe    \\[0.5ex]
21 & 22 36 4.3  +33 58 46.8             & 6809$\pm$4  & 1.2  & North lobe    \\[0.5ex]
22 & 22 36 4.3  +33 58 16.0             & 6662$\pm$17 & 0.2  & NGC7319 'Arm' \\[0.5ex]
23 & 22 36 4.3  +33 58 50.4             & 6868$\pm$6  & 0.7  & North lobe    \\[0.5ex]
24 & 22 36 4.0  +33 58 50.0             & 6769$\pm$7  & 1.3  & North lobe    \\[0.5ex]
25 & 22 36 3.8  +33 58 51.3             & 6779$\pm$5  & 0.9  & North lobe    \\[0.5ex]
26 & 22 36 3.4  +33 58 32.3             & 6768$\pm$9  & 3.8  & NGC7319 Nucl. \\[0.5ex]
27 & 22 36 3.3  +33 58 28.4             & 6586$\pm$20 & 3.9  & NGC7319 Nucl. \\[0.5ex]
28 & 22 36 3.3  +33 58 20.6             & 6715$\pm$7  & 0.5  & NGC7319 'Arm' \\[0.5ex]
29 & 22 36 3.3  +33 58 23.5             & 6355$\pm$11 & 0.3  & NGC7319 Nucl. \\[0.5ex]
30 & 22 36 3.1  +33 58 45.5             & 6750$\pm$8  & 0.5  & North lobe    \\[0.5ex]
31 & 22 36 3.1  +33 58 0.1              & 5863$\pm$12 & 0.2  & L1            \\[0.5ex]
32 & 22 36 2.9  +33 58 46.8             & 6799$\pm$12 & 0.2  & North lobe    \\[0.5ex]
33 & 22 36 1.4  +33 58 49.0             & 6049$\pm$7  & 0.5  & L2            \\[0.5ex]
34 & 22 36 1.4  +34 0 5.7               & 6198$\pm$14 & 0.2  & NW            \\[0.5ex]
35 & 22 36 1.0  +33 58 21.0             & 6381$\pm$10 & 1.0  & Bridge        \\[0.5ex]
36 & 22 36 0.9  +33 58 36.8             & 5988$\pm$13 & 0.2  & L2            \\[0.5ex]
37 & 22 36 0.7  +33 59 50.7             & 6028$\pm$6  & 0.6  & NW            \\[0.5ex]
38 & 22 36 0.6  +33 57 49.9             & 5849$\pm$9  & 0.3  & NI1           \\[0.5ex]
39 & 22 36 0.5  +33 58 6.3              & 5996$\pm$7  & 0.6  & L1            \\[0.5ex]
40 & 22 36 0.6  +33 58 17.9             & 6457$\pm$14 & 0.3  & Bridge        \\[0.5ex]
41 & 22 36 0.5  +33 59 2.1              & 6024$\pm$8  & 0.5  & L2            \\[0.5ex]
42 & 22 36 0.4  +33 57 55.3             & 5835$\pm$4  & 0.7  & NI1           \\[0.5ex]
43 & 22 36 0.4  +33 58 5.0              & 6003$\pm$11 & 0.2  & L1            \\[0.5ex]
44 & 22 36 0.4  +33 58 8.2              & 6057$\pm$12 & 0.2  & L1            \\[0.5ex]
45 & 22 36 0.3  +33 57 52.7             & 5809$\pm$6  & 1.0  & NI1           \\[0.5ex]
46 & 22 36 0.2  +33 57 57.5             & 5822$\pm$5  & 0.8  & NI1           \\[0.5ex]
47 & 22 36 0.2  +33 58 4.0              & 5913$\pm$9  & 0.2  & L1            \\[0.5ex]
48 & 22 36 0.2  +33 58 5.3              & 5898$\pm$12 & 0.3  & L1            \\[0.5ex]
49 & 22 36 0.2  +33 57 48.8             & 5797$\pm$10 & 0.6  & NI1           \\[0.5ex]
50 & 22 36 0.0  +33 58 4.2              & 5963$\pm$10 & 0.3  & L1            \\[0.5ex]
51$^{a}$ & 22 35 60.0 +33 57 11.1  & 6215$\pm$50 & 1.9  & Sh1           \\[0.5ex]
51$^{b}$ & 22 35 60.0 +33 57 11.1  & 6617$\pm$20 & 1.9  & Sh1           \\[0.5ex]
52 & 22 35 59.9 +33 57 51.4             & 5786$\pm$3  & 1.7  & NI1           \\[0.5ex]
53 & 22 36 0.0  +33 58 0.4              & 5907$\pm$10 & 0.2  & L1            \\[0.5ex]
54 & 22 36 0.1  +33 58 2.3              & 5924$\pm$10 & 0.2  & L1            \\[0.5ex]
55 & 22 36 0.0  +33 58 20.7             & 6093$\pm$11 & 0.2  & L1            \\[0.5ex]
56 & 22 35 60.0 +33 57 55.5             & 6846$\pm$13 & 0.2  & Sh3           \\[0.5ex]
57 & 22 35 60.0 +33 57 59.1             & 5997$\pm$18 & 0.2  & L1            \\[0.5ex]
\hline
\end{tabular}
\end{minipage}
\hspace{.1cm}
\begin{minipage}[t]{.48\linewidth}\vspace{.9cm}
\tiny
\centering
\begin{tabular}[t]{c | l l c c}
\hline\hline  \\[-2ex]
(1) & $\ \ \ \ \ \ \ \ \ \ \ \ \ \ \ $ (2) & (3) & (4) & (5) \\[0.5ex] 
Region & $\ \ \ \ \ \ \ \ \alpha\ \ \ \ \ \ \ \ \ \ \ \ \ \ \ \ \delta$ & Velocity & Area & Subzone\\[0.5ex]
ID & $\ \ $ (h m s) $\ \ \ \ \ \ $ ($^o\ ^\prime\ ^{\prime\prime}$) & ($km\, s^{-1}$) & (kpc$^2$) &  \\[0.5ex] 
\hline\\[-2ex]
58$^{a}$ & 22 35 59.9 +33 57 57.5 & 6037$\pm$14 & 0.6 & L1   \\[0.5ex]
58$^{b}$ & 22 35 59.9 +33 57 57.5 & 6325$\pm$30 & 0.6 & Sh3  \\[0.5ex]
59$^{a}$ & 22 35 59.9 +33 58 5.2  & 6165$\pm$16 & 1.6 & L1   \\[0.5ex]
59$^{b}$ & 22 35 59.9 +33 58 5.2  & 6621$\pm$32 & 1.6 & Sh3  \\[0.5ex]
60$^{a}$ & 22 35 59.9 +33 58 10.0 & 6073$\pm$14 & 0.3 & L1   \\[0.5ex]
60$^{b}$ & 22 35 59.9 +33 58 10.0 & 6612$\pm$17 & 0.3 & Sh3  \\[0.5ex]
61 & 22 35 59.9 +33 58 21.7            & 6046$\pm$12 & 0.2 & SSQA \\[0.5ex]
62 & 22 35 59.9 +33 57 13.3            & 6438$\pm$10 & 0.7 & Sh1  \\[0.5ex]
63 & 22 35 59.8 +33 57 22.8            & 6265$\pm$21 & 0.2 & Sh1  \\[0.5ex]
64 & 22 35 59.8 +33 57 36.6            & 6616$\pm$11 & 0.6 & Sh2  \\[0.5ex]
65 & 22 35 59.7 +33 57 45.8            & 5787$\pm$5  & 1.1 & NI1  \\[0.5ex]
66$^{a}$ & 22 35 59.8 +33 57 54.5 & 5946$\pm$12 & 0.5 & L1   \\[0.5ex]
66$^{b}$ & 22 35 59.8 +33 57 54.5 & 6362$\pm$14 & 0.5 & Sh3  \\[0.5ex]
67 & 22 35 59.8 +33 57 55.5            & 5955$\pm$13 & 0.5 & L1   \\[0.5ex]
68$^{a}$ & 22 35 59.7 +33 58 1.0  & 6025$\pm$9  & 1.6 & L1   \\[0.5ex]
68$^{b}$ & 22 35 59.7 +33 58 1.0  & 6574$\pm$37 & 1.6 & Sh3  \\[0.5ex]
69$^{a}$ & 22 35 59.8 +33 58 20.7 & 6081$\pm$20 & 0.3 & L1   \\[0.5ex]
69$^{b}$ & 22 35 59.8 +33 58 20.7 & 6452$\pm$32 & 0.3 & Sh4  \\[0.5ex]
70 & 22 35 59.7 +33 58 25.5            & 6016$\pm$4  & 0.7 & SSQA \\[0.5ex]
71 & 22 35 59.7 +33 58 27.4            & 6032$\pm$8  & 0.4 & SSQA \\[0.5ex]
72 & 22 35 59.8 +33 57 18.7            & 6189$\pm$15 & 0.3 & Sh1  \\[0.5ex]
73 & 22 35 59.7 +33 57 13.9            & 6352$\pm$10 & 0.7 & Sh1  \\[0.5ex]
74 & 22 35 59.7 +33 57 35.6            & 6589$\pm$10 & 0.7 & Sh2  \\[0.5ex]
75 & 22 35 59.6 +33 57 38.1            & 5728$\pm$5  & 1.5 & NI1  \\[0.5ex]
76$^{a}$ & 22 35 59.7 +33 58 19.7 & 6312$\pm$27 & 0.6 & Sh4  \\[0.5ex]
76$^{b}$ & 22 35 59.7 +33 58 19.7 & 6761$\pm$35 & 0.6 & Sh4  \\[0.5ex]
77 & 22 35 59.6 +33 57 35.2            & 6540$\pm$12 & 0.7 & Sh2  \\[0.5ex]
78 & 22 35 59.6 +33 57 28.5            & 6154$\pm$13 & 0.2 & Sh1  \\[0.5ex]
79$^{a}$ & 22 35 59.5 +33 58 5.1  & 6326$\pm$31 & 1.3 & Sh3  \\[0.5ex]
79$^{b}$ & 22 35 59.5 +33 58 5.1  & 6823$\pm$10 & 1.3 & Sh3  \\[0.5ex]
80$^{a}$ & 22 35 59.6 +33 58 17.4 & 6343$\pm$19 & 0.3 & Sh4  \\[0.5ex]
80$^{b}$ & 22 35 59.6 +33 58 17.4 & 6752$\pm$23 & 0.3 & Sh4  \\[0.5ex]
81$^{a}$ & 22 35 59.5 +33 58 30.3 & 6123$\pm$17 & 0.3 & L1   \\[0.5ex]
81$^{b}$ & 22 35 59.5 +33 58 30.3 & 6734$\pm$26 & 0.3 & Sh4  \\[0.5ex]
82 & 22 35 59.5 +33 59 39.6            & 6017$\pm$4  & 1.0 & NW   \\[0.5ex]
83$^{a}$ & 22 35 59.4 +33 57 33.3 & 6211$\pm$22 & 3.0 & Sh2  \\[0.5ex]
83$^{b}$ & 22 35 59.4 +33 57 33.3 & 6626$\pm$11 & 3.0 & Sh2  \\[0.5ex]
84 & 22 35 59.5 +33 57 50.6            & 6767$\pm$12 & 0.3 & Sh3  \\[0.5ex]
85 & 22 35 59.5 +33 57 55.4            & 6735$\pm$10 & 0.3 & Sh3  \\[0.5ex]
86$^{a}$ & 22 35 59.4 +33 58 7.0  & 6059$\pm$13 & 0.8 & L1   \\[0.5ex]
86$^{b}$ & 22 35 59.4 +33 58 7.0  & 6350$\pm$25 & 0.8 & Sh3  \\[0.5ex]
87$^{a}$ & 22 35 59.5 +33 58 11.5 & 6299$\pm$21 & 0.3 & Sh3  \\[0.5ex]
87$^{b}$ & 22 35 59.5 +33 58 11.5 & 6863$\pm$16 & 0.3 & Sh3  \\[0.5ex]
88 & 22 35 59.5 +33 58 13.8            & 6168$\pm$18 & 0.2 & L1   \\[0.5ex]
89$^{a}$ & 22 35 59.5 +33 58 26.4 & 6542$\pm$22 & 0.7 & Sh4  \\[0.5ex]
89$^{b}$ & 22 35 59.5 +33 58 26.4 & 6861$\pm$36 & 0.7 & Sh4  \\[0.5ex]
90 & 22 35 59.4 +33 58 29.9            & 6022$\pm$5  & 0.8 & SSQA \\[0.5ex]
91 & 22 35 59.4 +33 58 31.3            & 6026$\pm$10 & 0.4 & SSQA \\[0.5ex]
92 & 22 35 59.5 +33 58 34.2            & 5999$\pm$7  & 0.5 & SSQA \\[0.5ex]
93 & 22 35 59.4 +33 57 25.3            & 6148$\pm$11 & 0.2 & Sh1  \\[0.5ex]
94$^{a}$ & 22 35 59.5 +33 57 35.9 & 5749$\pm$15 & 0.3 & NI1  \\[0.5ex]
94$^{b}$ & 22 35 59.5 +33 57 35.9 & 6049$\pm$28 & 0.3 & Sh2  \\[0.5ex]
95$^{a}$ & 22 35 59.4 +33 57 42.9 & 6091$\pm$57 & 0.3 & Sh2  \\[0.5ex]
95$^{b}$ & 22 35 59.4 +33 57 42.9 & 6442$\pm$27 & 0.3 & Sh2  \\[0.5ex]
96 & 22 35 59.3 +33 58 24.4            & 6803$\pm$11 & 0.8 & H2   \\[0.5ex]
97 & 22 35 59.2 +33 59 34.6            & 5986$\pm$5  & 0.6 & NW   \\[0.5ex]
98 & 22 35 59.3 +33 58 27.0            & 6031$\pm$7  & 0.4 & SSQA \\[0.5ex]
99 & 22 35 59.4 +33 57 27.5            & 6166$\pm$23 & 0.2 & Sh1  \\[0.5ex]
\hline
\end{tabular}
\tablefoot{Continued.}
\end{minipage}
\end{table*}

\begin{table*}[t]
\begin{minipage}[t]{.48\linewidth}
\tiny
\begin{tabular}[t]{c | l l c c }
\hline\hline  \\[-2ex]
(1) & $\ \ \ \ \ \ \ \ \ \ \ \ \ \ \ $ (2) & (3) & (4) & (5) \\[0.5ex] 
Region & $\ \ \ \ \ \ \ \ \alpha\ \ \ \ \ \ \ \ \ \ \ \ \ \ \ \ \delta$ & Velocity & Area & Subzone\\[0.5ex]
ID & $\ \ $ (h m s) $\ \ \ \ \ \ $ ($^o\ ^\prime\ ^{\prime\prime}$) & ($km\, s^{-1}$) & (kpc$^2$) & \\[0.5ex] 
\hline\\[-2ex]
100$^{a}$ & 22 35 59.4 +33 57 41.0 & 6233$\pm$27 & 0.3 & Sh3  \\[0.5ex]
100$^{b}$ & 22 35 59.4 +33 57 41.0 & 6490$\pm$34 & 0.3 & Sh2  \\[0.5ex]
101 & 22 35 59.4 +33 57 53.2            & 6692$\pm$16 & 0.2 & Sh3  \\[0.5ex]
102$^{a}$ & 22 35 59.3 +33 58 1.2  & 6267$\pm$17 & 0.3 & Sh3  \\[0.5ex]
102$^{b}$ & 22 35 59.3 +33 58 1.2  & 6644$\pm$43 & 0.3 & Sh3  \\[0.5ex]
103$^{a}$ & 22 35 59.3 +33 58 5.7  & 6030$\pm$13 & 0.5 & L1   \\[0.5ex]
103$^{b}$ & 22 35 59.3 +33 58 5.7  & 6316$\pm$42 & 0.5 & Sh3  \\[0.5ex]
104$^{a}$ & 22 35 59.4 +33 58 13.4 & 6124$\pm$72 & 1.6 & Sh4  \\[0.5ex]
104$^{b}$ & 22 35 59.4 +33 58 13.4 & 6460$\pm$31 & 1.6 & Sh4  \\[0.5ex]
105 & 22 35 59.2 +33 58 25.7            & 6911$\pm$11 & 0.6 & H2   \\[0.5ex]
106 & 22 35 59.2 +33 58 32.2            & 6021$\pm$8  & 0.7 & SSQA \\[0.5ex]
107 & 22 35 59.2 +33 57 53.1            & 6707$\pm$13 & 0.8 & Sh3  \\[0.5ex]
108 & 22 35 59.2 +33 58 0.2             & 6558$\pm$27 & 0.2 & Sh3  \\[0.5ex]
109 & 22 35 59.2 +33 58 16.3            & 6532$\pm$19 & 0.2 & Sh4  \\[0.5ex]
110 & 22 35 59.2 +33 58 22.1            & 6817$\pm$8  & 0.8 & H2   \\[0.5ex]
111 & 22 35 59.2 +33 58 46.8            & 7000$\pm$6  & 0.5 & H1   \\[0.5ex]
112 & 22 35 59.1 +33 59 4.3             & 6032$\pm$4  & 1.9 & L2   \\[0.5ex]
113 & 22 35 59.1 +33 57 49.9            & 6768$\pm$9  & 0.3 & Sh3  \\[0.5ex]
114 & 22 35 59.1 +33 58 3.1             & 6427$\pm$14 & 0.3 & Sh3  \\[0.5ex]
115 & 22 35 59.1 +33 58 21.1            & 6797$\pm$13 & 0.4 & H2   \\[0.5ex]
116 & 22 35 59.1 +33 58 25.4            & 6885$\pm$8  & 1.4 & H2   \\[0.5ex]
117 & 22 35 59.0 +33 57 33.2            & 5675$\pm$1  & 2.2 & NI1  \\[0.5ex]
118 & 22 35 59.1 +33 58 1.1             & 5827$\pm$9  & 0.3 & NI1  \\[0.5ex]
119 & 22 35 59.1 +33 58 5.0             & 6469$\pm$18 & 0.2 & Sh3  \\[0.5ex]
120 & 22 35 59.0 +33 58 34.1            & 6031$\pm$3  & 1.5 & SSQA \\[0.5ex]
121 & 22 35 59.0 +33 58 45.7            & 6672$\pm$4  & 3.0 & SQA  \\[0.5ex]
122 & 22 35 58.8 +33 58 48.6            & 6680$\pm$2  & 4.9 & SQA  \\[0.5ex]
123 & 22 35 58.9 +33 58 58.1            & 6818$\pm$10 & 0.2 & H1   \\[0.5ex]
124 & 22 35 58.8 +33 59 35.5            & 6032$\pm$9  & 0.3 & NW   \\[0.5ex]
125 & 22 35 58.8 +33 57 36.0            & 5799$\pm$11 & 0.7 & NI2  \\[0.5ex]
126 & 22 35 58.6 +33 58 56.0            & 6881$\pm$8  & 1.1 & H1   \\[0.5ex]
127 & 22 35 58.8 +33 57 33.2            & 5744$\pm$4  & 1.4 & NI2  \\[0.5ex]
128 & 22 35 58.7 +33 57 34.1            & 5708$\pm$3  & 0.7 & NI2  \\[0.5ex]
129 & 22 35 58.4 +33 58 51.1            & 6017$\pm$2  & 2.3 & NSQA \\[0.5ex]
130 & 22 35 58.6 +33 57 47.5            & 5703$\pm$3  & 1.1 & NI3  \\[0.5ex]
131 & 22 35 58.6 +33 58 54.4            & 6856$\pm$11 & 0.6 & H1   \\[0.5ex]
132 & 22 35 58.4 +33 58 52.4            & 6009$\pm$2  & 1.4 & NSQA \\[0.5ex]
133 & 22 35 58.4 +33 57 32.8            & 5643$\pm$1  & 1.8 & NI2  \\[0.5ex]
134 & 22 35 58.4 +33 57 42.4            & 5624$\pm$5  & 1.0 & NI3  \\[0.5ex]
135 & 22 35 58.4 +33 58 48.5            & 6001$\pm$6  & 1.2 & NSQA \\[0.5ex]
136 & 22 35 58.1 +33 58 54.3            & 6013$\pm$3  & 1.8 & NSQA \\[0.5ex]
\hline
\end{tabular}
\end{minipage}
\hspace{.1cm}
\begin{minipage}[t]{.48\linewidth}
\tiny
\centering
\begin{tabular}[t]{c | l l c c }
\hline\hline  \\[-2ex]
(1) & $\ \ \ \ \ \ \ \ \ \ \ \ \ \ \ $ (2) & (3) & (4) & (5) \\[0.5ex] 
Region & $\ \ \ \ \ \ \ \ \alpha\ \ \ \ \ \ \ \ \ \ \ \ \ \ \ \ \delta$ & Velocity & Area & Subzone\\[0.5ex]
ID & $\ \ $ (h m s) $\ \ \ \ \ \ $ ($^o\ ^\prime\ ^{\prime\prime}$) & ($km\, s^{-1}$) & (kpc$^2$) & \\[0.5ex] 
\hline\\[-2ex]
137 & 22 35 58.1 +33 58 46.5            & 6001$\pm$3  & 2.1 & NSQA  \\[0.5ex]
138 & 22 35 58.2 +33 57 29.9            & 5630$\pm$2  & 1.6 & NI2   \\[0.5ex]
139 & 22 35 58.0 +33 58 59.1            & 5987$\pm$2  & 1.6 & NSQA  \\[0.5ex]
140 & 22 35 57.9 +33 58 50.3            & 5987$\pm$5  & 1.4 & NSQA  \\[0.5ex]
141 & 22 35 57.8 +33 58 56.5            & 5984$\pm$2  & 1.8 & NSQA  \\[0.5ex]
142 & 22 35 57.8 +33 58 54.2            & 5983$\pm$3  & 1.5 & NSQA  \\[0.5ex]
143 & 22 35 57.6 +33 58 45.1            & 5980$\pm$3  & 2.4 & NSQA  \\[0.5ex]
144 & 22 35 58.0 +33 57 8.8             & 5650$\pm$13 & 0.2 & SDR   \\[0.5ex]
145 & 22 35 57.7 +33 59 9.8             & 5994$\pm$11 & 0.2 & NSQA  \\[0.5ex]
146 & 22 35 57.5 +33 58 59.6            & 5981$\pm$2  & 2.2 & NSQA  \\[0.5ex]
147 & 22 35 57.4 +33 58 45.3            & 5974$\pm$8  & 0.3 & NSQA  \\[0.5ex]
148 & 22 35 57.4 +33 58 36.6            & 5994$\pm$5  & 0.6 & L4    \\[0.5ex]
149 & 22 35 57.4 +33 57 34.8            & 5613$\pm$7  & 0.5 & NI4   \\[0.5ex]
150 & 22 35 57.0 +33 57 34.8            & 5615$\pm$6  & 0.4 & NI4   \\[0.5ex]
151 & 22 35 57.0 +33 57 35.7            & 5626$\pm$7  & 0.4 & NI4   \\[0.5ex]
152 & 22 35 56.8 +33 57 36.6            & 5653$\pm$5  & 0.7 & NI4   \\[0.5ex]
153 & 22 35 56.6 +33 57 37.6            & 5684$\pm$3  & 1.4 & NI4   \\[0.5ex]
154 & 22 35 56.5 +33 58 17.0            & 5955$\pm$3  & 1.2 & L4    \\[0.5ex]
155 & 22 35 56.4 +33 58 19.9            & 5911$\pm$4  & 0.7 & L4    \\[0.5ex]
156 & 22 35 56.3 +33 57 6.7             & 5674$\pm$5  & 0.6 & SDR   \\[0.5ex]
157 & 22 35 56.3 +33 57 42.9            & 5717$\pm$4  & 1.0 & NI4   \\[0.5ex]
158 & 22 35 56.3 +33 57 19.4            & 5699$\pm$8  & 0.2 & SDR   \\[0.5ex]
159 & 22 35 56.2 +33 57 44.2            & 5727$\pm$5  & 1.4 & NI4   \\[0.5ex]
160 & 22 35 56.1 +33 57 7.0             & 5695$\pm$6  & 0.3 & SDR   \\[0.5ex]
161$^{a}$ & 22 35 55.9 +33 59 7.4  & 6108$\pm$26 & 1.9 & L3    \\[0.5ex]
161$^{b}$ & 22 35 55.9 +33 59 7.4  & 6473$\pm$68 & 1.9 & L3    \\[0.5ex]
162 & 22 35 56.0 +33 59 11.0            & 6064$\pm$9  & 0.3 & L3    \\[0.5ex]
163 & 22 35 56.1 +33 56 53.1            & 5625$\pm$10 & 0.2 & SDR   \\[0.5ex]
164 & 22 35 55.6 +33 57 37.4            & 5768$\pm$1  & 2.8 & NI5   \\[0.5ex]
165 & 22 35 55.5 +33 57 43.1            & 5776$\pm$1  & 2.5 & NI5   \\[0.5ex]
166 & 22 35 55.4 +33 57 44.7            & 5790$\pm$2  & 1.4 & NI5   \\[0.5ex]
167 & 22 35 55.3 +33 57 33.5            & 5751$\pm$2  & 1.7 & NI5   \\[0.5ex]
168 & 22 35 55.3 +33 57 10.3            & 5722$\pm$2  & 1.1 & SDR   \\[0.5ex]
169 & 22 35 55.2 +33 57 11.6            & 5750$\pm$3  & 0.7 & SDR   \\[0.5ex]
170 & 22 35 54.9 +33 57 9.7             & 5705$\pm$3  & 1.0 & SDR   \\[0.5ex]
171 & 22 35 54.3 +33 57 15.9            & 5768$\pm$3  & 1.3 & SDR   \\[0.5ex]
172 & 22 35 54.2 +33 57 29.5            & 5873$\pm$2  & 1.1 & SDR   \\[0.5ex]
173 & 22 35 54.2 +33 57 15.3            & 5786$\pm$2  & 1.0 & SDR   \\[0.5ex]
174 & 22 35 54.1 +33 57 16.2            & 5793$\pm$2  & 1.4 & SDR   \\[0.5ex]
175 & 22 35 53.9 +33 57 14.3            & 5797$\pm$5  & 0.7 & SDR   \\[0.5ex]
176 & 22 35 49.9 +33 56 36.5            & 6247$\pm$6  & 1.4 & NG    \\[0.5ex]
\hline
\end{tabular}
\tablefoot{The columns correspond to: 
(1) Identifier of the H$\alpha$ emission regions (ID); 
(2) Right ascension (hours, minutes, and seconds) and declination (degrees, arcminutes, and arcseconds);
(3) H$\alpha$ radial velocity ($km\, s^{-1}$);
(4) Area (kpc$^2$). 
(5) Subzone. '$\dagger$' (ID 1) denotes the information about NGC7320c. For regions with two velocity components we use '$^{a}$' and '$^{b}$' for low- and high-velocity components, respectively.
}
\end{minipage}
\end{table*}

\section{Results}
\label{sec:results}

\subsection{Kinematical properties of SQ H$\alpha$ emission regions}
\label{subsec:kinematics}
\begin{figure*}
    \centering
    \includegraphics[width=.88\textwidth]{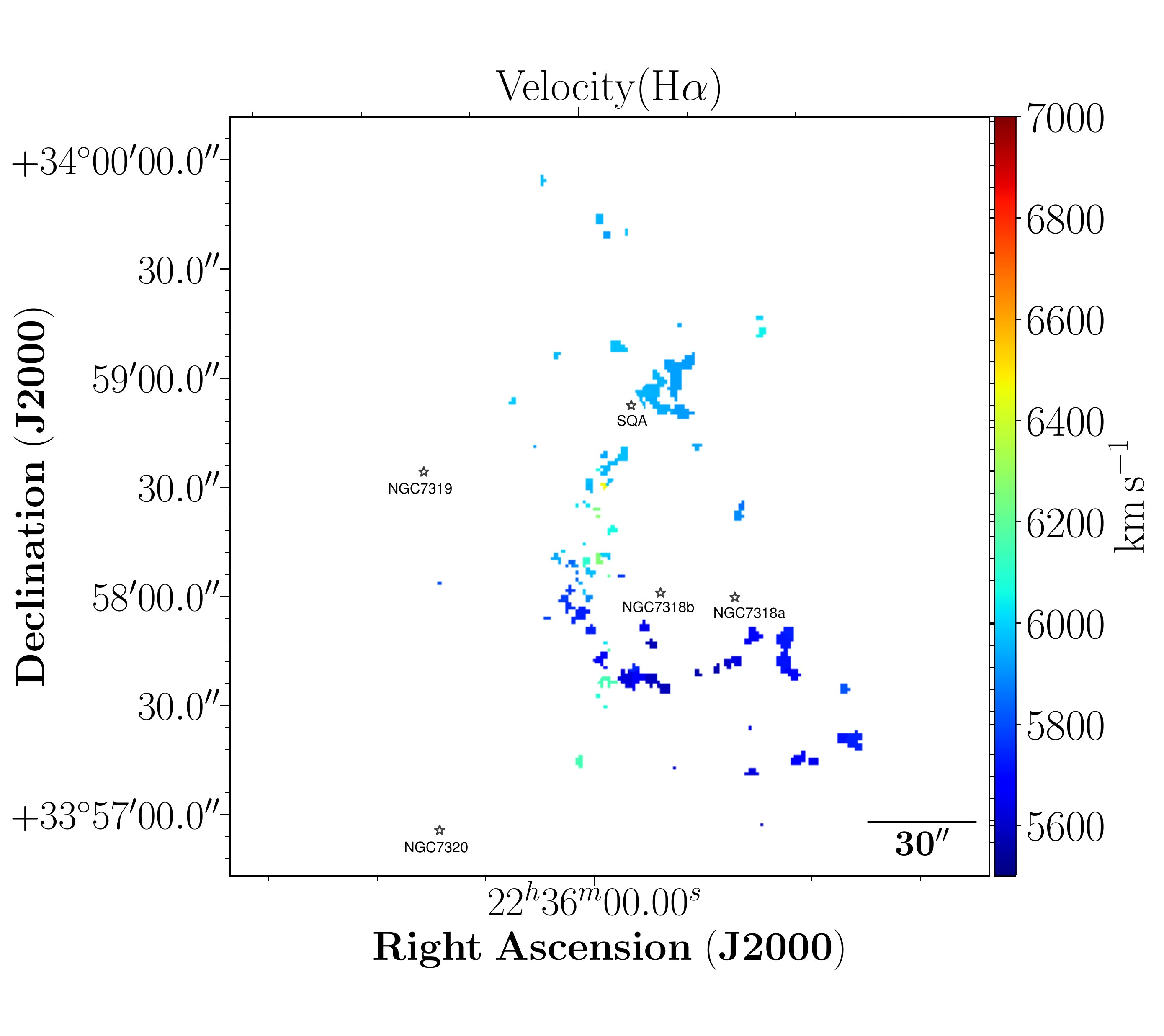}\\ \vspace{-0.5cm}
    \includegraphics[width=.88\textwidth]{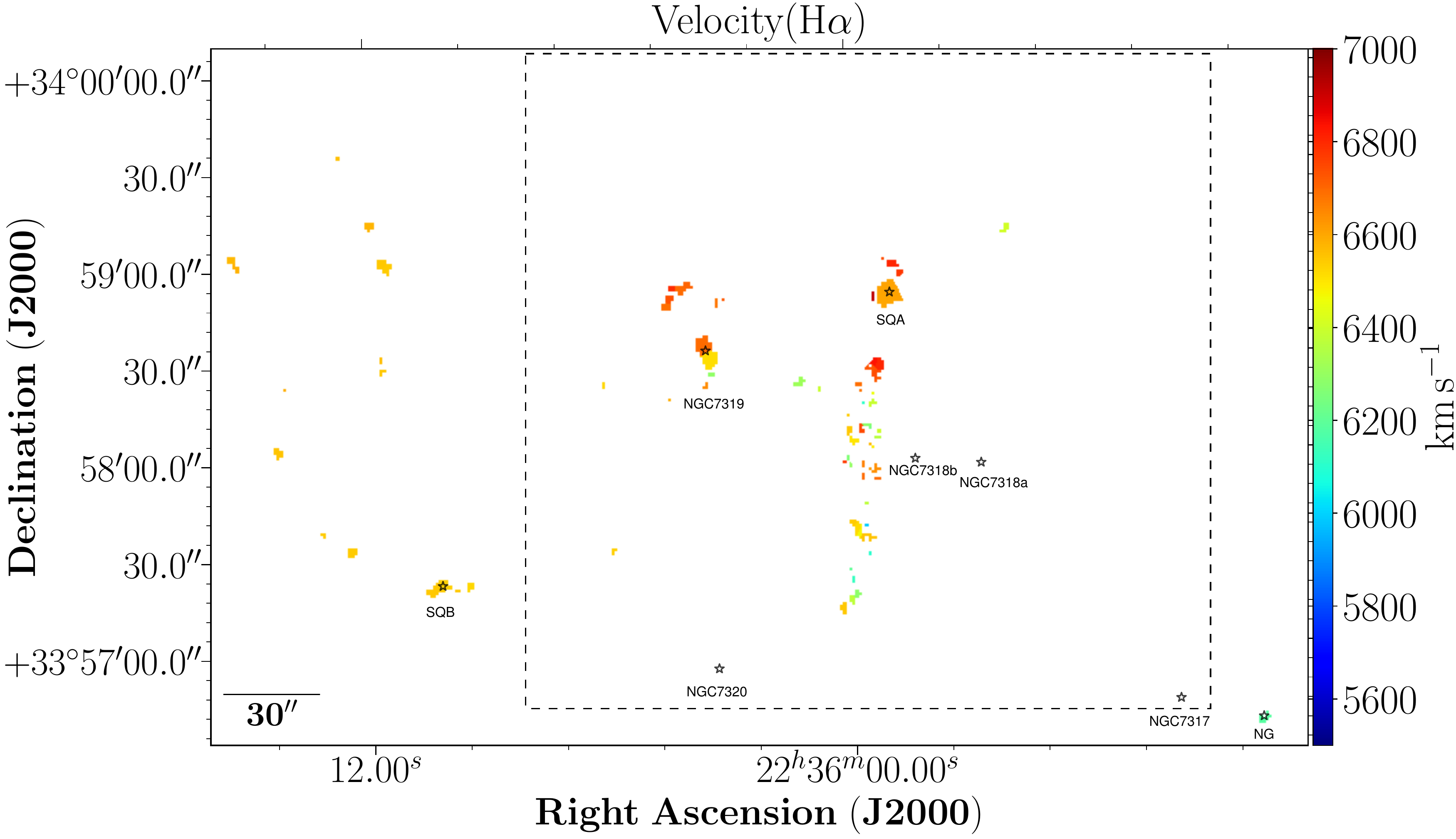}
    \caption{SQ spatial map, colour-coded according to radial velocity of H$\alpha$ for LV sample (upper panel) and HV sample (lower panel). The box in the lower panel indicates the zoomed zone shown in the upper panel.}
    \label{fig:velocity_map_Ha_ranges}
\end{figure*}

Figure~\ref{fig:velocity_map_Ha_ranges} shows the radial velocity maps for the 175 SQ H$\alpha$ regions. The SQ H$\alpha$ regions have been catalogued in two velocity sets, LV and HV, according to the definitions given in Sect.~\ref{subsec:reduc}. Traditionally, two velocity ranges were observed in the shock zone. However, \cite{2012A&A...539A.127I} showed an additional velocity range in the shock region. As can be seen in Fig.~\ref{fig:velocity_map_Ha}, at least five different velocity ranges are clearly distinguished, therefore, we can see two more velocity components. For low radial velocity components, two radial velocity ranges are shown: i) v=[5600-5900] $km\, s^{-1}$, associated with the NIs and the SDR; ii) v=[5900-6100] $km\, s^{-1}$, associated with the north of NSQA, SSQA, NW, and the Ls. For high radial velocity components, three radial velocity ranges are shown: i) v=[6100-6600] $km\, s^{-1}$, associated with the Shs, connecting the part of SQ with high radial velocities with the lower radial velocities; ii) v=[6600-6800] $km\, s^{-1}$, associated with the YTT, SQA, and AGN north lobe; iii) v=[6800-7000] $km\, s^{-1}$, associated with the Hs, the strands that connect Shs with SQA, and those regions of strands Sh3 and Sh4 (e.g. ID 56, 80b, 85, 89b, and 113) that are closest to the Hs strands.

The upper panel of Figure~\ref{fig:velocity_map_Ha_ranges} shows a radial velocity gradient that was found in the SQ H$\alpha$ emission regions from NI2 to SSQA, NI2 to NI3, and NI2 to NI5, increasing northwards up to 6100 $km\, s^{-1}$ (the minimum radial velocity is in region 149, radial velocity $\sim$5610 $km\, s^{-1}$). Located north of NSQA we can see the tidal tail NW with radial velocity $\sim$6000 $km\, s^{-1}$. According to the simulations from \cite{2010ApJ...724...80R}, NW could have been formed by the merger between NGC7318B and NGC7318A. We have no clear evidence of this, but we believe that NSQA and SSQA have participated in this formation. This possible large-scale structure, which could be linked with the region 145 (v=5994 $km\, s^{-1}$), is revealed. NI strands and SDR have lower radial velocities than $\sim$5900 $km\, s^{-1}$. Moreover, regions with radial velocities higher than 6160 $km\, s^{-1}$ are presented in Fig.~\ref{fig:velocity_map_Ha_ranges} (lower panel). We detected several strands (Shs) that connect HV with LV parts covering a range of radial velocities from $\sim$6100 to $\sim$6600 $km\, s^{-1}$. SQA has radial velocity $\sim$6670 $km\, s^{-1}$. Additionally, the north lobe has radial velocities from 6750 to 6870 $km\, s^{-1}$. We found two SQ regions in the H$\alpha$ 'bridge' (radial velocity $\sim$6400 $km\, s^{-1}$, see Sect.\ref{subsec:diff} for more information). The strands H1 and H2 connect Shs with SQA. The radial velocity and the Dec increase in H1 up to region 111 (radial velocity $\sim$7000 $km\, s^{-1}$). From region 111 to SQA (in H2) the radial velocity decreases. The NGC7319 nucleus (regions 26, 27, and 29) shows radial velocities between $\sim$6350 and 6760 $km\, s^{-1}$ as a result of the outflow from the nucleus \citep{2014MNRAS.442..495R}. It is noteworthy that the NGC7319 'arm' has regions with radial velocities from $\sim$6710 (in region 28) to $\sim$6600 $km\, s^{-1}$ (in region 18). We do not discard the possibility that it is produced by the interaction between NGC7319 and another galaxy and it will remain as a conatus of tidal tail. The YTT has an average radial velocity of $\sim$6620 $km\, s^{-1}$. YTT increases its radial velocity as we move away from NGC7319. Another interesting finding is the NG (see Sect.~\ref{subsec:glx_SQ}).

\subsection{Regions with broad line profile}
\label{subsec:broad}
\begin{figure}
    \centering
    \includegraphics[width=\columnwidth]{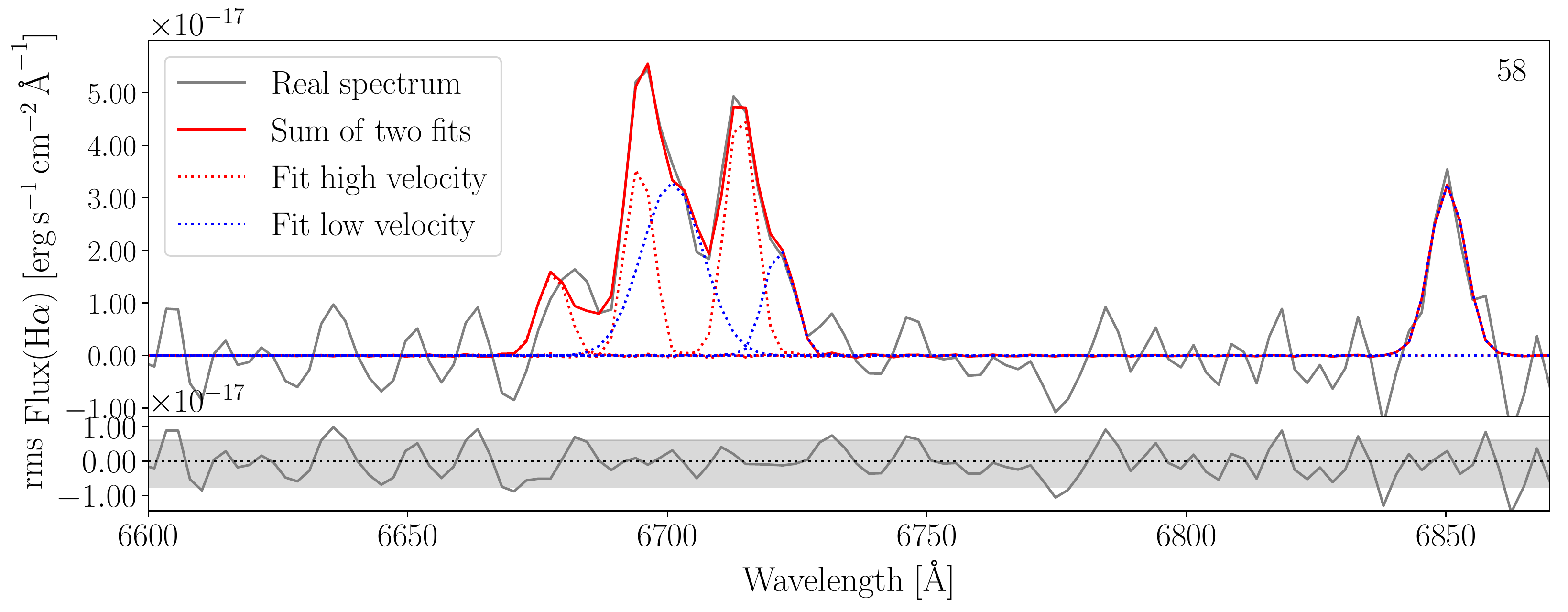}\\ \vspace{-0.6cm}
    \includegraphics[width=\columnwidth]{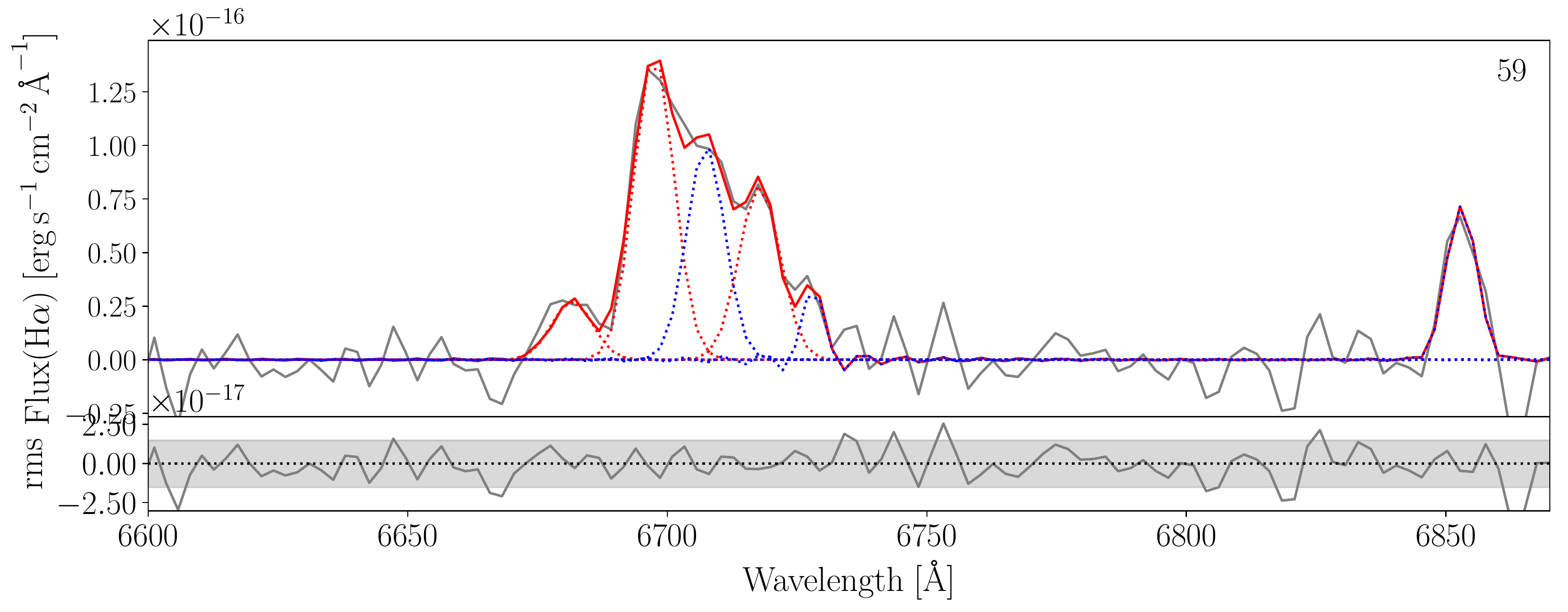}\\ \vspace{-0.6cm}
    \includegraphics[width=\columnwidth]{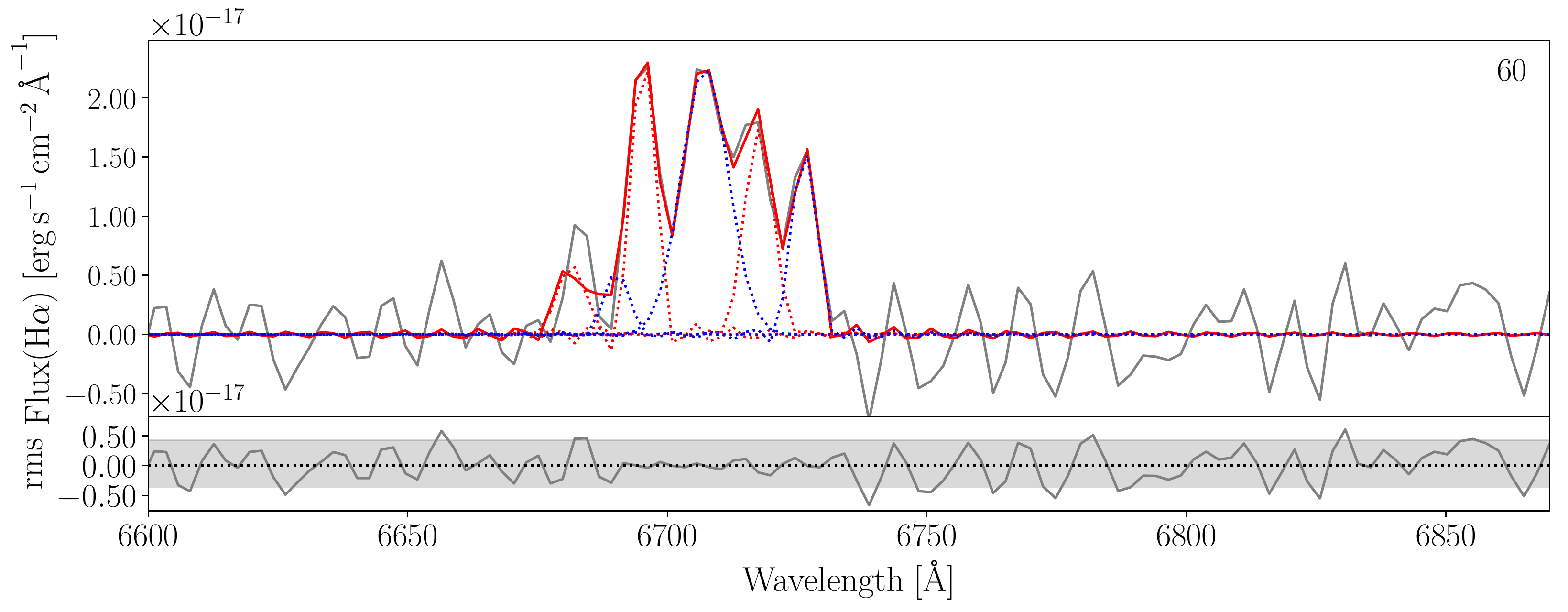}\\ \vspace{-0.6cm}
    \includegraphics[width=\columnwidth]{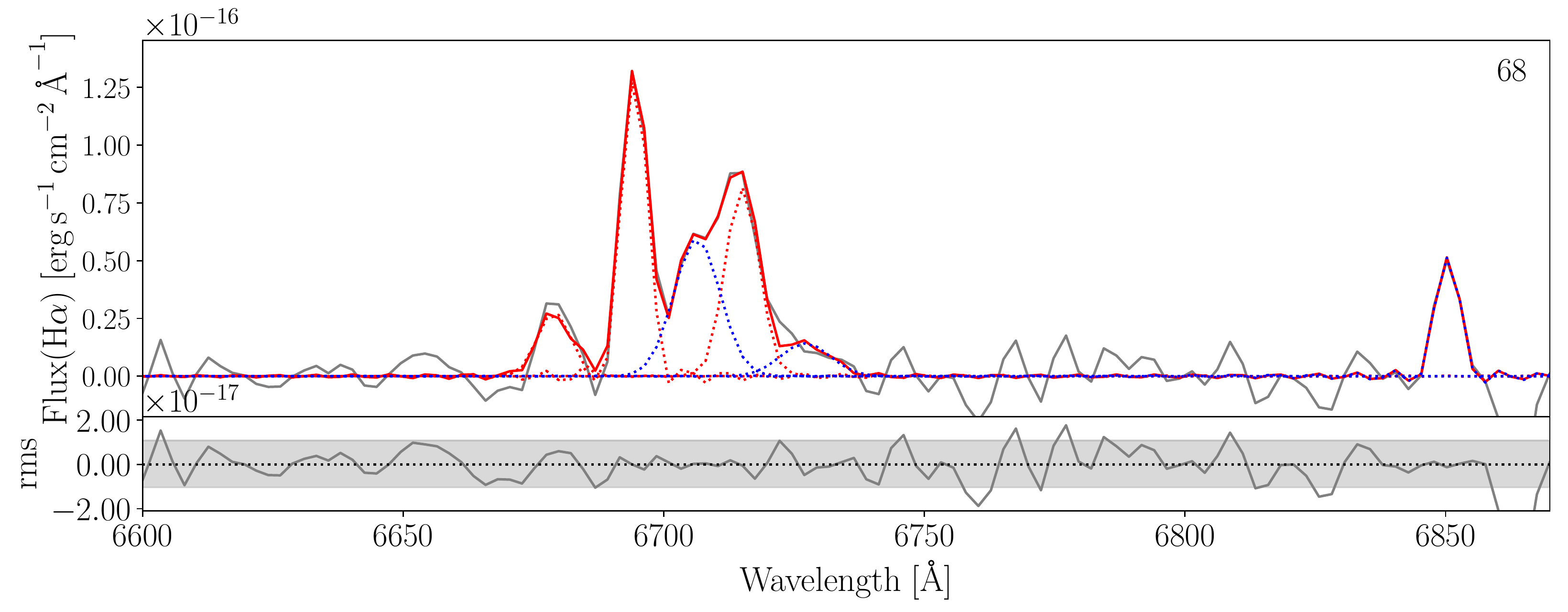}\\ \vspace{-0.6cm}
    \includegraphics[width=\columnwidth]{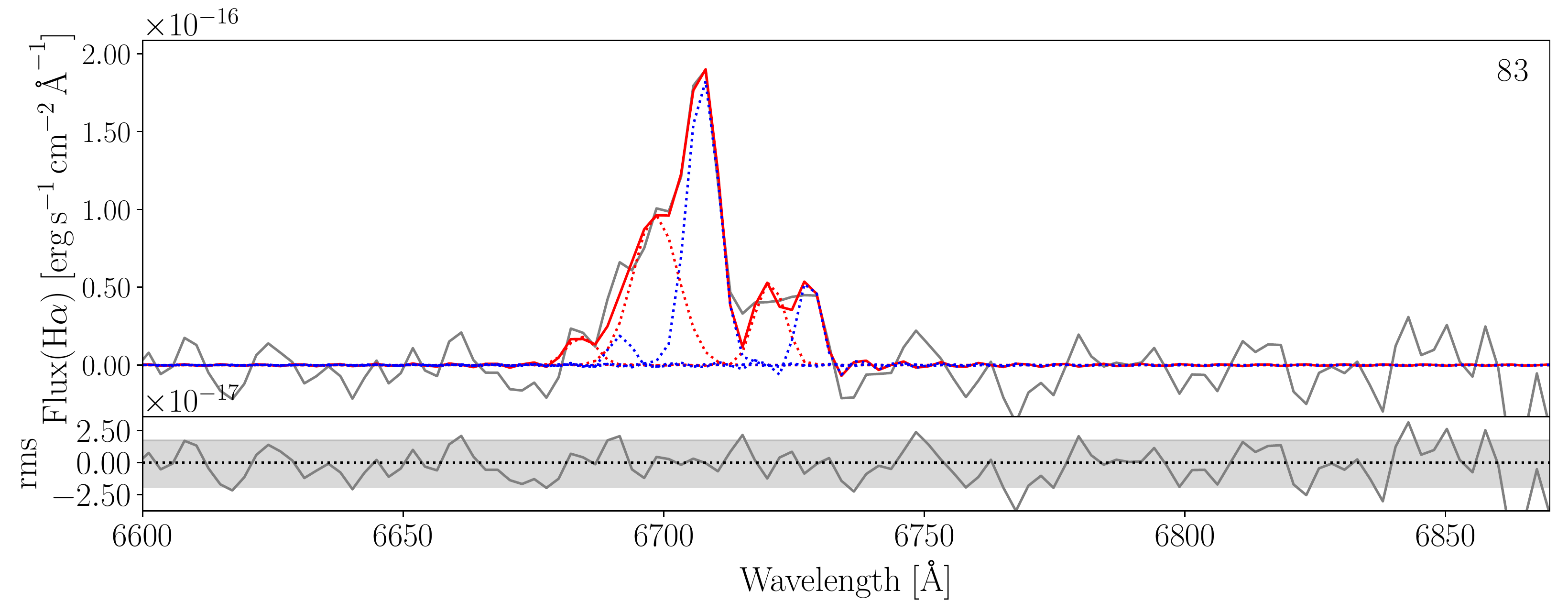}
    \caption{From top to bottom: examples of the line fits from regions 58, 59, 60, 68, and 83 with a broad line profile. In each upper panel, the grey line shows the real spectrum, and the red and blue dotted lines correspond to the fit of the low and high velocity components. The red solid line shows the sum of the two velocity component fits. The name of the region is indicated in each upper panel (top right). In the lower panels, the grey line shows the residual after the fit. The horizontal grey band indicates the 3$\sigma$ scatter.}
    \label{fig:spectrobroad}
\end{figure}

As we can see in Sect.~\ref{subsec:sample}, SQ presents regions with broad and narrow line profiles. The regions showing a broad line profile are located in the LSSR, the AGN nucleus of NGC7319, and the L1 and L4 zones. In Fig.~\ref{fig:spectrobroad} we show several examples of these individual regions (noted as ID 58, 59, 60, 68, 83 in Table~\ref{table:table2}) across the LSSR, and their corresponding fits using two sincgaussian functions, as explained in Sect.~\ref{subsec:sample}. A close visual inspection has also been carefully performed of all the fits of the broad line profile.

\begin{figure}
    \centering
    \includegraphics[width=\columnwidth]{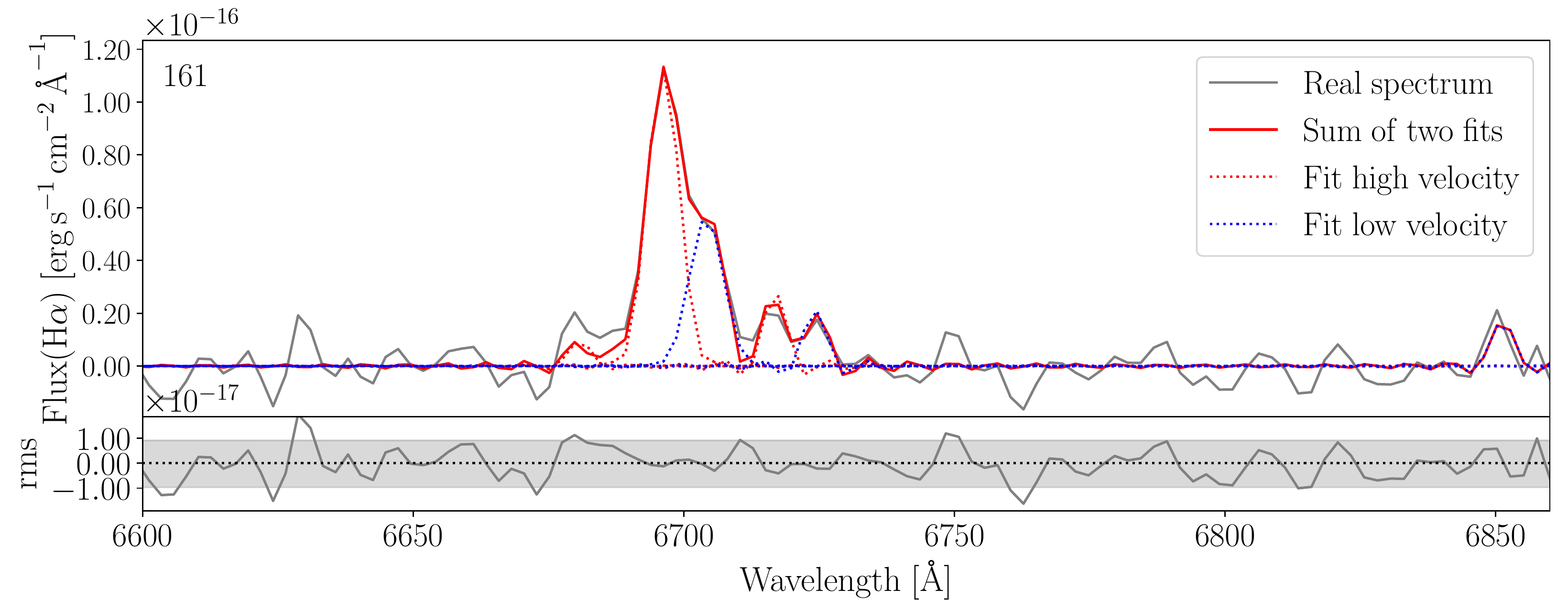}
    \caption{Same as Fig.\ref{fig:spectrobroad} but for region 161.}
    \label{fig:spectrobroad_161}
\end{figure}

In Fig.~\ref{fig:spectrobroad_161} we present the broad H$\alpha$ profile we found for region 161 located in the L4 area, a somewhat outlying zone (see Fig.~\ref{fig:regions}), and its two velocity component fits. As far as we know, this broad profile is presented here for the first time for region 161. It is tempting to associate this broad profile region with a nearby secondary maximum apparent in the X-Ray maps of \cite{2005A&A...444..697T}. However, the precise nature and the excitation of this gas still has not been ascertained.

\begin{figure}
    \centering
    \includegraphics[width=\columnwidth]{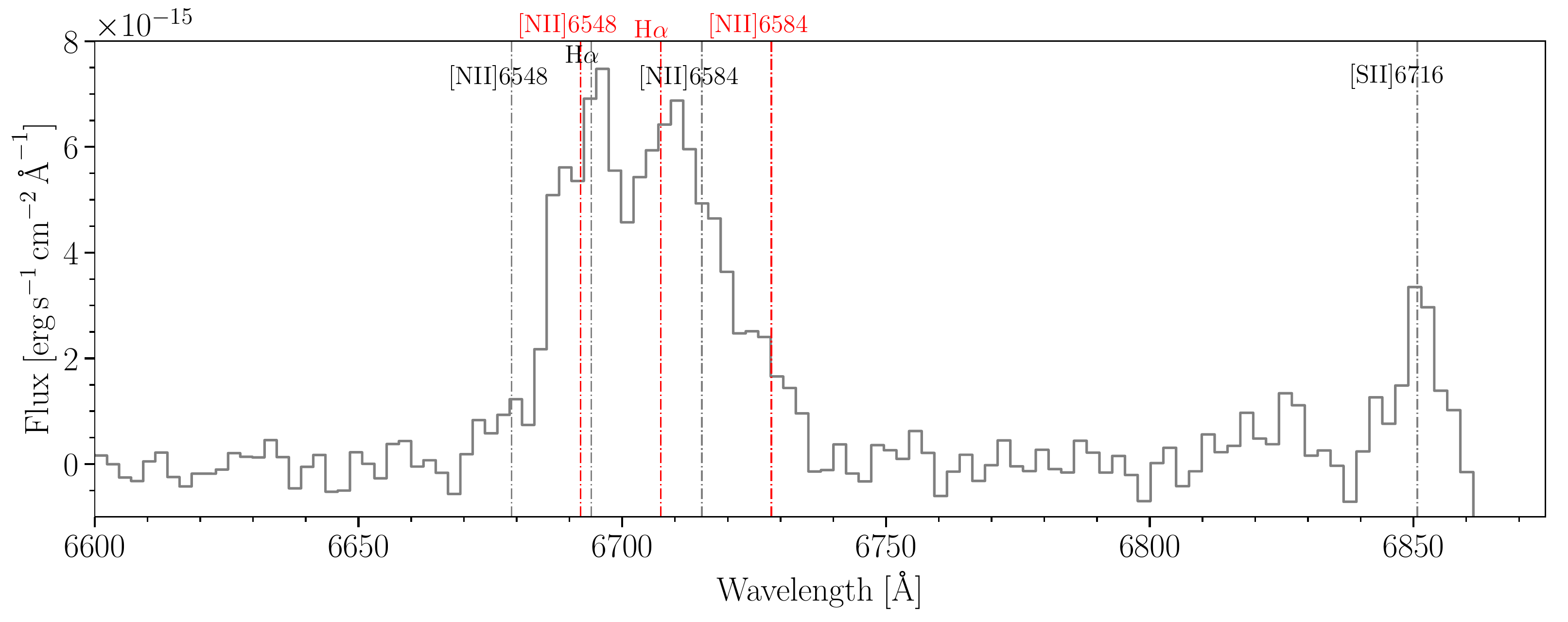}\\
    \includegraphics[width=\columnwidth]{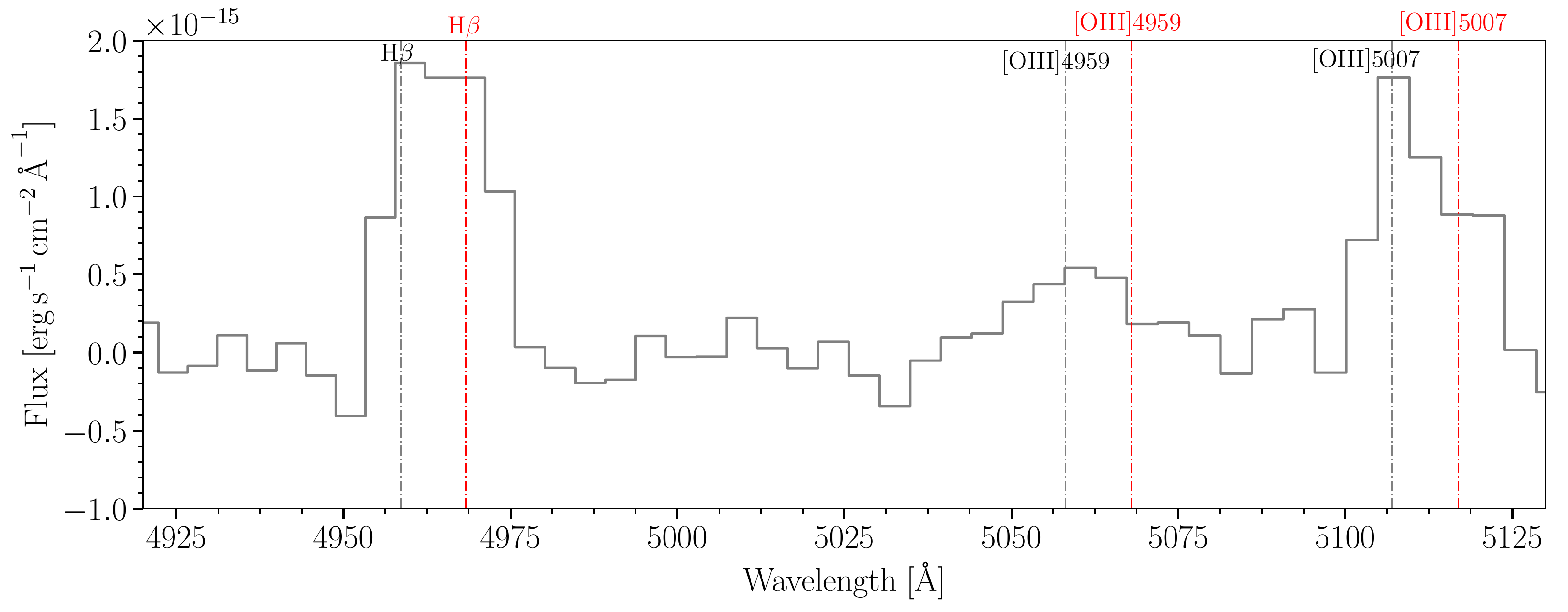}\\
    \includegraphics[width=\columnwidth]{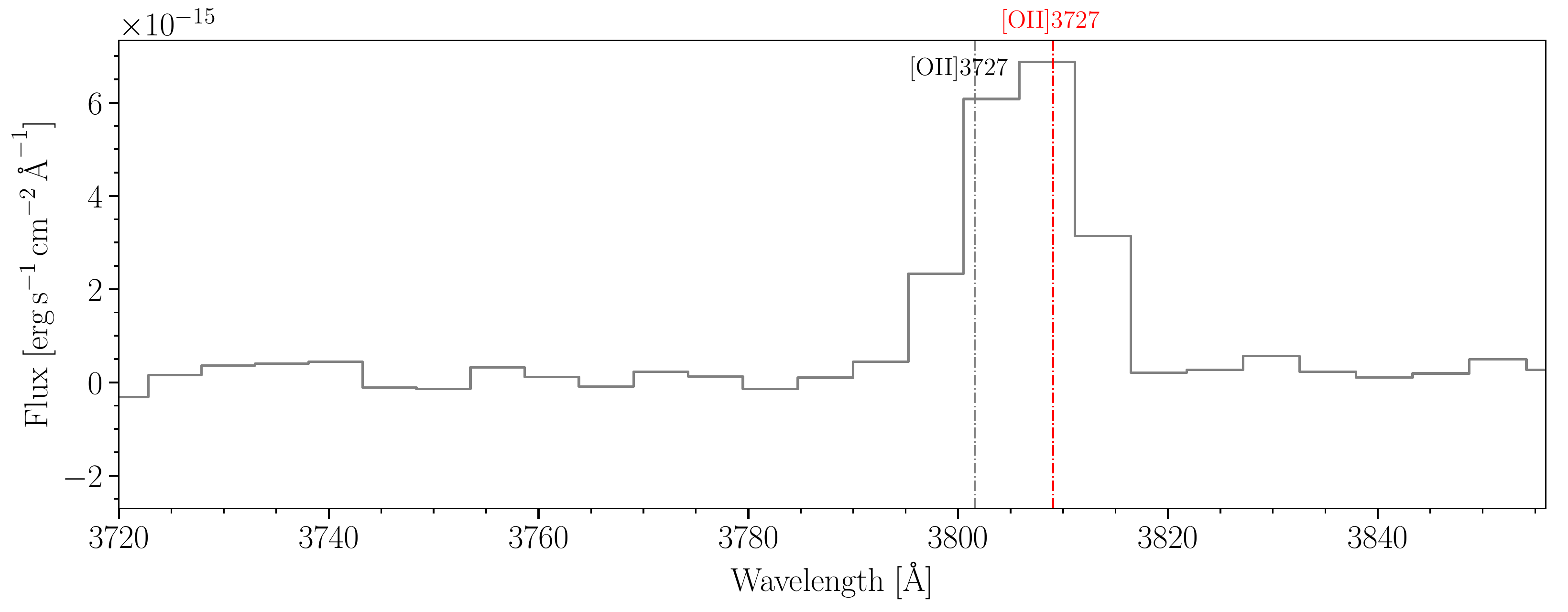}
    \caption{Integrated spectra from large scale shock region in SN3, SN2, and SN1 data cubes. The black and red vertical dashed lines show the position of the emission lines studied at H$\alpha$ radial velocities of 6000 $km\, s^{-1}$ and 6600 $km\, s^{-1}$, respectively.}
    \label{fig:shock}
\end{figure}

\begin{figure}
    \centering
    \includegraphics[width=\columnwidth]{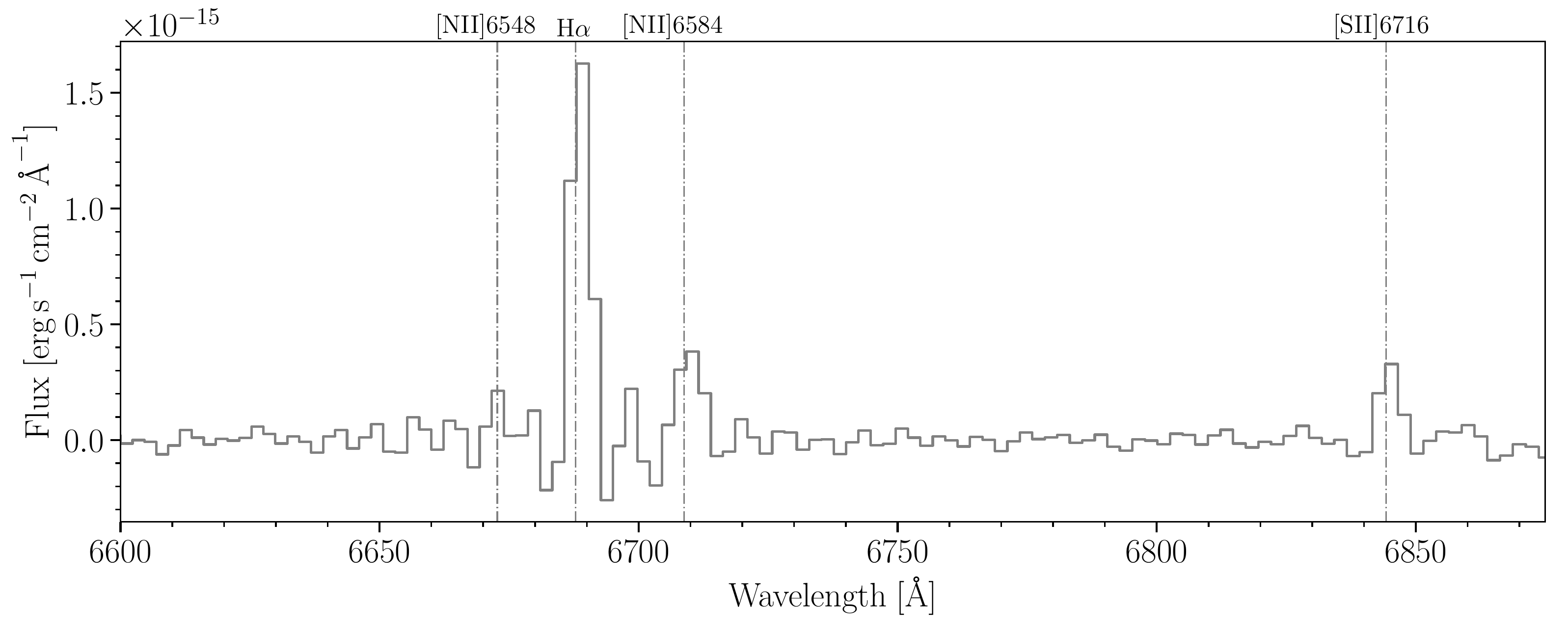}\\
    \includegraphics[width=\columnwidth]{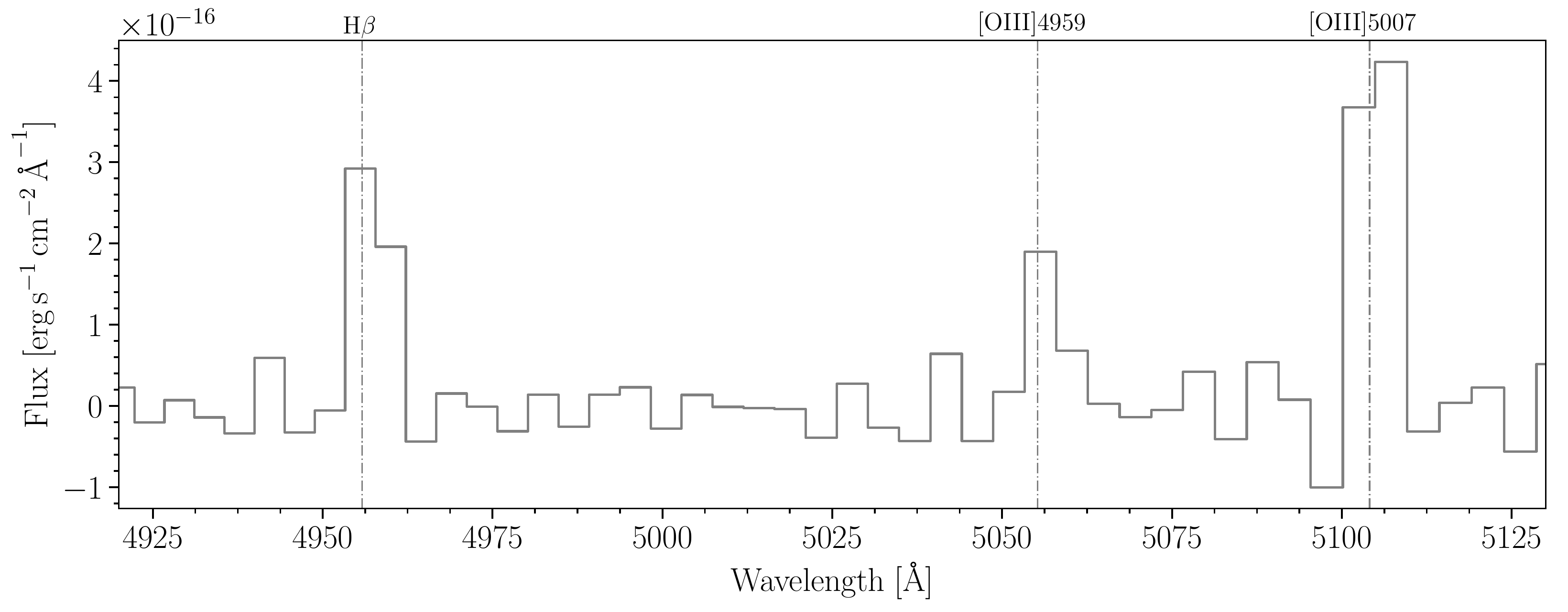}\\
    \includegraphics[width=\columnwidth]{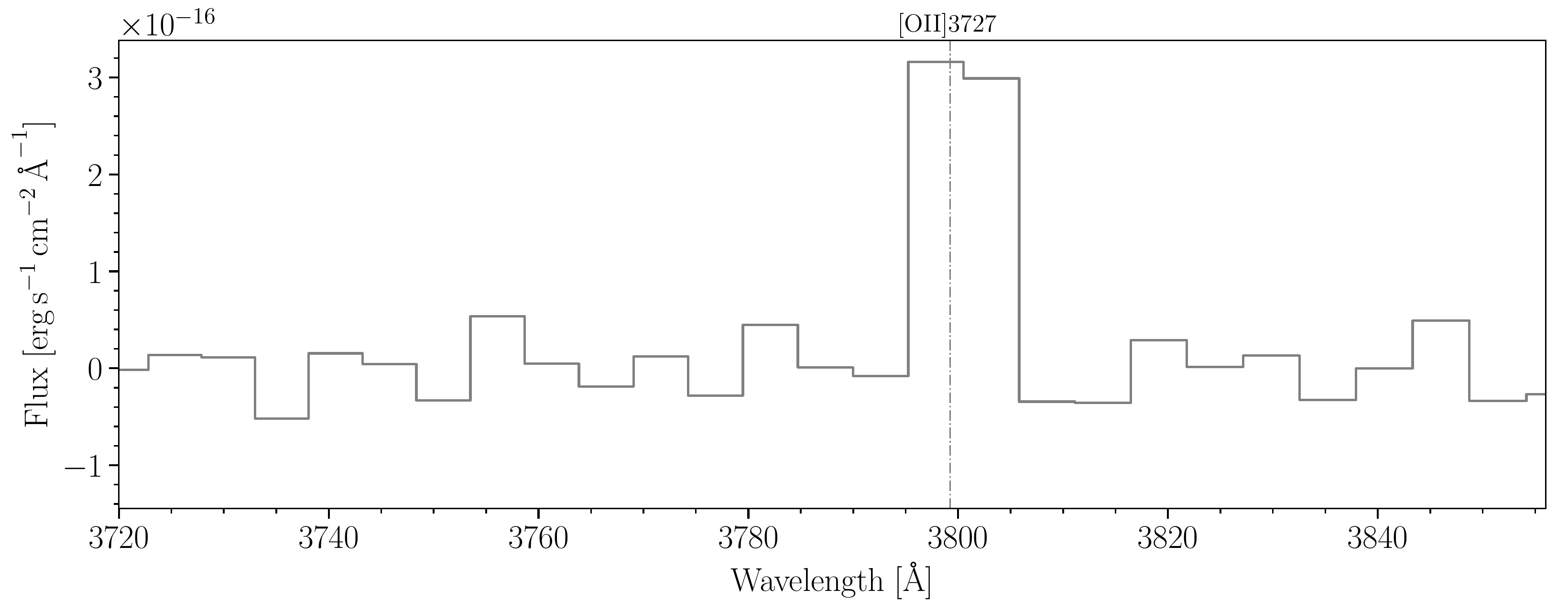}
    \caption{Integrated spectra from SQ H$\alpha$ region 164 in SN3, SN2, and SN1 data cubes.}
    \label{fig:narrow}
\end{figure}

In order to show the wide range of velocities encompassed within LSSR, its integrated emission is presented in Fig.~\ref{fig:shock} in the SN1, SN2, and SN3 filter windows. To obtain these spectra, we considered all the H$\alpha$ emission detected (binning 6x6 and contrast $\geq$ 2) with H$\alpha$ radial velocities $\geq$ 5000 $km\, s^{-1}$ within a rectangle centred on the coordinate RA=339 deg and Dec=33.966 deg, with a width of $\Delta$RA= 27.786 arcsec and a height of $\Delta$Dec=99.512 arcsec (defined according to the definition of the flux, as determined in the next section). As expected, the integrated spectra from LSSR show broad lines and several predominant peaks (e.g. H$\alpha$ and [\ion{N}{ii}]$\lambda$6584). For illustrative purposes, these can be compared with a typical narrow-lined H$\alpha$ emission region (i.e. ID 164; Fig.~\ref{fig:narrow}).

\subsection{Unveiling low surface brightness H$\alpha$ emission}
\label{subsec:diff}

\begin{figure*}
    \centering
    \includegraphics[width=\textwidth]{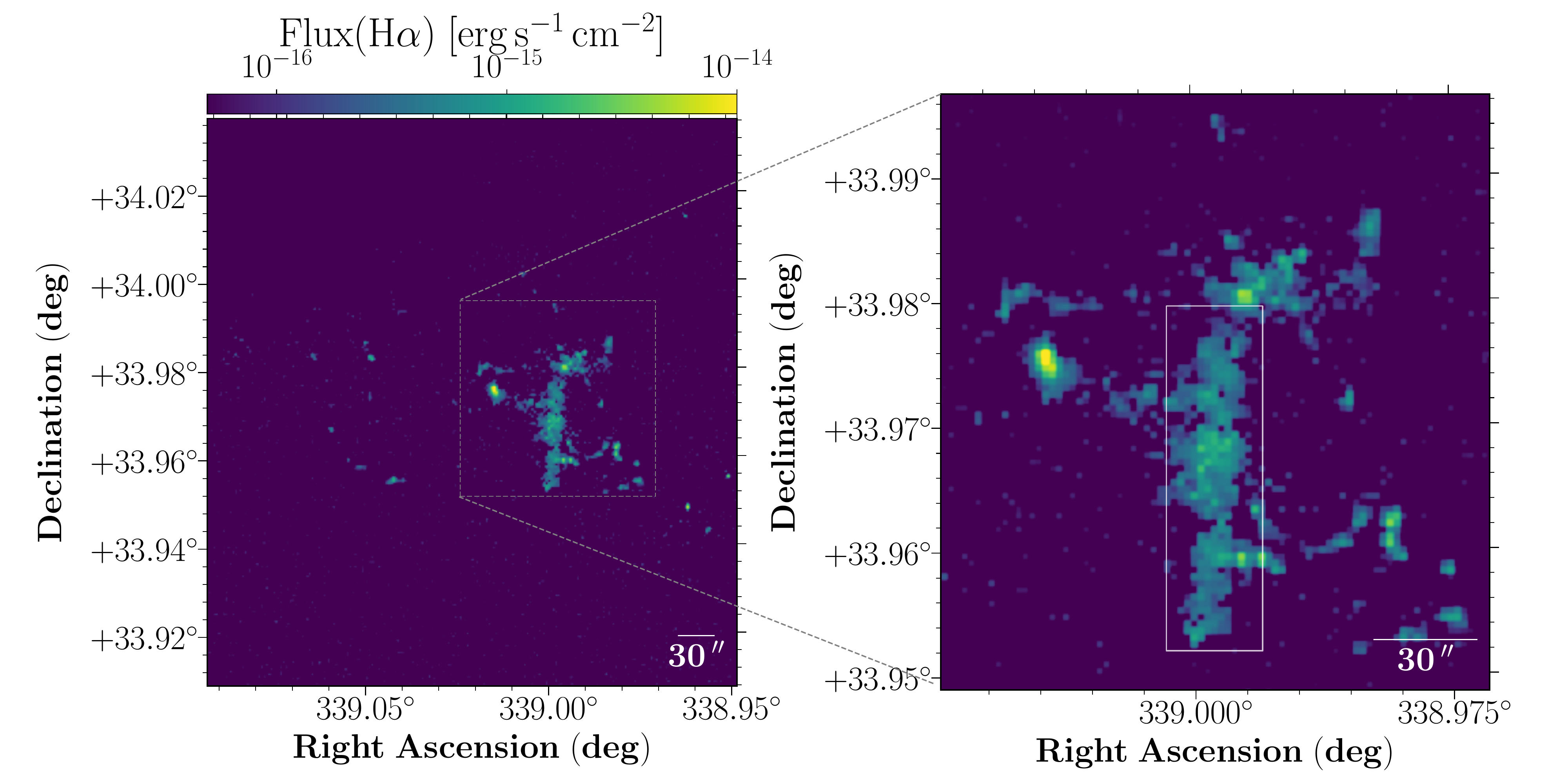}\\
    \includegraphics[width=\textwidth]{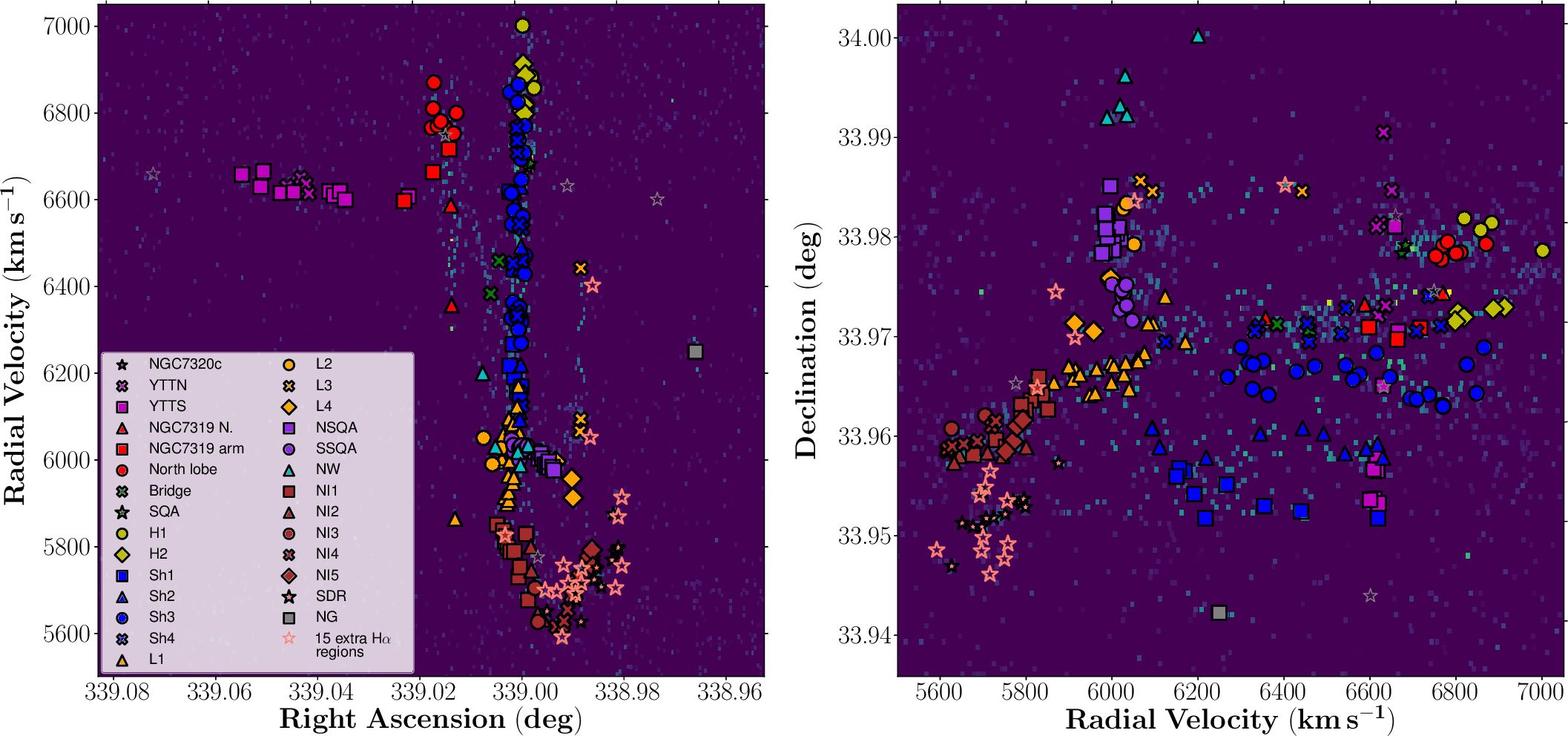}
    \caption{H$\alpha$ flux spatial map of entire SQ (upper left panel), H$\alpha$ flux spatial map zoomed in to shock region (upper right panel). The lower left panel shows the radial velocity versus RA diagram and the lower right panel shows Dec versus radial velocity diagram. All the points in the lower panels of the figures have the same colours and markers as Fig.~\ref{fig:regions}. The white rectangle centred on LSSR is overplotted with a solid line in the upper right panel. This rectangle is centred on the coordinate RA=339 deg and Dec=33.966 deg, with a width of $\Delta$RA=27.786 arcsec and a height of $\Delta$DEC=99.512 arcsec.}
    \label{fig:halphaSQ}
\end{figure*}

In order to detect the low surface brightness H$\alpha$ emission regions, we produced a deep overview of all the H$\alpha$ emission of the SQ system in Fig.~\ref{fig:halphaSQ}. This was produced using a binning 6x6, contrast(H$\alpha$)$\geq$2, and radial velocities in the range of the SQ, in order to increase the detectability of the low surface brightness emission, as explained in Sect.~\ref{subsec:reduc}. From a statistical analysis of our H$\alpha$ flux measurements, we find that below a value of H$\alpha$ flux per pixel of 6$\times$10$^{-18}$ erg/s/cm$^2$ we can distinguish a very low surface brightness H$\alpha$ component besides the SQ H$\alpha$ emission regions measured. The upper right panel of Fig.~\ref{fig:halphaSQ} shows a zoomed-in view of the LSSR, whereas the lower panels, as in Fig.~\ref{fig:velocity}, shows an overlapped view of the symbols corresponding to the very low surface brightness regions detected with the SQ H$\alpha$ emission regions found in Sect.\ref{subsec:sample}. This low surface brightness gas emission represents different components spread throughout the whole system. This gives us important information about the connections between the different SQ zones (see Sect.~\ref{subsec:sample}) and helps us to better understand the complexity of SQ.

At lower radial velocities, v$\leq$6160 $km\, s^{-1}$ (see lower panels in Fig.~\ref{fig:halphaSQ}), we observe at least two subsystems that were probably in collision. On one hand, NI (brown symbols)\footnote{All symbols referenced in this subsection come from Fig.~\ref{fig:regions}.} colliding with the debris field in the shock path at L1 strand (orange triangles), on the other hand, the interaction between NSQA and SSQA (violet squares and circles, respectively). While the collision between NI and the debris field in the LSSR appears to be ongoing at this moment \citep[e.g.][]{2012A&A...539A.127I}, our observations indicate that the interaction between NSQA and SSQA has had at least one previous collision. Several strands of ionised gas emerging from NSQA are found here: to the north of NSQA a tidal tail, NW (cyan triangles) is indicated, while on the right we see the strand L3 (which increases towards higher radial velocities, $\sim$6470 $km\, s^{-1}$, orange crosses) and the strand L4 (that goes southwards and could link with NI, orange diamonds). The strand L2 (orange circles) connects NSQA with SSQA, which reinforces the idea that NSQA and SSQA have had at least one previous interaction. The strand L1 joins the system formed by NSQA and SSQA with NI1. We must bear in mind that the majority of the regions from the strand L1 are compatible with fast shocks without precursor for solar metallicity and low density, with velocities between 175 $km\, s^{-1}$ and 300 $km\, s^{-1}$ from \cite{2008ApJS..178...20A} models (Duarte Puertas et al. paper II, in prep).

At intermediate radial velocities (i.e. from v$\sim$6160 to $\sim$6660 $km\, s^{-1}$), several strands can be seen connecting Hs with Ls through the Shs strands (namely Sh1, Sh2, Sh3, and Sh4 represented by blue squares, triangles, circles, and crosses, respectively; see Figs.~\ref{fig:velocity}, upper right panel, and \ref{fig:halphaSQ}). It seems clear that this gas is a product of the interaction processes that occurred in SQ between NI and the debris field in the shock, although understanding these processes becomes complex. We also see at radial velocities $\sim$6360 $km\, s^{-1}$ H$\alpha$ flux that connects the shock (Sh4 strand) with NGC7319, into an H$\alpha$ 'bridge' (green crosses). This bridge has also been observed at other wavelength ranges \citep[e.g. molecular hydrogen bridge detected:][]{2010ApJ...710..248C,2010A&A...518A..59G,2013ApJ...777...66A,2017ApJ...836...76A}. In Fig.~\ref{fig:halphaSQ} (lower left panel) we can appreciate that this connection extends several hundred $km\, s^{-1}$.

At higher radial velocities, v$>$6660 $km\, s^{-1}$ we distinguish several gas strands. On one hand, from the NGC7319 nucleus (red triangles) an outflow emerges, from radial velocities $\sim$6760 $km\, s^{-1}$ to $\sim$6350 $km\, s^{-1}$, in agreement with \cite{2014MNRAS.442..495R}. On the other hand, the NGC7319 'arm' (red squares) emerges up to radial velocities $\sim$6600 $km\, s^{-1}$ westward towards YTT (magenta crosses and squares). The AGN north lobe zone (red circles) covers velocities higher and lower than that of the NGC7319 nucleus ($\Delta\,v\sim$200 $km\, s^{-1}$). The right part of the north lobe seems to reach the NGC7319 'arm'. Conversely, in the shock part we see the strands H1 and H2 (yellow circles and triangles). These strands connect Shs with SQA (green stars) in the following ways: i) from Shs (radial velocity $\sim$6660 $km\, s^{-1}$) to the most distant region detected (region 111, radial velocity $\sim$7000 $km\, s^{-1}$) through the strand H2; and ii) from region 111 to SQA (radial velocity $\sim$6670 $km\, s^{-1}$) through strand H1.

We fail to detect ionised gas neither in the old tail, nor in the vicinity of NGC7318A in agreement with \cite{1998A&A...334..473M}, nor in NGC7317 in agreement with \cite{2018MNRAS.475L..40D}. Neither do we find evidence that there is a gas connection between the NGC7319 north lobe and SQA. Finally, we do not detect gas connecting the LSSR and NI5, nor in SDR (salmon stars). 

\subsection{The other galaxies in SQ}
\label{subsec:glx_SQ}

\begin{figure*}
    \centering
    \includegraphics[width=.47\textwidth]{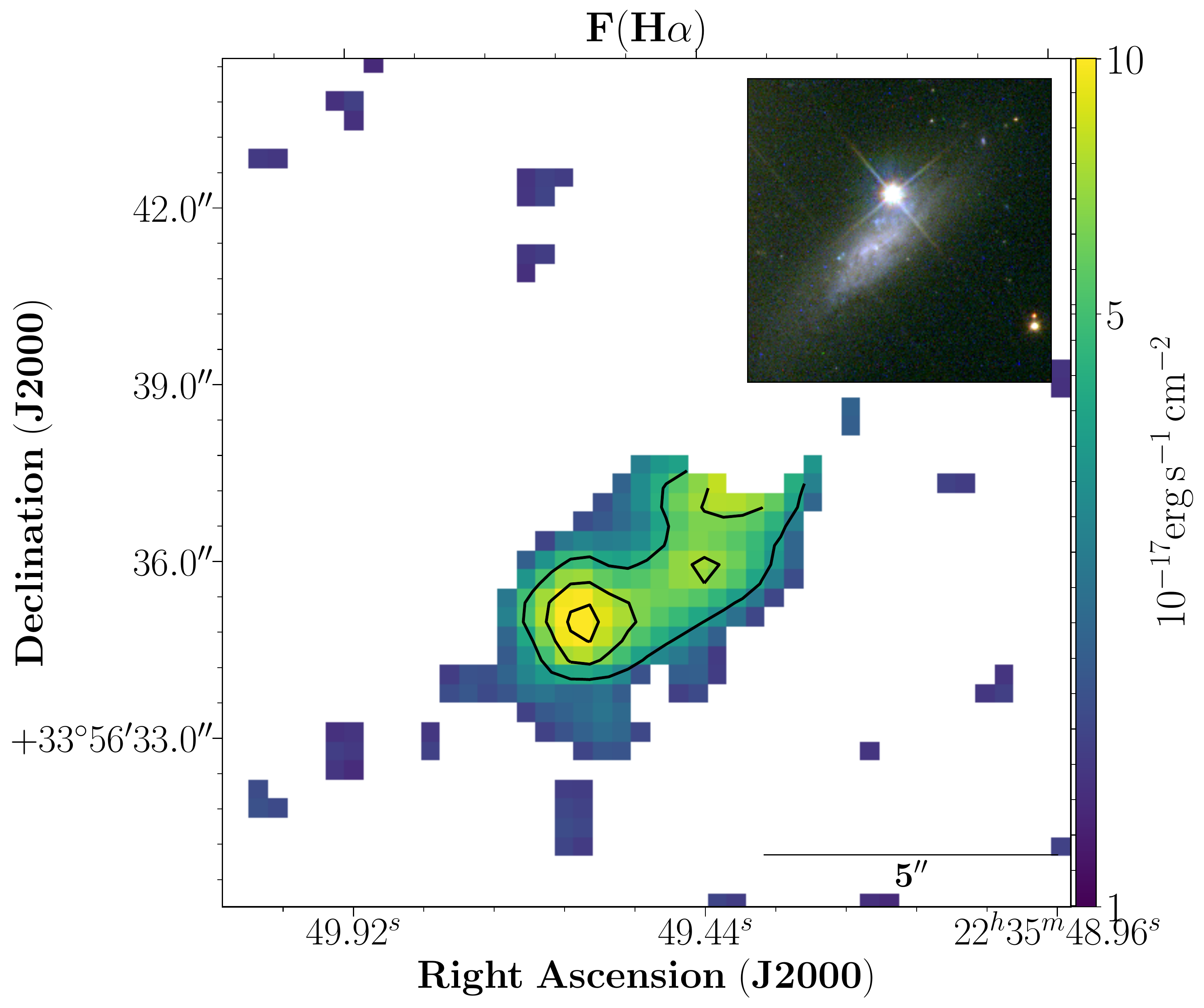}
    \includegraphics[width=.49\textwidth]{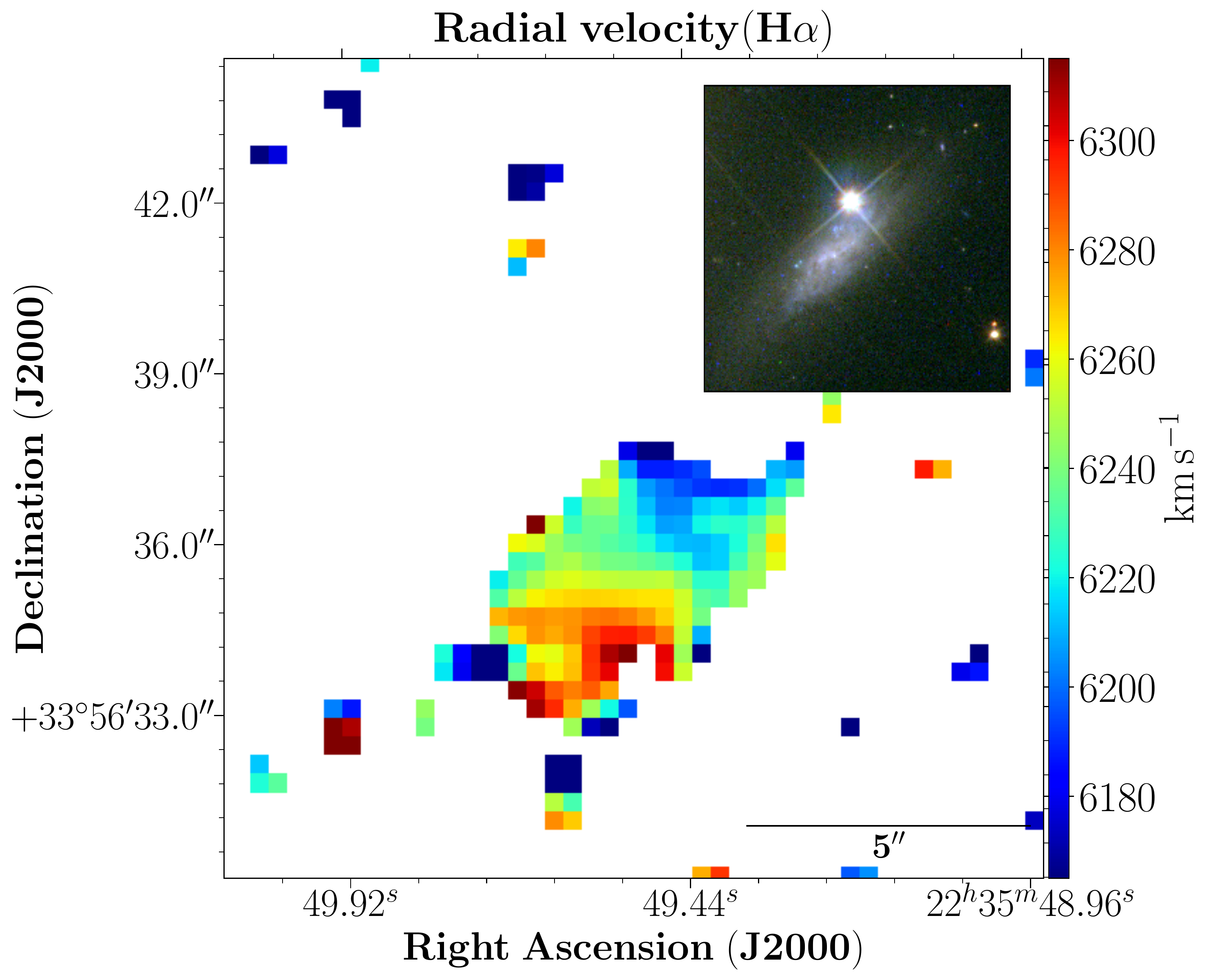}
    \caption{NG H$\alpha$ flux spatial map (left panel) and H$\alpha$ radial velocity map (right panel). The inner panels show the HST/WFC3 optical imaging for NG. The black contours are $1 \times 10^{-17}$, $4 \times 10^{-17}$, and $7 \times 10^{-17}\ erg\,s^{-1}\,cm^{-2}$.}
    \label{fig:NG}
\end{figure*}

\begin{figure*}
    \centering
    \includegraphics[width=.47\textwidth]{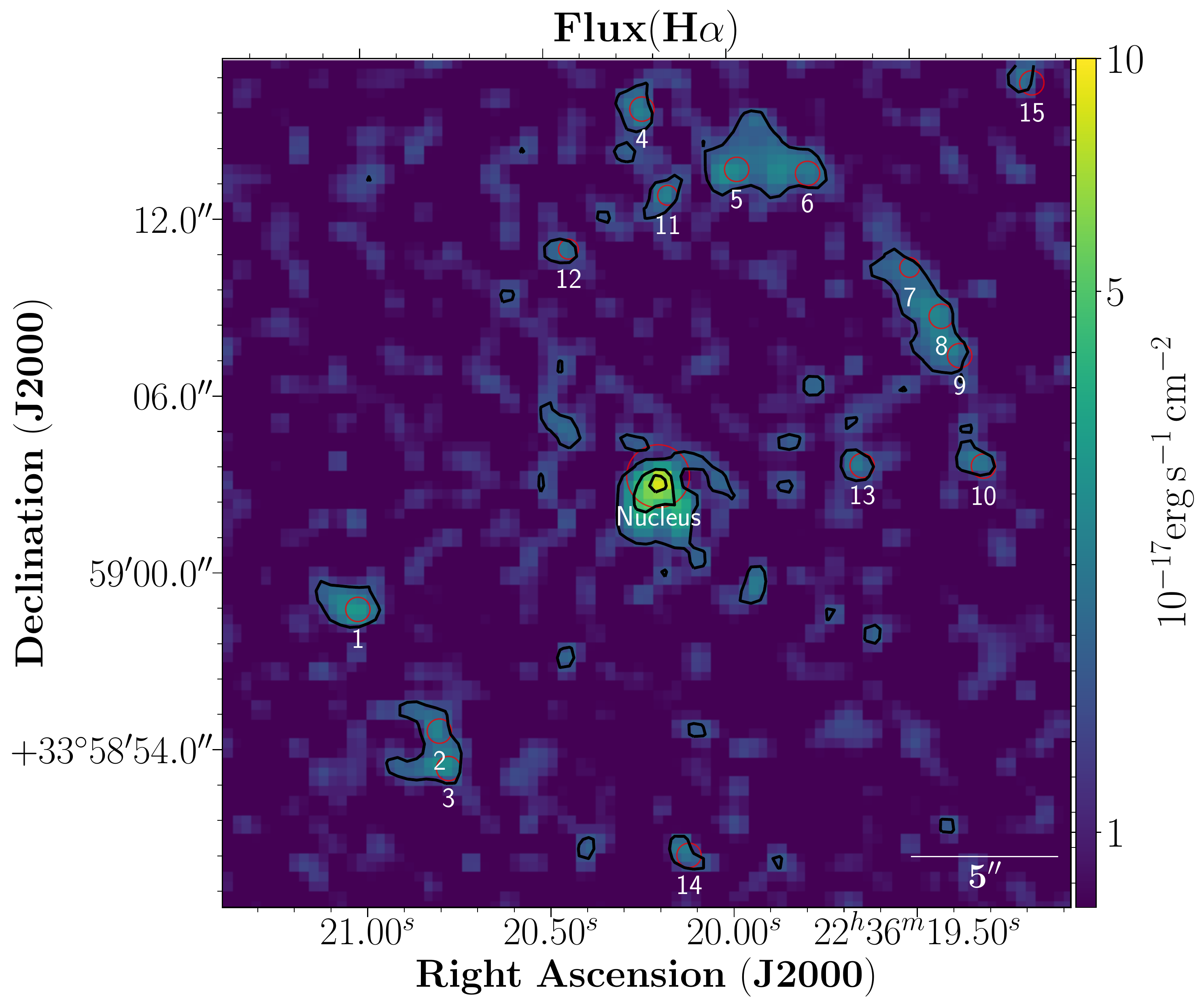}
    \includegraphics[width=.47\textwidth]{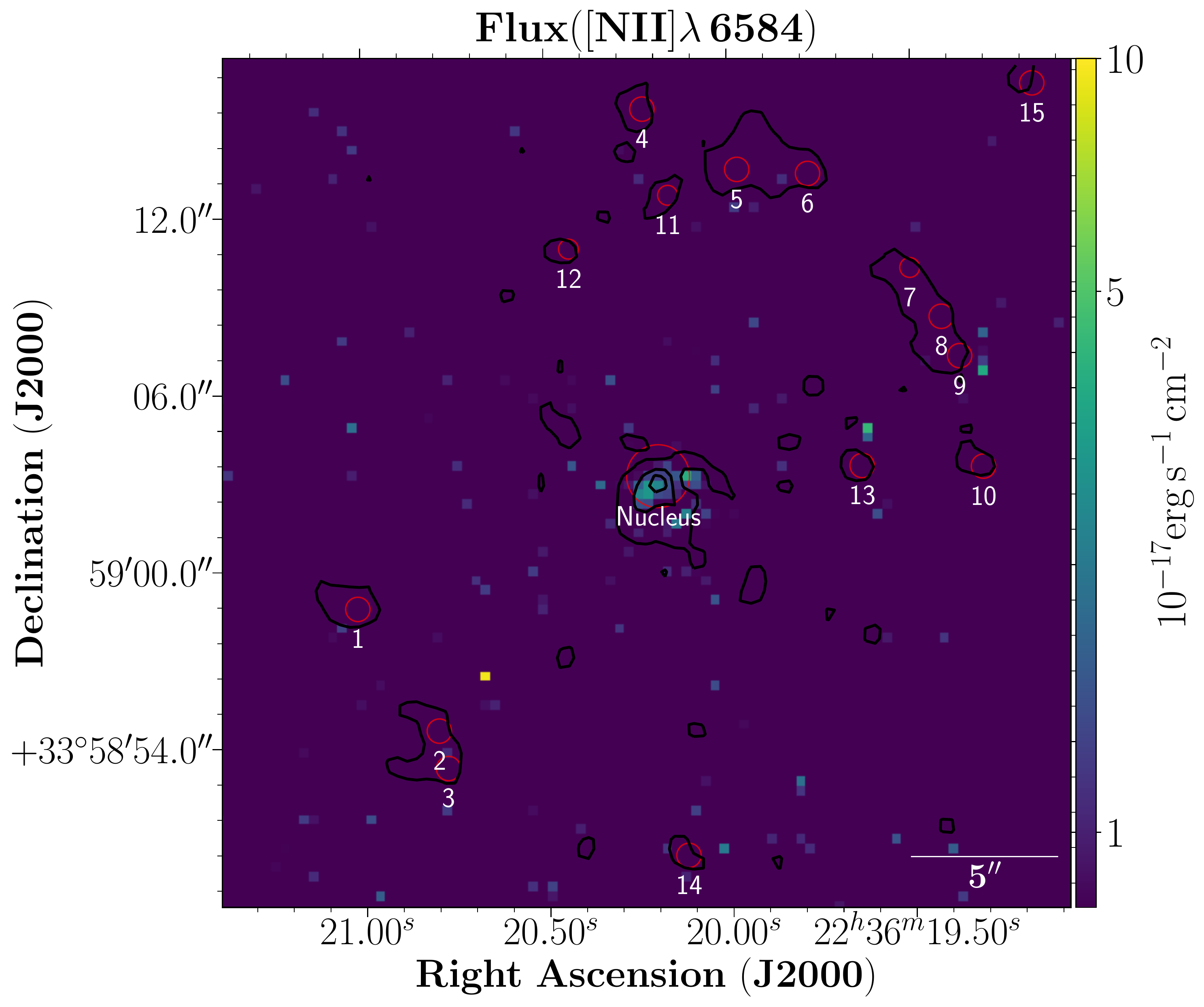}\\
    \includegraphics[width=.49\textwidth]{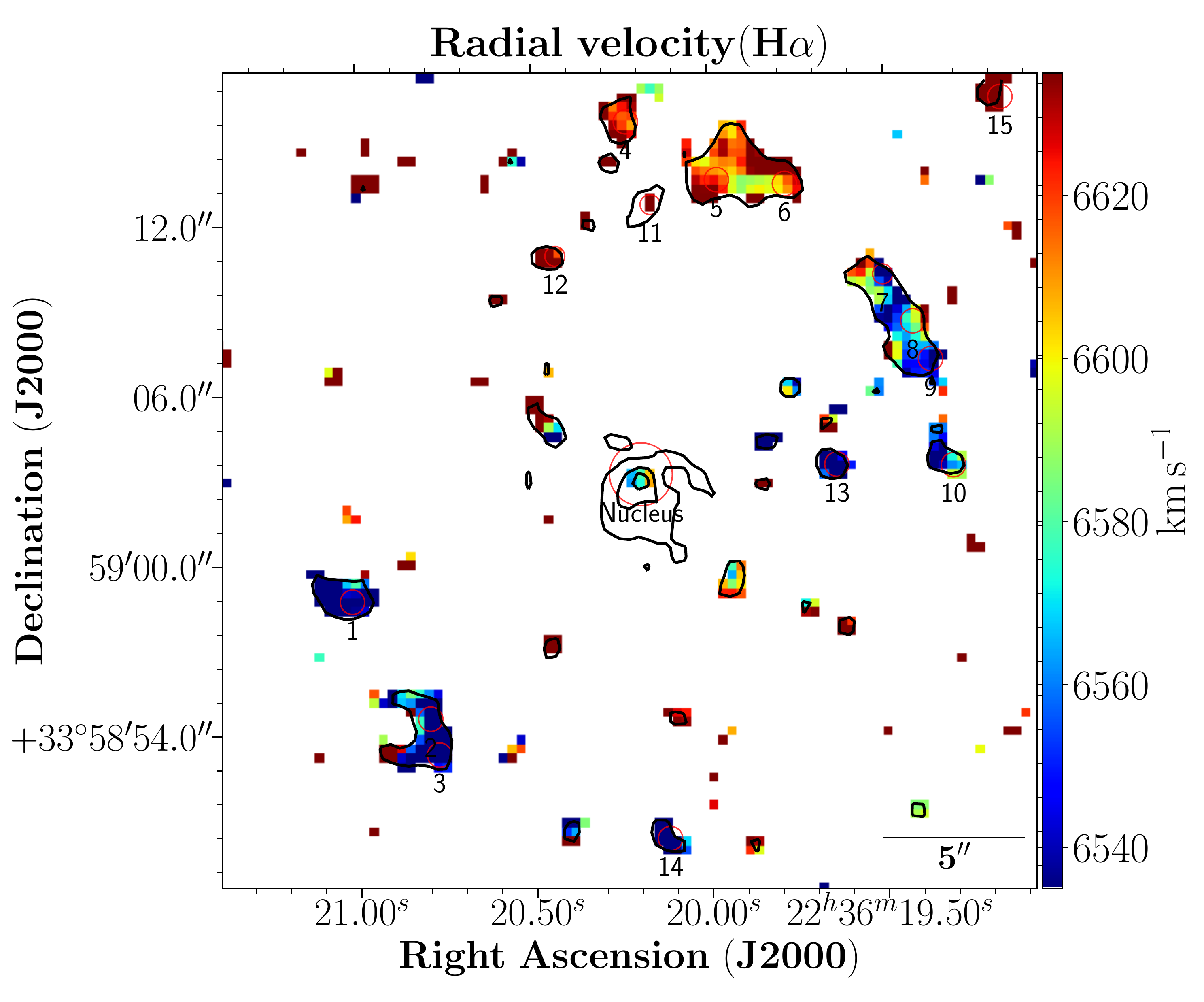}
    \includegraphics[width=.42\textwidth]{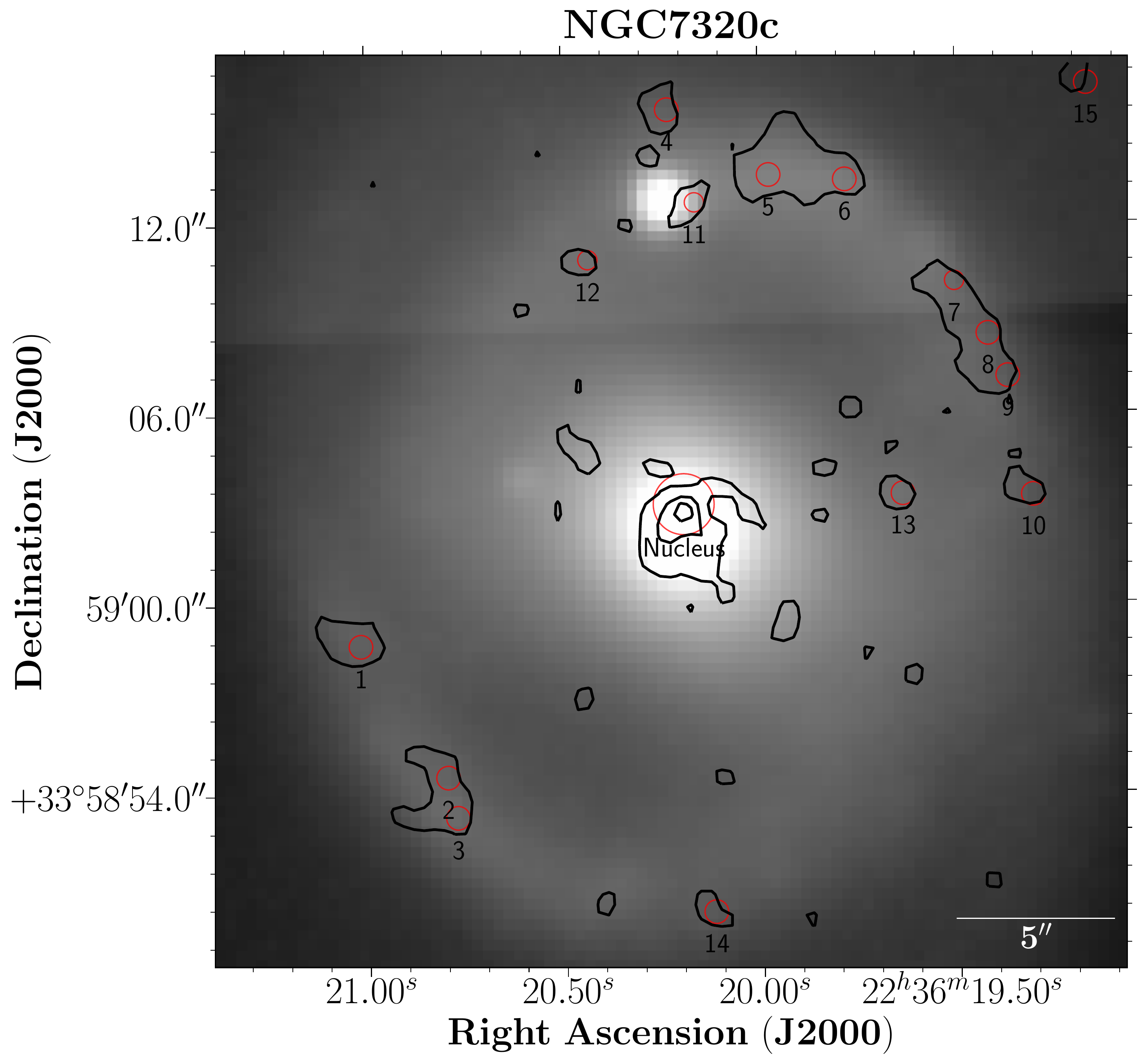}
    \caption{NGC7320c H$\alpha$ flux spatial map (upper left panel), [\ion{N}{ii}]$\lambda$6584 flux spatial map (upper right panel), H$\alpha$ radial velocity map considering pixels with contrast (H$\alpha$) $\geq$ 2.5 (lower left panel), and deep image from SN2 filter (lower right panel). Red circles represent the locations of the NGC7320c nucleus and 15 H$\alpha$ emission regions from its spiral arms. The black contours are $1.5 \times 10^{-17}$, $4 \times 10^{-17}$, and $7 \times 10^{-17}\ erg\,s^{-1}\,cm^{-2}$.}
    \label{fig:NG7320c}
\end{figure*}

In most articles that study SQ, the galaxies that are studied in detail are NGC7317, NGC7318a, NGC7318b, and NGC7319, as well as NGC7320. In this work, thanks to the large SITELLE FoV, we were able to add to the study of the galaxy NGC7320c (the OI) and the NG. In Figs.~\ref{fig:NG} and \ref{fig:NG7320c} we present the spatially resolved H$\alpha$ flux and radial velocity maps for NG and NGC7320c, respectively.

We detect H$\alpha$ emission in NG. This, combined with the fact that NG is extended and presents a galactic rotation of $\pm$ 60 $km\, s^{-1}$ as shown in Fig.~\ref{fig:NG} \citep[e.g.][]{2002A&A...390..829S,2013MNRAS.428.1743W}, indicates that NG is consistent with a dwarf galaxy that will end up interacting with the SQ system in the future, similar to M82 (according
to the Tully-Fisher relation). A more detailed study has been made for NG. In Fig.~\ref{fig:NG} we show the H$\alpha$ flux (left panel) and the radial velocity map (right panel) of NG. The inner panel shows the Hubble Space Telescope optical imaging of NG. The velocity gradient is clearly seen from the south to the north (v$\sim$6300 to $\sim$6150 $km\, s^{-1}$). As we can see in Table~\ref{table:table2}, region 176, the average radial velocity for NG is 6246 $km\, s^{-1}$.

In Fig.~\ref{fig:halphaSQ} (upper left panel) we show the low surface brightness H$\alpha$ emission in the strongly stripped galaxy NGC7320c. Past interaction processes have stripped much of the gas in NGC7320c. For a better understanding of this galaxy, we studied the H$\alpha$ and [\ion{N}{ii}]$\lambda$6584 fluxes for this galaxy. Figure~\ref{fig:NG7320c} shows the H$\alpha$ flux map (upper left panel), the [\ion{N}{ii}]$\lambda$6584 flux map (upper right panel), the H$\alpha$ radial velocity map (lower left panel), and the deep image from the SN2 filter (lower right panel) for NGC7320c. In this figure, we see the presence of H$\alpha$ and [\ion{N}{ii}]$\lambda$6584 emission in the galactic nucleus, and H$\alpha$ emission in several regions located in the spiral arms. Figure~\ref{fig:spectroNGC7320c} shows examples of spectra for several regions with H$\alpha$ emission found in NGC7320c. Observing the fit for the galactic nucleus emission line (see upper panel in Fig.~\ref{fig:spectroNGC7320c}), we ascertain that the spectrum has an AGN profile (and found a strong emission of [\ion{N}{ii}]$\lambda$6584). We barely observe [\ion{N}{ii}]$\lambda$6584 emission (or it may be very small) in the rest of the regions located in the NGC7320c spiral arm. The radial velocity map covers a range of radial velocities from $\sim$6540 to $\sim$6640 $km\, s^{-1}$.

\section{Discussion and final remarks}
\label{sec:conclu}
A sample of 175 SQ H$\alpha$ regions, 22 of them with two velocity components, has been made according to the following criteria: i) the radial velocity of the region is within the range of SQ (between $\sim$5600 and $\sim$7000 $km\, s^{-1}$); ii) the detection of at least one additional emission line besides H$\alpha$. Additionally, in Appendix~\ref{append:app1} we add the relevant information of 15 extra H$\alpha$ emitters that we have found in the southern debris region, close to NI, and to the west of SQ (regions from 177 to 191 in Table~\ref{table:table3}).

\begin{figure}[h!]
    \centering
    \includegraphics[width=.8\columnwidth]{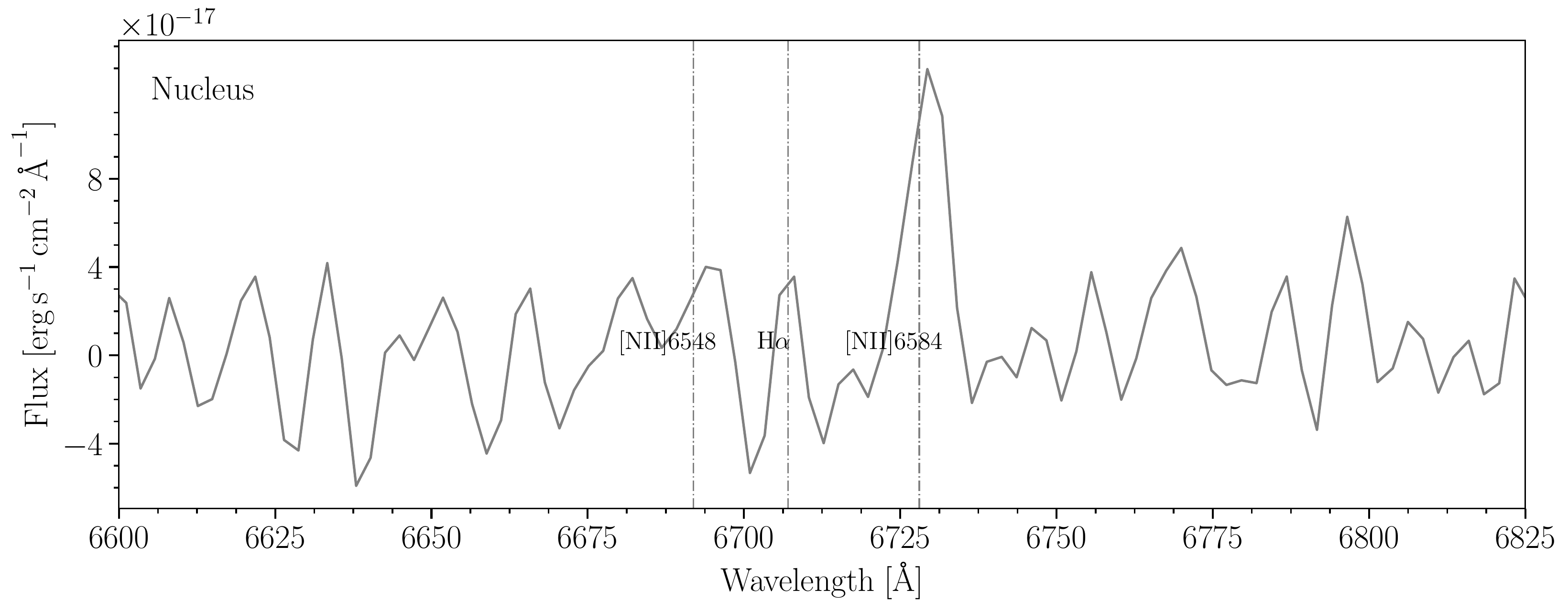}\\ \vspace{-0.6cm}
    \includegraphics[width=.8\columnwidth]{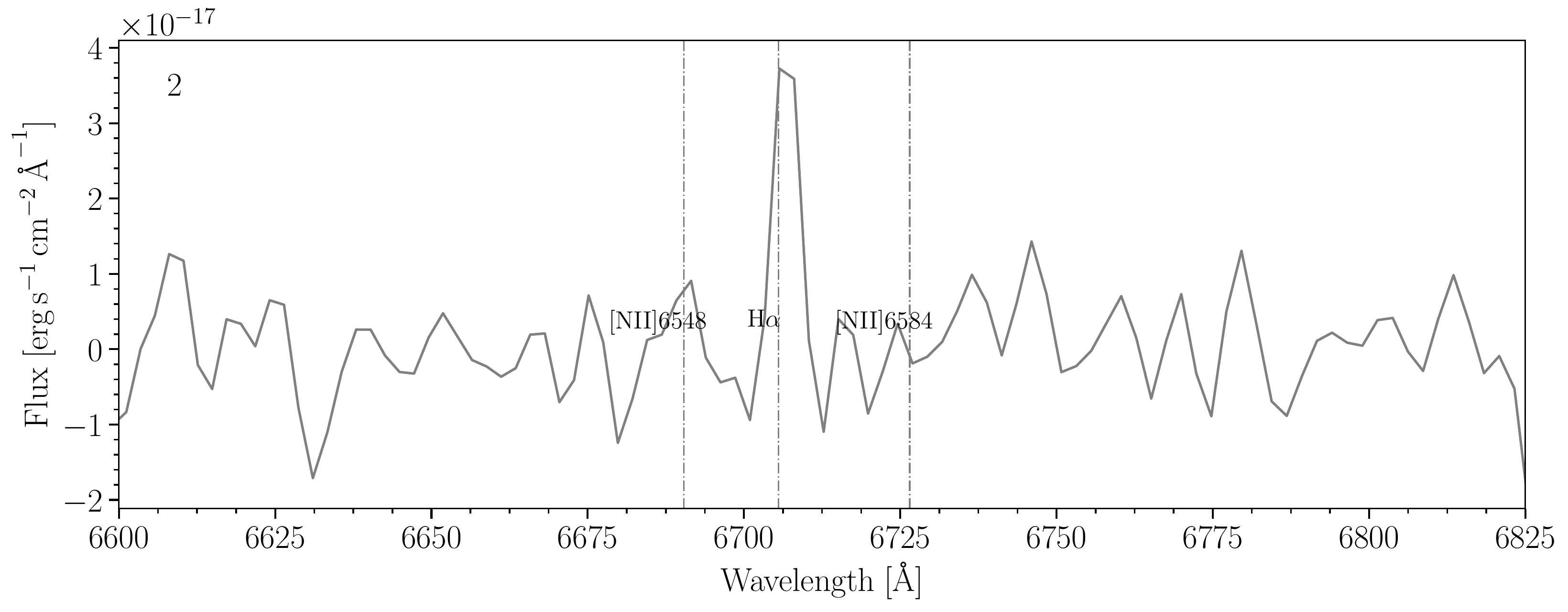}\\ \vspace{-0.6cm}
    \includegraphics[width=.8\columnwidth]{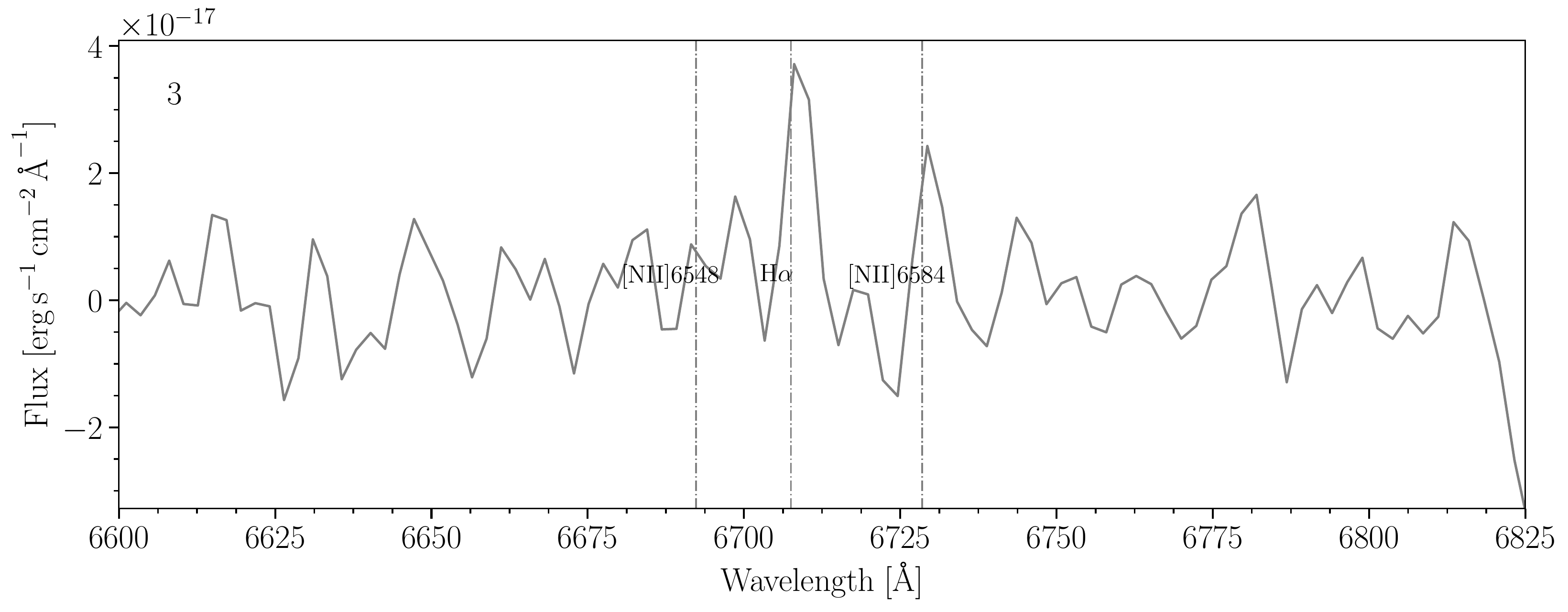}\\ \vspace{-0.6cm}
    \includegraphics[width=.8\columnwidth]{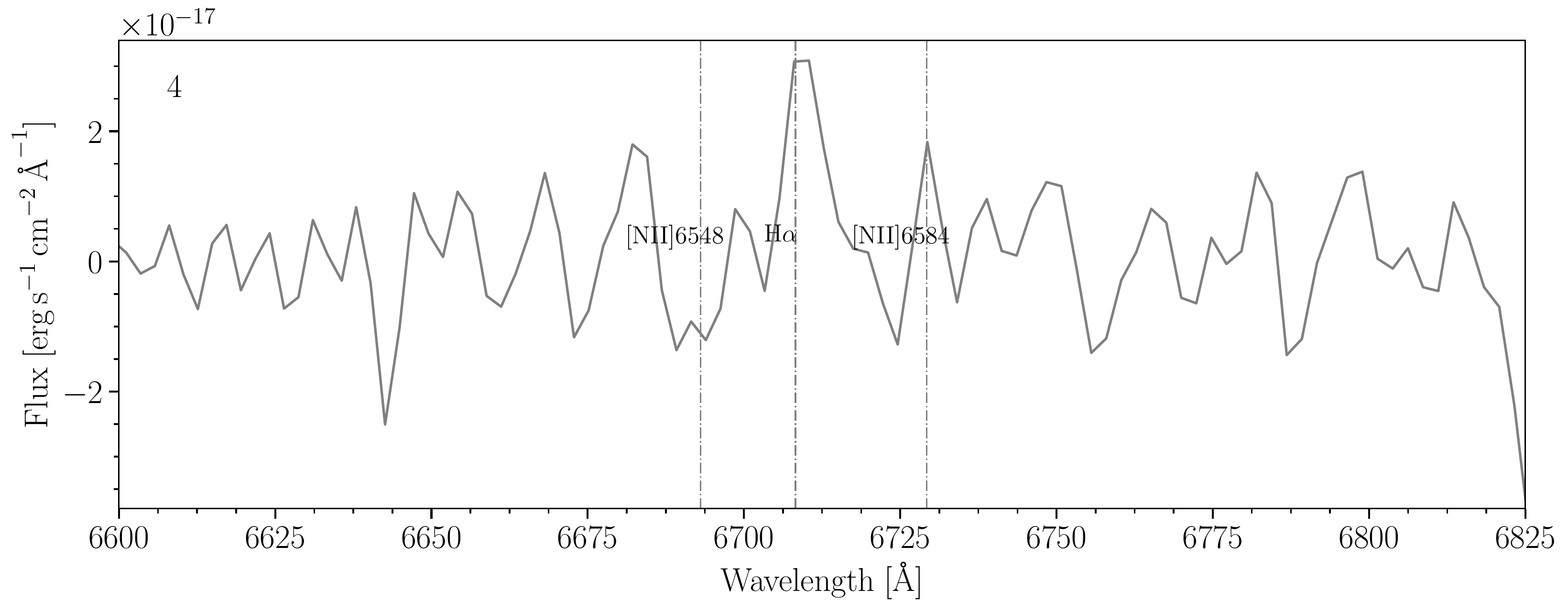}\\ \vspace{-0.6cm}
    \includegraphics[width=.8\columnwidth]{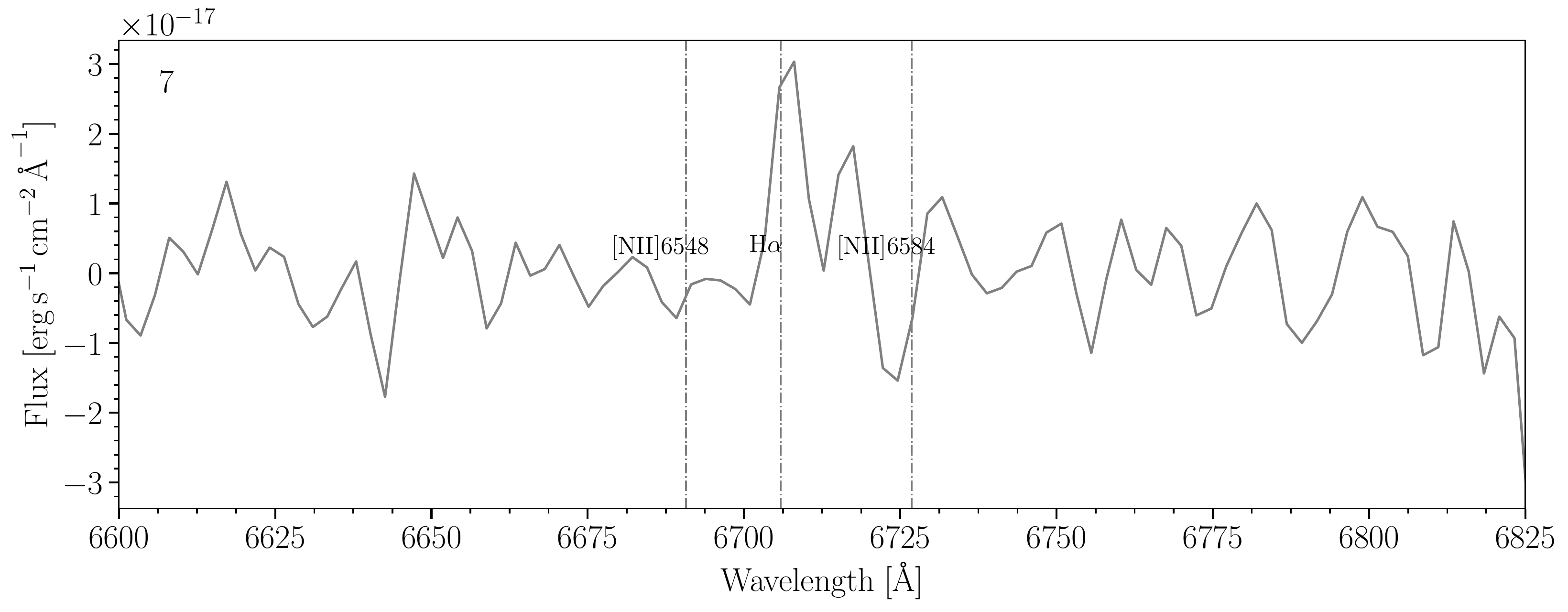}\\ \vspace{-0.6cm}
    \includegraphics[width=.8\columnwidth]{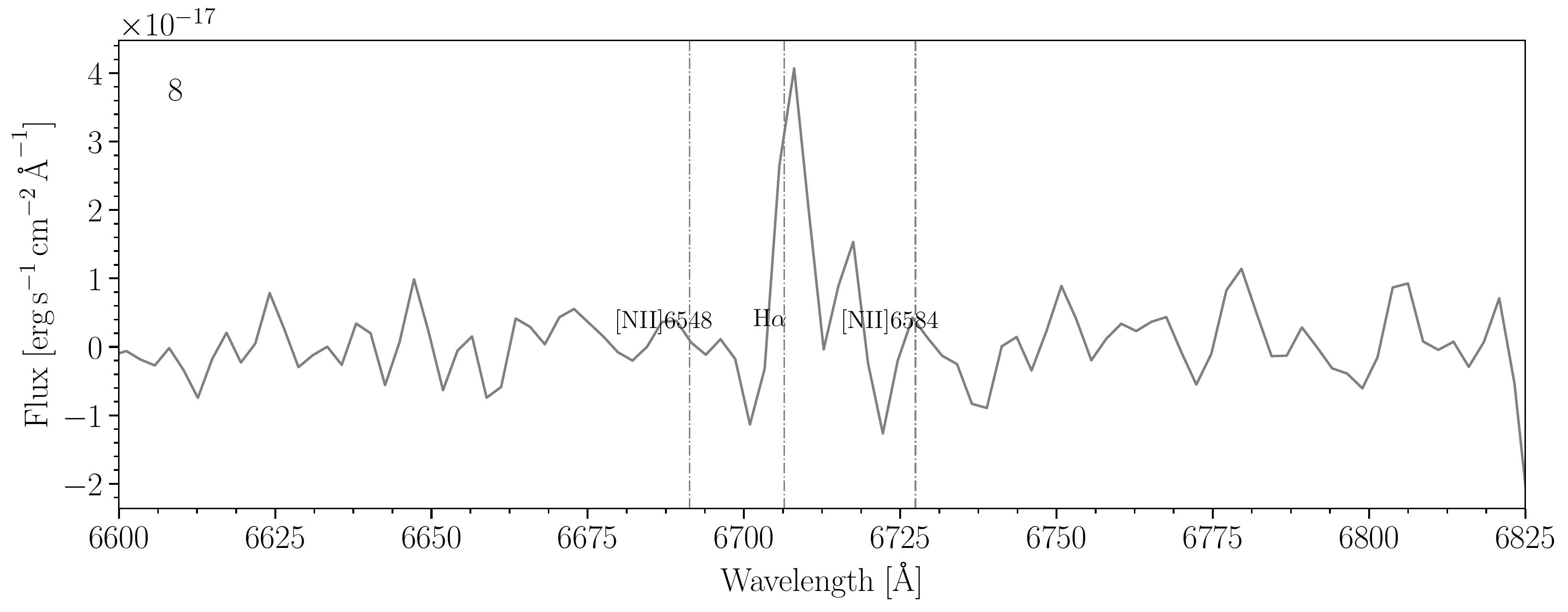}\\
    \caption{Spectra from NGC7320c nucleus, and several H$\alpha$ regions in its galactic disc. Shown from top to bottom are the NGC7320c galactic nucleus and regions 2, 3, 4, 7, and 8 (see Fig.~\ref{fig:NG7320c}). The black vertical dashed lines show the position of the emission lines studied for the SN3 filter at an H$\alpha$ radial velocity of $\sim$6590 $km\, s^{-1}$.}
    \label{fig:spectroNGC7320c}
\end{figure}

Two large groups of radial velocities were found (radial velocities $\leq$ 6160 $km\, s^{-1}$ and > 6160 $km\, s^{-1}$). We found five radial velocity components in SQ: i) v=[5600-5900] $km\, s^{-1}$, associated with NI and SDR; ii) v=[5900-6100] $km\, s^{-1}$, associated with the NSQA and SSQA and the strands that are connected to these zones; iii) v=[6100-6600] $km\, s^{-1}$, associated with the strands located in the shock; iv) v=[6600-6800] $km\, s^{-1}$, associated with YTT, SQA, NGC7319, and the north lobe; and v) v=[6800-7000] $km\, s^{-1}$, associated with the strands that connect the shock with SQA.

When considering a contrast (H$\alpha$)$\geq$2 and binning 6x6, we do not detect gas in the old tail, in the vicinity of NGC7318A nor in NGC7317. We also do not have evidence that there is any connection between i) the north lobe in NGC7319 and the left side of SQA; or ii) the shock and the right part of NI nor SDR. Conversely, we find a link between the shock and the NGC7319 AGN nucleus (H$\alpha$ bridge). This bridge was detected in other wavelengths \citep[e.g.][]{2001ApJ...550..204I,2001AJ....122.2993S,2005A&A...444..697T,2010ApJ...710..248C,2010A&A...518A..59G,2012ApJ...749..158G,2013ApJ...777...66A,2017ApJ...836...76A}.

A region of remarkable line emission is located between NGC7319 and the NI galaxy (NGC7318B). A rectangular area has been defined (see Sect.~\ref{subsec:broad}, shown in Fig.~\ref{fig:halphaSQ}) to include this emission which is associated with the LSSR. This has allowed us to investigate the complex kinematical nature of this emitting gas, making use of detailed spatially resolved multi-sincgaussian fitting. The LSSR has revealed a rich knotty substructure, spatially distributed, and also a set of kinematical properties of the H$\alpha$ emitting knots. These knots show line profiles that have been fitted by two sincgaussian functions, with central velocities clustering around two main radial velocity values ($\sim$6000 $km\, s^{-1}$, $\sim$6600 $km\, s^{-1}$), and a different broadness in each case. Overall, taking into account: i) an oblique interaction over a large region, and new and preceding gas \citep[e.g.][]{2009ApJ...701.1560O,2013A&A...550A.106L,2017ApJ...836...76A}; and ii) that the gas is excited by shocks (corresponding to a shock radial velocity $\sim$300 $km\,s^{-1}$ according to \citealt{2008ApJS..178...20A}, see Duarte Puertas et al. paper II, in prep.), we suggest that a variety of interaction processes occurred previous to the intrusion of NI in the main group. We found sub-structures $\sim$400pc or larger in size, typical of giant HII extragalactic regions.

A possible outermost tail of H$\alpha$ emitting regions is suggested here, which would be in agreement with \cite{2010ApJ...724...80R}. We provide an observational clue of this prediction. We found that this tidal tail connects with NSQA (reg 145). Other strands connected with NSQA have been found: i) the L2 strand that connects NSQA and SSQA; ii) the L3 strand that connects NSQA with region 161 up to radial velocities $\sim$6470 $km\, s^{-1}$; and iii) the L4 strand that connects NSQA with regions 154 and 155. This last connection could be extended up to the NI5 strand according to the diffuse gas show in the low surface brightness gas emission.

From our sample of SQ H$\alpha$ emission regions we have studied the region 176 (i.e. NG) in detail (see Sect.~\ref{subsec:glx_SQ}). This region was previously observed by \cite{2009A&A...507..723T} (see their Fig. 14, ID 35) and by \cite{2014MNRAS.442..495R}. In this paper we have made a spatially resolved study of the H$\alpha$ radial velocity and H$\alpha$ flux in this region. The data points out that this region is in fact a dwarf galaxy with a velocity gradient across its disc ($\pm$60 $km\, s^{-1}$, see Fig.~\ref{fig:NG}), which was not considered by \cite{2014MNRAS.442..495R}. These authors found an age of $\sim$10$^{10}$ yr for 80\% of the stellar mass in this region, assuming that this region is a star cluster separated from a galaxy by some previous interactions. According to the Tully-Fisher relation, we found that NG would be similar to M82. Additionally, we found H$\alpha$ emission in the NGC7320c galactic disc and a significant presence of broad [\ion{N}{ii}]$\lambda$6584 emission in the NGC7320c nucleus. This leads us to propose that NGC7320c hosts an AGN (see Fig~\ref{fig:spectroNGC7320c}). 

\begin{figure}[h]
    \centering
    \includegraphics[width=\columnwidth]{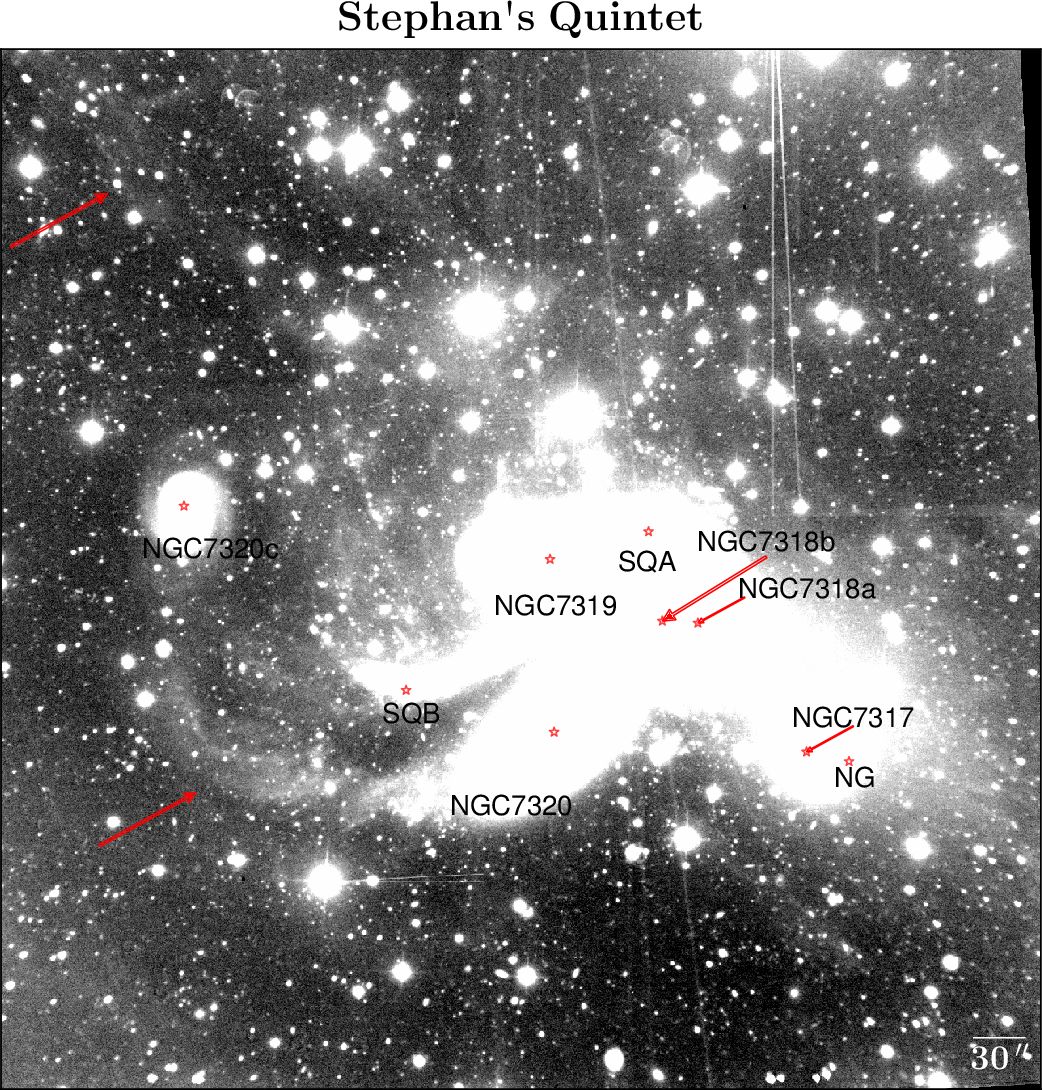}
    \caption{Deep imaging from SN2 filter of SQ. This image displays the full extension of the entire system, including the diffuse ionised gas halo around the SQ. The most relevant components are indicated by their labels and red arrows. See the text for details.}
    \label{fig:Deep_ima_SQ}
\end{figure}

A panoramic view of the full extent of the low surface brightness emission of the SQ is illustrated in Fig.~\ref{fig:Deep_ima_SQ}, which shows a deep imaging of SQ obtained from the SITELLE SN2 filter. This figure displays the full extension and complexity of the entire system (including the diffuse ionised gas halo around the SQ, the old tail, the halo around NGC7317, and a conspicuously low surface brightness extension to the north-east of the whole system). For details, the reader is referred to the recent work by \cite{2018MNRAS.475L..40D}, which offers an interesting study of the complete system.

\begin{figure*}[h]
    \centering
    \includegraphics[width=\textwidth]{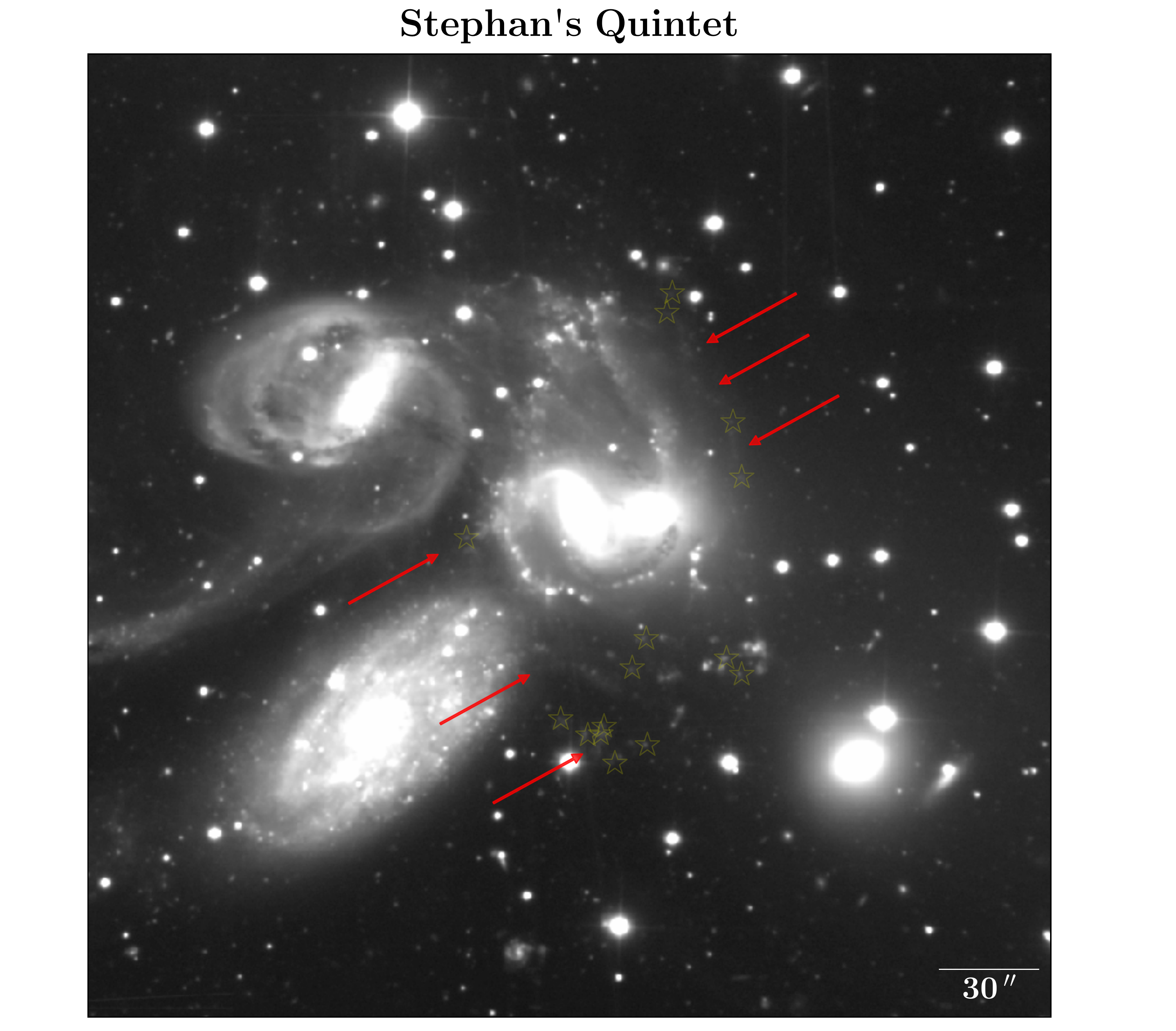}
    \caption{Deep imaging from SN2 filter of SQ. Red arrows point out the possible outer structures that can be connected. The yellow stars show the location of the 15 extra H$\alpha$ emitter regions presented in Table~\ref{table:table3}.}
    \label{fig:Deep_ima2_SQ}
\end{figure*}

Figure~\ref{fig:Deep_ima2_SQ} shows the deep image from the SN2 filter. In this figure, we suggest a possible tidal structure (delineated by an outer rim to the north-west of NGC7318B/A) with diffuse gas emission that could connect the south part of L3 and NSQA with NI4, NI5, and SDR (marked by a green arrow). The detection of the 15 additional low surface brightness H$\alpha$ emission regions (i.e. IDs from 177 to 191 in Table~\ref{table:table3}, located in: SDR, close to NI, and to the west of SQ) with the expected radial velocities in this zone ($\sim$5600 to $\sim$5900 $km\, s^{-1}$) has allowed us to support this possible connection. This outer rim is almost parallel with another inner rim (i.e. strand L4) that shows H$\alpha$ emitting regions. Conversely, we do not discard the possibility of a connection between NI1 (i.e. region 38) and the left side of SDR going behind NGC7320 (indicated by red arrows). Also this interaction could be supported by the existence of region 31 (v$\sim$5860 $km\, s^{-1}$) presenting the same radial velocity as the strand NI1 some 13 kpc away to the east. The interaction suffered, which might explain these rims, appears complex, although we believe that it must have been produced by the interaction between NI and the group. The large-scale outer rim winding the NGC7318B/A system clockwise north-west to south-east has been highlighted in continuum and from H$\alpha$. This structure may be reminiscent of a previously proposed scenario for SQ as a sequence of individual interactions \citep{2010ApJ...724...80R}. Further observations are needed with higher spectral resolution to distinguish the complex kinematics that is taking place in the LSSR.

\begin{acknowledgements}
We thank the anonymous referee for very constructive suggestions that have improved this manuscript. Based on observations obtained with SITELLE, a joint project of Universit\'e Laval, ABB, Universit\'e de Montr\'eal, and the Canada-France-Hawaii Telescope (CFHT), which is operated by the National Research Council of Canada, the Institut National des Sciences de l'Univers of the Centre National de la Recherche Scientifique of France, and the University of Hawaii. The authors wish to recognise and acknowledge the very significant cultural role that the summit of Mauna Kea has always had within the indigenous Hawaiian community. We are most grateful to have the opportunity to conduct observations from this mountain. SDP, JIP, JVM, and CK acknowledge financial support from the Spanish Ministerio de Econom\'ia y Competitividad under grants AYA2013-47742-C4-1-P and AYA2016-79724-C4-4-P, from Junta de Andaluc\'ia Excellence Project PEX2011-FQM-7058, and also acknowledge support from the State Agency for Research of the Spanish MCIU through the 'Center of Excellence Severo Ochoa' award for the Instituto de Astrof\'isica de Andaluc\'ia (SEV-2017-0709). LD is grateful to the Natural Sciences and Engineering Research Council of Canada, the Fonds de Recherche du Qu\'ebec, and the Canada Foundation for Innovation for funding. \\

This research made use of Python ({\tt \href{http://www.python.org}{http://www.python.org}}) and IPython \citep{PER-GRA:2007}; Numpy \citep{2011arXiv1102.1523V}; Pandas \citep{mckinneyprocscipy2010}; of Matplotlib \citep{Hunter:2007}, a suite of open-source Python modules that provides a framework for creating scientific plots. This research made use of Astropy, a community-developed core Python package for Astronomy \citep{2013A&A...558A..33A}. The Astropy web site is {\tt \href{http://www.astropy.org/}{http://www.astropy.org}}. This research made use of astrodendro, a Python package to compute dendrograms of astronomical data ({\tt \href{http://www.dendrograms.org/}{http://www.dendrograms.org/}})\\

\end{acknowledgements}

\bibliography{SQ_publication.bib}

\begin{appendix}
\section{15 H$\alpha$ emitters in outer debris regions}
\label{append:app1}

A close visual inspection, guided by the HST/WFC3 images (PID 11502, PI Keith S. Noll) of the SQ outskirts of our deep SITELLE images has allowed us to define a set of 15 extra H$\alpha$ regions (located in SDR, near NI, and west of SQ). In Table~\ref{table:table3} we show the relevant information for these H$\alpha$ emitters for the regions from ID 177 to 191.

\begin{table}[t]
\tiny
\caption{Positional information of 15 extra H$\alpha$ emitter regions in the southern debris region, close to the NI, and to the west of SQ.}
\label{table:table3}
\centering
\begin{tabular}[t]{c | l l }
\hline\hline  \\[-2ex]
(1) & $\ \ \ \ \ \ \ \ \ \ \ \ \ \ \ $ (2) & (3) \\[0.5ex] 
Region & $\ \ \ \ \ \ \ \ \alpha\ \ \ \ \ \ \ \ \ \ \ \ \ \ \ \ \delta$ & Velocity \\[0.5ex]
ID & $\ \ $ (h m s) $\ \ \ \ \ \ $ ($^o\ ^\prime\ ^{\prime\prime}$) & ($km\, s^{-1}$) \\[0.5ex] 
\hline\\[-2ex]
177 & 22 35 58.8        +33 56 55.4 & 5698 \\[0.5ex]
178 & 22 35 58.1        +33 56 50.5 & 5695 \\[0.5ex]
179 & 22 35 57.8        +33 56 53.0 & 5756 \\[0.5ex]
180 & 22 35 57.8        +33 56 50.8 & 5590 \\[0.5ex]
181 & 22 35 57.5        +33 56 42.3 & 5714 \\[0.5ex]
182 & 22 35 56.7        +33 56 47.5 & 5748 \\[0.5ex]
183 & 22 35 57.1        +33 57 10.0 & 5690 \\[0.5ex]
184 & 22 35 56.7        +33 57 18.5 & 5714 \\[0.5ex]
185 & 22 35 54.8        +33 57 12.5 & 5703 \\[0.5ex]
186 & 22 35 54.5        +33 57 7.7 & 5754 \\[0.5ex]
187 & 22 35 54.4        +33 58 5.7 & 5912 \\[0.5ex]
188 & 22 35 56.0        +33 59 0.9 & 6401 \\[0.5ex]
189 & 22 35 56.1        +33 58 55.1 & 6051 \\[0.5ex]
190 & 22 35 54.6        +33 58 22.1 & 5867 \\[0.5ex]
191 & 22 36 1.0         +33 57 49.0 & 5824 \\[0.5ex]
\hline
\end{tabular}
\tablefoot{The columns correspond to: 
(1) Identifier of the H$\alpha$ emission regions (ID); 
(2) Right ascension (hours, minutes, and seconds) and declination (degrees, arcminutes, and arcseconds);
(3) Velocity ($km\, s^{-1}$). Information about the 15 extra H$\alpha$ emitter regions is given in SDR, close to the NI, and to the west of SQ.
}
\end{table}
\end{appendix}

\end{document}